%% file: main.tex
\documentclass{aa}  

\usepackage{subfig}
\usepackage{xspace}
\usepackage{amsmath}
\usepackage{graphicx}
\usepackage{natbib}
\usepackage{lscape}

\usepackage{multirow}
\usepackage{todonotes}

\begin{document}
\newcommand{\rr}[1]{{\textcolor{red}{[rogerio: #1]}}}

\newcommand{\um}{\ensuremath{\mu\mathrm{m}}}
\newcommand{\lnir}{\ensuremath{L_{\rm AGN}^{\rm NIR}}}
\newcommand{\lmir}{\ensuremath{L_{\rm AGN}^{\rm MIR}}}
\newcommand{\lx}{\ensuremath{L_{\rm AGN}^{\rm X}}}
\newcommand{\oiv}{\ensuremath{{\rm [O IV]} \lambda 25.89 \um{}}}
\newcommand{\loiv}{\ensuremath{L_{\rm AGN}^{\rm [OIV]}}}
\newcommand{\fnir}{\ensuremath{f_{\rm AGN}^{\rm NIR}}}
\newcommand{\brg}{\ensuremath{{\rm Br}\gamma}}
\newcommand{\halpha}{\ensuremath{{\rm H}\alpha}}
\newcommand{\hbeta}{\ensuremath{{\rm H}\beta}}
\newcommand{\STARLIGHT}{\textsc{STARLIGHT}}

\title{LLAMA: Stellar populations in the nuclei of ultra hard X-ray selected AGN and matched inactive galaxies}
\titlerunning{Nuclear stellar populations in local, luminous AGNs}

   \author{L. Burtscher\inst{1}
            \and
            R. I. Davies\inst{2}
            \and
            T. T. Shimizu\inst{2}
            \and
            R. Riffel\inst{3}
            \and
            D. J. Rosario\inst{4}
            \and
            E. K. S. Hicks\inst{5}
            \and
            M.-Y. Lin\inst{6}
            \and
            R. A. Riffel\inst{7}
            \and
            M. Schartmann\inst{8}
            \and
            A. Schnorr-M\"{u}ller\inst{3}
            \and
            T. Storchi-Bergmann\inst{3}
            \and
            G. Orban de Xivry\inst{9}
            \and
            S. Veilleux\inst{10}
            }

    \institute{Leiden Observatory, PO Box 9513, 2300 RA, Leiden,  \\ 
                The Netherlands \email{burtscher@strw.leidenuniv.nl}
                \and Max-Planck-Institut f\"ur extraterrestrische Physik, Postfach 1312, 85741, Garching, Germany 
                \and Departamento de Astronomia, Universidade Federal do Rio Grande do Sul, IF, CP 15051, 91501-970 Porto Alegre, RS, Brazil 
                \and Centre for Extragalactic Astronomy, Department of Physics, Durham University, South Road, Durham DH1 3LE, UK
                \and Department of Physics \& Astronomy, University of Alaska Anchorage, AK 99508-4664, USA
                \and Institute of Astronomy and Astrophysics, Academia Sinica, 11F of AS/NTU Astronomy-Mathematics Building, No.1, Sec. 4, Roosevelt Rd, Taipei 10617, Taiwan
                \and Departamento de F\'{i}sica/CCNE, Universidade Federal de Santa Maria, 97105-900, Santa Maria, RS, Brazil
                \and OmegaLambdaTec GmbH, Lichtenbergstr. 8, 85748 Garching
                \and Space Sciences, Technologies, and Astrophysics Research Institute, Universit\'{e} de Li\'{e}ge, 4000 Sart Tilman, Belgium
                \and Department of Astronomy and Joint Space-Science Institute, University of Maryland, College Park, Maryland 20742, USA
            }

\authorrunning{L.~Burtscher et al.}

\date{Draft version, \today}


\abstract{The relation between nuclear ($\lesssim$ 50 pc) star formation and nuclear galactic activity is still elusive: theoretical models predict a link between the two, but it is unclear whether active galactic nuclei (AGNs) should appear at the same time, before or after nuclear star formation activity is ongoing. We present a study of this relation in a complete, volume-limited sample of nine of the most luminous ($\log L_{\rm 14-195 keV} > 10^{42.5}$ erg/s) local AGNs (the LLAMA sample), including a sample of 18 inactive control galaxies (6 star-forming; 12 passive) that are matched by Hubble type, stellar mass (9.5 $\lesssim \log M_{\star}/M_{\sun} \lesssim$ 10.5), inclination and distance. This allows us to calibrate our methods on the control sample and perform a differential analysis between the AGN and control samples.\\
We perform stellar population synthesis on VLT/X-SHOOTER spectra in an aperture corresponding to a physical radius of $\approx$ 150 pc. We find young ($\lesssim$ 30 Myr) stellar populations in seven out of nine AGNs and in four out of six star-forming control galaxies. In the non-star-forming control population, in contrast, only two out of twelve galaxies show such a population. We further show that these young populations are not indicative of ongoing star-formation, providing evidence for models that see AGN activity as a consequence of nuclear star formation.\\
Based on the similar nuclear star-formation histories of AGNs and star-forming control galaxies, we speculate that the latter may turn into the former for some fraction of their time. Under this assumption, and making use of the volume-completeness of our sample, we infer that the AGN phase lasts for about 5 \% of the nuclear starburst phase.}

\keywords{Galaxies: active -- Galaxies: evolution -- Galaxies: nuclei -- Galaxies: Seyfert -- Galaxies: stellar content -- Techniques: spectroscopic}

\maketitle

\input{sections/1-introduction.tex}
\input{sections/2-sample_obs_reduction.tex}
\input{sections/3-sps}
\input{sections/4-sps-results}

\input{sections/5-emission-lines}
\input{sections/6-sample-discussion}
\input{sections/7-conclusions.tex}

\begin{acknowledgements}
The authors would like to thank the anonymous referee for comments that improved the paper.\\
Based on observations made with ESO Telescopes at the La Silla Paranal Observatory under programme IDs 092.B-0083 and 095.B-0059.\\
L.B. would like to thank Jarle Brinchmann, Anna Feltre, Reynier Peletier, Almudena Prieto, Marc Sarzi, and Scott Trager for helpful discussions and Roberto Cid Fernandes for discussions and for providing a special version of STARLIGHT that allowed us to investigate the posterior probability distribution functions ourselves. L.B. has been partly supported by a DFG grant at the Max-Planck-Institute for Extraterrestrial Physics within the SPP 1573 ``Physics of the interstellar medium''.
RR thanks CNPq, CAPES and FAPERGS for partial financial support.
R.A.R. acknowledges financial support from CNPq (302280/2019-7 ) and FAPERGS (17/2551-0001144-9).
DJR acknowledges support from STFC (ST/T000244/1).
ASM thanks CNPq for partial financial support.
\end{acknowledgements}

\appendix{}
\input{sections/a1-ssp_fits_appendix.tex}

\clearpage
\bibliographystyle{aa}
\bibliography{apj-jour,references}

\clearpage

\end{document}

%% file: sections/1-introduction.tex
\section{Introduction}
One of the most controversial topics in galaxy evolution is the so-called ``co-evolution'' \citep[e.g.][]{heckman2014} between star formation in galaxies and the accretion activity onto their central super-massive black hole (SMBH). We know that the two processes are related on cosmological timescales \citep{madau2014}. In addition, most cosmological models require ``feedback'', i.e. regulation of the molecular mass reservoir from the central Active Galactic Nucleus (AGN), in order to reproduce observed properties of our universe such as the cosmic star formation history, the galaxy luminosity function, and the low baryon fraction in massive dark matter halos \citep{somerville2008,vogelsberger2014,su2019}, but also the obscuration statistics of AGN \citep{hopkins2016}.

The products of both AGN accretion and star-formation are related over many orders of magnitude, as evidenced in the well known relation between the mass of the SMBH and the velocity dispersion of stars in the bulge \citep{magorrian1998,ferrarese2000,gebhardt2000,kormendy2013} and this may be further evidence of a direct link between the two processes \citep[but see][for an alternative explanation]{jahnke2011}.

Observers have struggled for many years, however, to pin down the specific impact that AGN activity might have on star formation or vice versa. This is partly because galaxy mass is the most significant driver of both AGN activity \citep[e.g.][]{kauffmann2003c} and star-formation rate \citep[e.g.][]{speagle2014} and also because the apparent relation between AGN and starburst activity may in fact be driven by the presence of large gas reservoirs, which are relevant for both processes \citep[e.g.][]{heller1994}. This is seen particularly when comparing star formation enhancement in AGNs against mass-matched {\em star-forming} non-AGNs. There does not seem to be a significant difference between these systems in terms of star-formation rate \citep{silverman2009,rosario2012,rovilos2012,santini2012}. And while AGNs clearly drive outflows \citep[e.g.][]{fabian2012}, it is unclear if and how these affect the molecular gas reservoirs \citep{schulze2019}. ``Positive'' AGN feedback (e.g. radio jets triggering star formation) may also be possible \citep[e.g.][]{zinn2013,zubovas2013,maiolino2017,gallagher2019}. For further discussion on this topic we refer to the recent review by \citet{harrison2017}.

It is also still unclear what triggers nuclear activity in galaxies \citep[e.g.][]{alexander2012} -- and this is not due to a lack of ideas: Mergers may \citep[e.g.][]{ramosalmeida2012,treister2012,gao2020} or may not \citep[e.g.][]{cisternas2011,kocevski2012,marian2019} trigger AGNs, but are in any case only part of the answer \citep{marian2020}. The galactic environment may play a role, too, for AGN fueling. \citet{davies2017} found that the group environment is most conducive to X-ray selected AGNs. Bar instabilities were thought to play a role for AGN feeding in the past \citep[e.g.][]{knapen2000}, but more recent work seems to not find this relation \citep{cisternas2015,goulding2017}. Dust lanes correlate with AGN activity, at least in gas-poor galaxies \citep{martini2003b,simoeslopes2007}, and may show accretion structures from kiloparsec down to parsec scales \citep{prieto2019}.

The reason why it remains difficult to pin down which, if any, of these processes is the most relevant fueling mechanism for AGNs is due to their very different timescales compared to AGN activity. AGN activity is likely intermittent on timescales of $10^5$ -- $10^6$ years as predicted by simulations \citep[e.g.][]{novak2011} and also evidenced by e.g. the number statistics of ``starting'' AGNs, so-called X-ray bright, optically normal galaxies \citep{schawinski2015} or from observations of intermittent outflows \citep{lutz2020}. The timescale for star formation in the host galaxy, however, is much longer. Typically one assumes a galaxy's rotation period or dynamical time of $\sim 10^8$ years. This mismatch in timescales makes it hard to relate nuclear accretion to global star formation \citep{hickox2014,volonteri2015}.

A much better correlation is expected on smaller scales, however, from simulations \citep{thompson2005,hopkins2010b} as well as observationally: \citet{diamondstanic2012a} find that black hole growth strongly correlates with nuclear ($r < 1$ kpc) star formation, but only weakly with extended ($r > 1$ kpc) star formation. Star formation in ``nuclear rings'' appears on radii of several hundred pc up to $\sim$ 1.5 kpc in some 20\% of disk galaxies \citep{regan2003,comeron2010}. It is often said that the processes of star formation and AGN activity take place on scales of many kpc and sub-pc scales, respectively. This is an exaggeration, however, as the gravitational potential of the black hole becomes dominant already at scales of $\sim$ 1--10 pc \citep[e.g.][]{schartmann2010}.

These are the scales where some 80\% of both early and late type galaxies exhibit ``Nuclear Stellar Clusters'' (NSCs) which can also contain young stars \citep{neumayer2020}. Their relation to AGN feeding is still an active matter of research.

The circum-nuclear environment is also the location of the AGN ``torus'' as postulated in the original AGN unification model \citep{antonucci1985}. Due to its geometrically thick structure, it is supposed to explain the dichotomy in observed AGN types -- broad-line (type 1) and narrow-line (type 2) AGNs. It had been realised early on, however that puffing up a thick dusty structure, and keeping it geometrically thick, is quite difficult \citep{krolik1988}. This ``scale height problem'' is solved in more recent unification models by constructing the torus as a disk-wind puffed up by infrared radiation pressure on smallest scales only \citep[e.g.][]{elvis2000,hoenig2019}, and the disk-wind driven outflow also explains the polar-elongated mid-IR morphologies of the AGN heated dust on parsec scales \citep{burtscher2013,hoenig2013,lopezgonzaga2016a}.

There is plenty of material in the ``torus'' plane as well, however, and it can certainly contribute to obscuring the AGN \citep[e.g.][]{hicks2009}. Sub-mm interferometric observations further show that this region is often characterised by counter-rotating gas disks \citep{combes2019}, consistent with simulations that expect various instabilities on these scales \citep[e.g.][]{hopkins2012}. It is clear that the dusty material surrounding the central engine is not just a mere ``by-stander'' as the ``torus'' picture implied, but very likely the reservoir and therefore an active player in fueling the AGN.

A causal link between AGN activity and {\em nuclear} star formation might exist on these scales if infalling gas, stalling at its angular momentum barrier, first forms stars and then fuels the AGN through stellar mass loss \citep{norman1988,vollmer2008,wada2009}, possibly contributing to AGN obscuration in this phase \citep{wada2002}. This fueling might work best {\em after} the violent phase of the starburst (including type II supernovae) has died off and the colliding slow winds of stars in their post-main-sequence evolution can efficiently cancel angular momentum and thus feed the AGN \citep{schartmann2009,schartmann2010}, but in some models the nuclear starburst and AGN accretion happens at the same time \citep[e.g.][]{kawakatu2008}.

Circum-nuclear star formation in AGNs is well established observationally \citep[e.g.][]{terlevich1990,cidfernandes2001b,storchibergmann2000,riffel_r2009,storchibergmann2012,esquej2014,lin2018}, but observational efforts to time nuclear starburst vs. AGN activity are still scarce and inconclusive. \citet{davies2007} presented observational evidence that young stars hinder accretion, and that gas can only accrete efficiently to smaller scales after these early turbulent phases of stellar feedback. While other observations are partially consistent with this, young stars may sometimes also be present during the active phase \citep{cidfernandes2004,sarzi2007b,kauffmann2009}. To resolve this, one must take into account not only the location of the stellar populations \citep{riffel_ra2010a,riffel_r2011,diniz2017}, but also the AGN luminosity (i.e. mass accretion rate). The result of \citet{davies2007} implies that the age of the stellar population is important only for Seyferts with $L_{\rm AGN} \gtrsim 10^{43}$ erg/s, i.e. for accretion at levels $\gtrsim 10^{-3} M_{\odot}$/yr. Other (less efficient) processes may be sufficient to supply lower accretion rates.

It is, however, difficult to characterise the nuclear stellar populations of powerful AGNs in the optical directly since the glare of the AGN reduces the equivalent width of stellar absorption features below the detectable (or calibratable) accuracy. Past studies of star formation or stellar populations in the nuclei of AGNs have therefore either relied on indirect tracers, such as PAHs \citep{imanishi2011,esquej2014,jensen2017,esparzaarredondo2018} or the far-IR \citep{melendez2014}, or restricted themselves to type 2 AGNs \citep[e.g.][]{sarzi2007b}. An interesting alternative is to look in the near-IR where both the glare of the AGN as well as the extinction of the nucleus are significantly reduced compared to the optical. The $J,H$ and $K$ bands offer a number of useful stellar features, particularly of stars in their post-AGB phase which may be well suited to fuel the AGN \citep{riffel_r2009,riffel_r2015,riffel_r2019}.

Finally, in order to calibrate the results, it is important to not just look at the nuclear stellar populations of AGNs, but also -- and with the same resolution and method -- at inactive galaxies that are matched especially in stellar mass. Such a comparison is the main aim of the {\em Local Luminous AGNs and Matched Analogs} project \citep{davies2015} that has already seen a number of publications, e.g. on nuclear obscuration \citep{burtscher2015,burtscher2016,schnorrmueller2016}, nuclear kinematics and luminosity distributions \citep{lin2018}, the molecular gas content \citep{rosario2018}, the environment \citep{davies2017}, the density of AGN-driven outflows \citep{davies2020}, and the $M_{\rm BH}-\sigma_{\star}$ relation of LLAMA galaxies \citep{caglar2020}. In addition a few ``special'' sources have also been studied in more detail: NGC~5728 \citep{shimizu2019} and NGC~2110 \citep{rosario2019}.

In this article, we analyse the nuclear star formation histories of AGNs and matched inactive galaxies of the LLAMA sample, based on stellar population synthesis of VLT/X-SHOOTER spectra. After describing the sample, the observations and the data reduction in Section~\ref{sec:sample}, we present the stellar population synthesis in Section~\ref{sec:sps}. We analyse its results in Section~\ref{sec:sps_results} and discuss the emission line diagnostics in Section~\ref{sec:bpt_analysis}. In Section~\ref{sec:discussion}, we interpret our results in terms of the sample statistics and compare them to previous works. We end with conclusions and an outlook in Section~\ref{sec:conclusions}. The Appendix~\ref{sec:appendix:fits} shows all individual galaxy fits.

%% file: sections/2-sample_obs_reduction.tex
\section{Sample, Observations, and Data Reduction}
\label{sec:sample}
\subsection{The LLAMA sample of type 2 AGNs and control galaxies}

\begin{table*}
\caption{Definition of our sample}
\label{tab:sample}
\centering
\begin{tabular}{cccccccc}
\hline
\hline
Name & optical type & $\log L_{14-195}$ & Hubble stage & axis ratio & distance & redshift \\
&&[erg s$^{-1}$]&&&[Mpc]&\\
\hline
NGC 2110 & Sy 2 (1h) & 43.64 & -3.0 & 0.74 & 34 & 0.007789 \\
NGC 2992 & Sy 1.8 & 42.62 & 1.0 & 0.3 & 36 & 0.00771 \\
MCG-05-23-016 & Sy 1.9 & 43.47 & -1.0 & 0.45 & 35 & 0.008486 \\
NGC 3081 & Sy 2 (1h) & 43.06 & 0.0 & 0.77 & 34 & 0.007976 \\
ESO 021-G004 & Sy 2 & 42.49 & -0.4 & 0.45 & 39 & 0.009841 \\
NGC 5728 & Sy 2 & 43.21 & 1.0 & 0.57 & 39 & 0.009353 \\
ESO 137-G034 & Sy 2 & 42.62 & 0.0 & 0.79 & 35 & 0.009144 \\
NGC 7172 & Sy 2 (1i) & 43.45 & 1.4 & 0.56 & 37 & 0.008683 \\
NGC 7582 & Sy 2 (1i) & 42.67 & 2.0 & 0.42 & 22 & 0.005254 \\
\hline{}
NGC 718 & inactive & -- & 1.0 & 0.87 & 23 & 0.005781 \\
NGC 1079 & inactive & -- & 0.0 & 0.6 & 19 & 0.004843 \\
NGC 1315 & inactive & -- & -1.0 & 0.89 & 21 & 0.005387 \\
NGC 1947 & inactive & -- & -3.0 & 0.87 & 19 & 0.003669 \\
ESO 208-G021 & inactive & -- & -3.0 & 0.7 & 17 & 0.003619 \\
NGC 2775 & inactive & -- & 2.0 & 0.77 & 21 & 0.004503 \\
NGC 3175 & inactive & -- & 1.0 & 0.26 & 14 & 0.003673 \\
NGC 3351 & inactive & -- & 3.0 & 0.93 & 11 & 0.002595 \\
ESO 093-G003 & inactive & -- & 0.3 & 0.6 & 22 & 0.006106 \\
NGC 3717 & inactive & -- & 3.0 & 0.18 & 24 & 0.005781 \\
NGC 3749 & inactive & -- & 1.0 & 0.25 & 42 & 0.009013 \\
NGC 4224 & inactive & -- & 1.0 & 0.35 & 41 & 0.008683 \\
NGC 4254 & inactive & -- & 5.0 & 0.87 & 15 & 0.008029 \\
NGC 5037 & inactive & -- & 1.0 & 0.32 & 35 & 0.006351 \\
NGC 5845 & inactive & -- & -4.6 & 0.63 & 25 & 0.00491 \\
NGC 5921 & inactive & -- & 4.0 & 0.82 & 21 & 0.004937 \\
IC 4653 & inactive & -- & -0.5 & 0.63 & 26 & 0.00517 \\
NGC 7727 & inactive & -- & 1.0 & 0.74 & 26 & 0.006231 \\
\hline
\end{tabular}
\tablefoot{
The table shows name, optical spectral type, observed 14–195 keV luminosity, Hubble stage, axis ratio, distance and redshift from NED. Hard X-ray fluxes are from Swift-BAT \citep{baumgartner2013}. For references and discussion on the latter four columns, see \citet{davies2015}. The optical spectral type is from our own analysis and follows the classification scheme of \cite{veroncetty2010}.
}
\end{table*}

\begin{figure}
\includegraphics[width=\columnwidth]{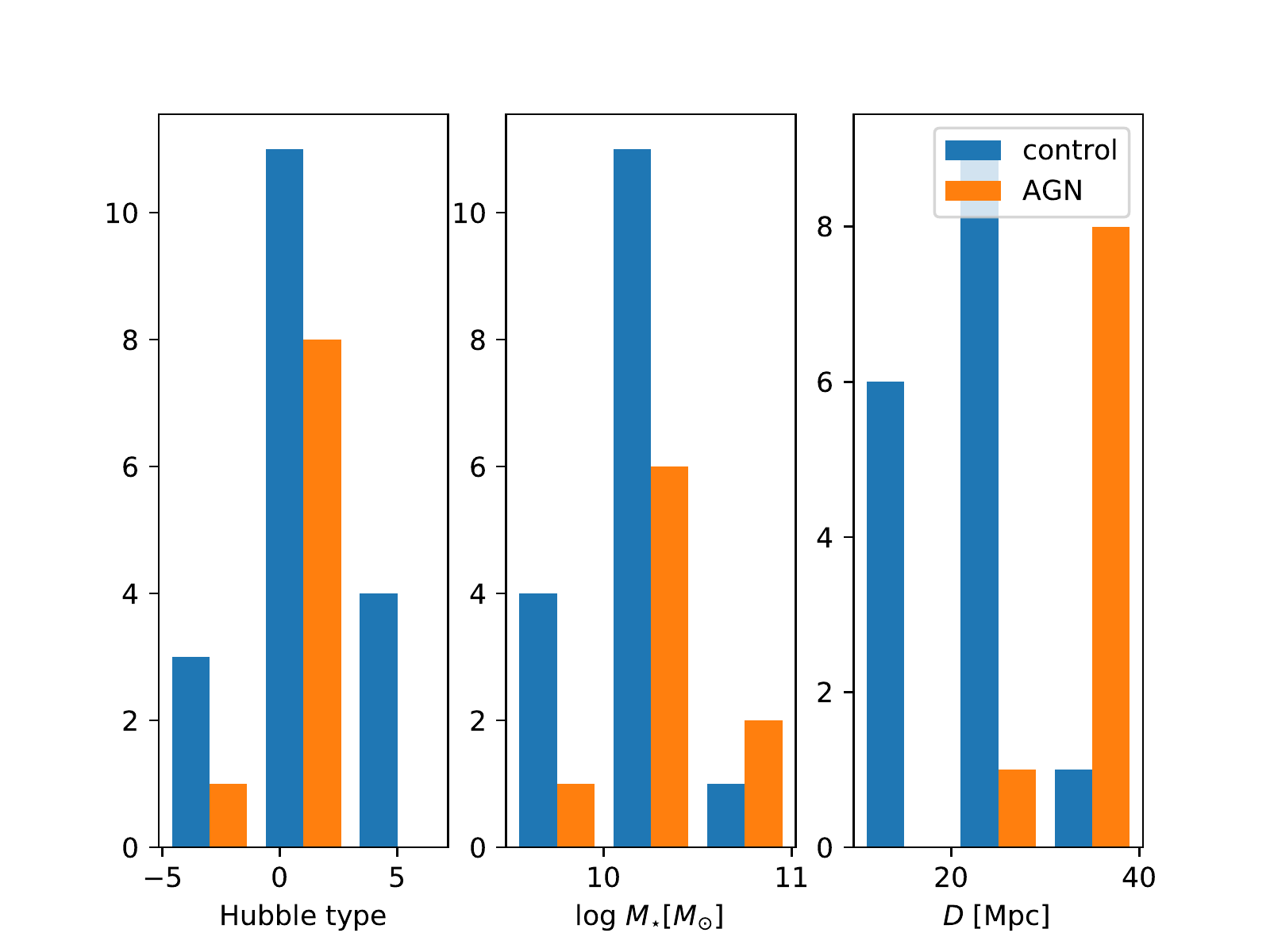}
\caption{\label{fig:sample_compare}Comparison of the most relevant host galaxy properties of our AGN and control sample. From left to right: numerical Hubble type (morphology), log stellar mass $M_{\star}$ in units of solar masses $M_{\odot}$, and distance. The AGN sample is shown in orange, the control sample in blue.}
\end{figure}

The sample -- and control sample -- selection is one of the most important advances of this work, in our view. This work is based on the LLAMA sample \citep{davies2015} which includes 20 AGNs that have been selected from the {\em Switft}/BAT all-sky ultra-hard X-ray (14-195 keV) survey in its 58-month edition. Selecting by ultra-hard X-rays has the advantage of being mostly insensitive to obscuration, except for the most Compton-thick sources. We select the most luminous ($\log L_{\rm X} / ({\rm erg/s}) > 42.5$) sources to limit ourselves to highly accreting ``bona-fide'' AGNs and because they are more likely to be fueled by a single cohesive mechanism. The lower the luminosity, the more options for AGN fueling there are \citep{martini2004}. By further selecting only local ($z < 0.01$, corresponding to distances of $\lesssim 40$ Mpc) AGNs, we are able to distinguish between the nuclear and circum-nuclear stellar populations. In addition, LLAMA includes 19 control galaxies which have been selected in order to match the AGN sample in distance, Hubble type, stellar mass ($H$ band luminosity) and axis ratio (inclination)\footnote{The axis ratio (inclination) match was performed for the dynamical analyses done in other LLAMA projects. For the sample used here the median axis ratio of AGNs and control galaxies is 0.56 and 0.67, respectively. For the stellar population fitting the inclination of the host galaxy does not play a major role, except perhaps for obscuration effects which are discussed separately.}. We consider this matched control sample an integral part of LLAMA as it allows us to calibrate our results against a similar sample of galaxies which only differ from the AGN sample in terms of their nuclear accretion rate. Fig.~\ref{fig:sample_compare} shows the galaxy properties of our AGN and control samples.

\begin{figure}
\includegraphics[width=\columnwidth]{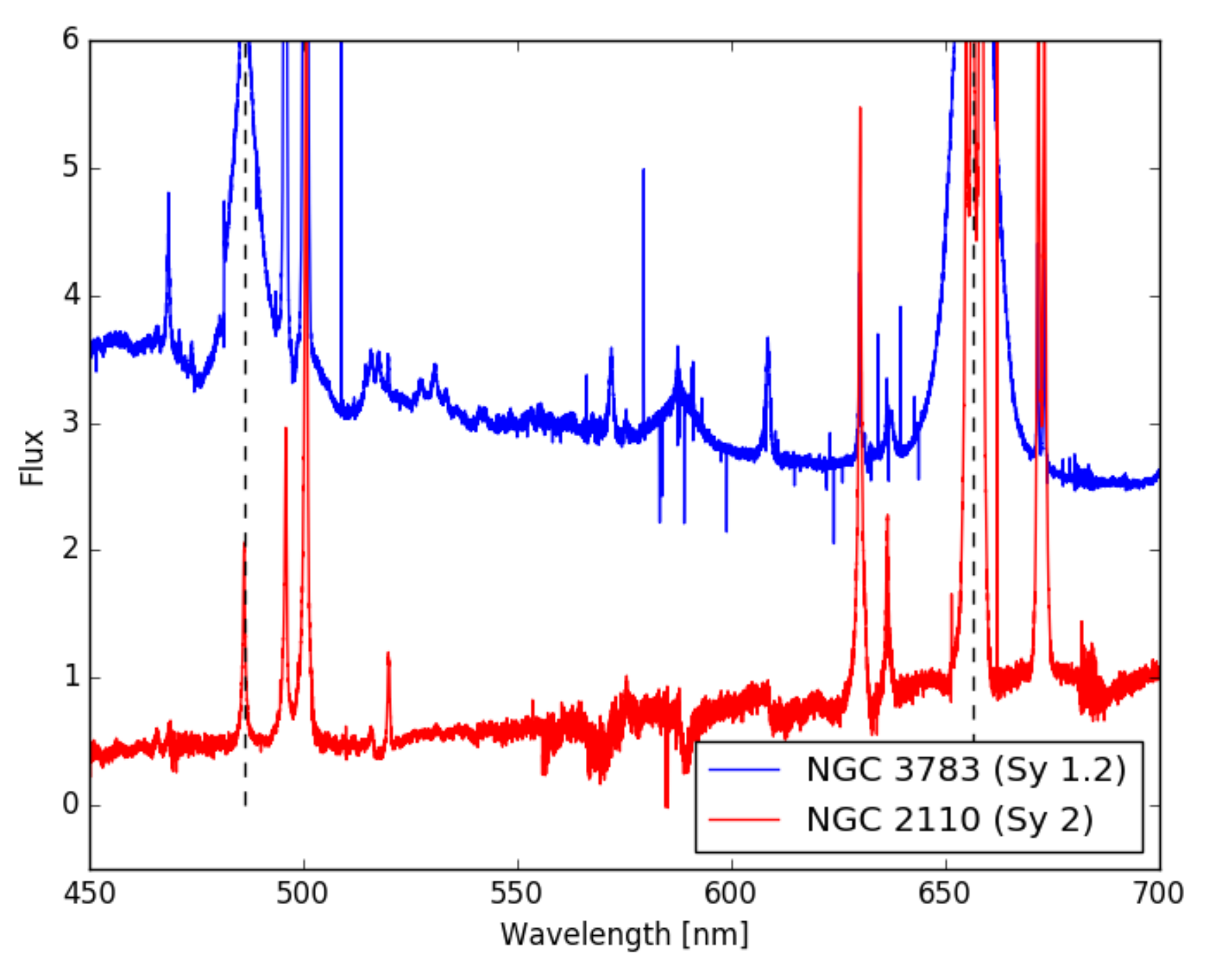}
\caption{\label{fig:Sy1Sy2}Comparison of spectra for a Seyfert 1 and a Seyfert 2 galaxy from the LLAMA sample. A number of very broad emission features can be seen in the Seyfert 1 galaxy that outshine any underlying stellar absorption lines. The Balmer emission lines \halpha{} and \hbeta{} at 6562.8 \AA{} and 4861.3 \AA{}, respectively, are indicated with dashed lines.}
\end{figure}

For this work we limit the AGN-subsample of LLAMA to only the Seyfert 1.8 and more highly obscured AGNs (i.e. Seyfert class $\geq$ 1.8) which we will collectively refer to as type~2 AGNs. This sub-selection ensures that we can cleanly separate the stellar absorption features from AGN emission (see Fig.~\ref{fig:Sy1Sy2}). Our attempts to do this in the type 1 AGNs of LLAMA were not successful. We also had to remove the type~2 AGN NGC~5128 (Centaurus A) since its nucleus is so highly obscured that the nuclear stellar population cannot be determined in the optical. We further excluded the type~2 AGN NGC~1365 since it shows so many emission lines that no good stellar population fit could be obtained either.

\subsection{Observations}
\begin{table}[h!]
\caption{List of all observations}
\label{tab:OBs}
\centering
\begin{tabular}{ccccc}
\hline\hline
OB & night & seeing & par. angle & flag \\
   &       & [''] & [degrees] & \\
\hline
NGC2110\_1 & 2013-11-24 & 0.58 & -148.4 & 1 \\
NGC2110\_2 & 2013-11-24 & 0.5 & 161.8 & 0 \\
NGC2992\_1 & 2014-02-26 & 0.73 & -111.1 & 1 \\
NGC2992\_2 & 2014-02-25 & 0.76 & -111.2 & 0 \\
MCG523\_1 & 2014-01-21 & 1.06 & -77.4 & 1 \\
MCG523\_2 & 2014-01-21 & 0.95 & -26.3 & 0 \\
NGC3081\_1 & 2014-02-19 & 0.58 & -100.0 & 0 \\
NGC3081\_2 & 2014-02-19 & 0.8 & 99.4 & 1 \\
ESO021\_1 & 2016-08-01 & 0.41 & 50.0 & 1 \\
NGC5728\_1 & 2015-05-12 & 0.81 & 134.7 & 1 \\
NGC5728\_2 & 2015-07-15 & 0.78 & 139.4 & 0 \\
ESO137\_1 & 2015-05-18 & 1.03 & -5.8 & 0 \\
ESO137\_2 & 2015-05-20 & 1.54 & -3.9 & 0 \\
ESO137\_3 & 2015-06-23 & 0.8 & -1.8 & 1 \\
NGC7172\_1 & 2015-08-11 & 1.13 & -68.9 & 1 \\
NGC7172\_2 & 2015-08-11 & 1.44 & 11.8 & 0 \\
NGC7582\_1 & 2016-07-26 & 0.38 & -67.8 & 0 \\
NGC7582\_2 & 2016-08-08 & 0.72 & -39.7 & 1 \\
\hline{}
NGC718\_1 & 2015-12-04 & 0.6 & 144.7 & 1 \\
NGC1079\_1 & 2013-11-22 & 0.8 & 79.1 & 1 \\
NGC1315\_1 & 2013-12-10 & 0.89 & -171.5 & 1 \\
NGC1315\_2 & 2013-12-10 & 0.93 & 104.8 & 0 \\
NGC1947\_1 & 2013-12-22 & 0.67 & 48.6 & 1 \\
NGC1947\_2 & 2014-02-07 & 1.21 & 42.5 & 0 \\
ESO208\_1 & 2013-12-11 & 0.61 & -5.2 & 1 \\
ESO208\_2 & 2014-01-21 & 1.05 & -12.9 & 0 \\
NGC2775\_1 & 2015-11-14 & 0.69 & -132.2 & 1 \\
NGC2775\_2 & 2015-12-07 & 0.73 & -149.5 & 0 \\
NGC3175\_1 & 2014-03-08 & 1.49 & -92.5 & 1 \\
NGC3175\_2 & 2014-03-09 & 0.63 & -89.5 & 0 \\
NGC3351\_1 & 2014-02-20 & 1.08 & -150.2 & 1 \\
NGC3351\_2 & 2014-02-20 & 1.06 & -176.5 & 0 \\
ESO093\_1 & 2014-01-21 & 1.44 & -2.2 & 0 \\
ESO093\_2 & 2014-03-20 & -1.0 & -36.0 & 1 \\
NGC3717\_1 & 2014-03-21 & 1.37 & -90.8 & 1 \\
NGC3717\_2 & 2014-03-21 & 1.22 & -78.0 & 0 \\
NGC3749\_1 & 2014-03-21 & 1.58 & -16.9 & 1 \\
NGC3749\_2 & 2014-03-31 & -1.0 & -90.3 & 0 \\
NGC4224\_1 & 2015-05-12 & 0.79 & -169.6 & 1 \\
NGC4254\_1 & 2016-06-01 & 0.42 & -178.1 & 1 \\
NGC5037\_1 & 2015-05-12 & 1.18 & 127.7 & 1 \\
NGC5037\_2 & 2016-02-03 & 0.76 & -115.6 & 0 \\
NGC5845\_1 & 2015-05-22 & -1.0 & -173.0 & 0 \\
NGC5845\_2 & 2016-03-15 & 0.91 & -142.5 & 1 \\
NGC5921\_1 & 2015-06-15 & 0.82 & 147.8 & 1 \\
NGC5921\_2 & 2015-06-23 & 0.86 & -177.1 & 0 \\
IC4653\_1 & 2015-05-18 & 0.93 & 3.8 & 0 \\
IC4653\_3 & 2016-06-03 & 0.55 & -49.7 & 1 \\
NGC7727\_1 & 2015-08-24 & 0.7 & -139.9 & 1 \\
\hline
\end{tabular}
\tablefoot{
Name of ESO observing block (OB), date of night begin, seeing (-1 indicates that the measurement is not available) and parallactic angle at the start of the observation. The last column indicates whether this OB was used (flag = 1) for the further analysis or not (flag = 0). Top: AGNs, bottom: inactive galaxies. Please see Section~\ref{sec:ob_selection} for more on the OB selection.
}
\end{table}

Except for the Seyfert~2 galaxy NGC~4388 and the control galaxy NGC~4260 all galaxies from our volume-complete sample of luminous local type~2 AGNs and control galaxies have been observed at least once with VLT/X-SHOOTER. Together with the explicit exclusions mentioned above, this leaves us with nine type~2 AGNs and 18 control galaxies to be analysed in this work (see Table~\ref{tab:sample}).

For the work presented here, we use UV and optical spectra obtained with X-SHOOTER at the VLT \citep{vernet2011}. The spectra from the near-IR arm of X-SHOOTER require a different treatment in terms of data reduction (e.g. sky emission line correction is non-trivial in the IFU mode of X-SHOOTER) and analysis (e.g. requiring different stellar population models) and will be analysed in a separate project. X-SHOOTER is a long-slit and integral-field unit (IFU) spectrograph which simultaneously records spectra in a wide wavelength range from the atmospheric UV cut-off at $\sim$ 3000\AA{} to the near-infrared $K$ band. It consists of three arms to maximise the sensitivity in the UV, the visible and the near-infrared wavelength range, respectively.

Since we wanted to focus on the nuclear environment of nearby galaxies, we chose to observe in the IFU mode of X-SHOOTER. In this mode, an input field of 4\farcs0 $\times$ 1\farcs8 is re-imaged onto a pseudo-slit of 12\farcs0 $\times$ 0\farcs6. The spectral resolving power $R \equiv \lambda/\Delta \lambda$ in this mode is 7900, 12600 and 8100 in the three arms respectively \citep{vernet2009}. We oriented the slit along the parallactic angle to ensure that atmospheric dispersion does not introduce wavelength-dependent flux losses. This means, however, that multiple Observing Blocks (OBs) will in general have been observed at different position angles on sky. We therefore quote the parallactic angle at the begin of each observation in Tab.~\ref{tab:OBs}. We also checked the difference in parallactic angle between start and end of each observation and it is $\lesssim 3 \deg$ in all but three cases (MCG523\_2: 6 \degr, NGC7172\_2: 6 \degr, NGC1315\_1: 14 \degr). Since we chose a quadratic extraction aperture (see below), the impact of this slight rotation of PA during the observation did not impact the quality of the spectra, even for the source with the largest rotation of the parallactic angle during the observation, NGC~1315.

We integrated on average about two hours per target. For most targets this led to two independent observations (with their own preset and calibration stars) which provided a useful test for systematic errors in the later analysis.

\subsection{Data Reduction}
\label{subsec:data_reduction}

\subsubsection{Raw Data Reduction}
We used the {\em reflex} pipeline provided by ESO in version 2.6.8 together with the Kepler GUI interface \citep{modigliani2010}. Several improvements over previous versions of the pipeline have been triggered by our use of X-SHOOTER in the IFU-mode. We essentially then used the pipeline in its default configuration\footnote{We found, however, that we had to set the property {\tt GlobalCutUVBSpectrum} to {\tt False} in order to retrieve a correct wavelength calibration in the UVB arm of X-SHOOTER.}.

\subsubsection{Differential atmospheric refraction and the Extraction Aperture}
\label{sec:aperture}
The ESO {\em reflex} pipeline produces a data cube for each observation and for each of the UVB, VIS and NIR arms of X-SHOOTER. From these raw-reduced data we extract 1D spectra for each of the science, telluric calibrator stars and flux calibrator stars. The data cubes from our X-SHOOTER IFU observations consist of three spaxels in the $x$ direction of 0\farcs6 each and several spaxels in $y$ direction. The total extent in the $y$ direction is 6\farcs0.

For the extraction, we first determine the centroid position in the $J$ and $H$ bands. From there we propagate the (expected) centroid position to the VIS and UVB arms using the revised \'Edlen's formula from \citet{boensch1998}; specifically we used Eqs. (6a,7,8,9a) to compute the refractive index for Paranal reference conditions, with ${\rm CO}_2$ content adjusted to the current value and a correction for moist air. The so-derived atmospheric dispersion was converted to detector pixel coordinates to provide a moving extraction window with wavelength. We verified our procedure for standard stars where it resulted in an extraction accuracy of better than 1 pixel in the $y$ direction (0\farcs16) relative to the actual position of the star in each slice.

We then use a rectangular extraction aperture of 1\farcs8, i.e. we used all three slitlets in the $x$ direction and a box with a length of 1\farcs8 in the along-slit direction $y$. We also tried to use a smaller square aperture of 0\farcs6 $\times$ 0\farcs6, but rejected this smaller aperture for two reasons: Firstly, the position of the nucleus is not well defined in the UVB/VIS arms for even only moderately obscured sources: The brightest region would often be so far off from the expected position of the nucleus in the UVB/VIS arms, that it would lie outside this compact aperture. About a third of our sample would be affected by this if we had chosen a smaller extraction aperture. Secondly, our observations are seeing-limited and while the seeing has been smaller than 0\farcs6 in some cases, it has been up to 1\farcs6 in other observations. By choosing a square aperture of 1\farcs8 $\times$ 1\farcs8, the extraction aperture is unaffected by seeing variations.

This extraction aperture corresponds to a physical radius of 150 (80) pc at the median distances of our AGN (control) samples which is sufficient to isolate the nuclear stellar population from the circum-nuclear population (that can be found in nuclear rings)\footnote{The median distance of the AGN and control galaxy sub-samples is 35 Mpc and 21.5 Mpc, respectively.}. An analysis of more than 100 galaxies in our redshift range \citep{comeron2010} has shown that $\gtrsim$ 90\% of nuclear rings have radii larger than 150 pc. Of the five sources we have in common with the sample of \citet{comeron2010}, all have ring radii well outside our extraction aperture. Additionally, we have analysed the LLAMA VLT/SINFONI data for 19 out of the 27 objects presented in this paper and only one source (NGC~7582) shows strong evidence for a nuclear ring within the $3\farcs0 \times 3\farcs0$ field-of-view of SINFONI (but outside the aperture used for this work). In all other sources there does not appear to be a circum-nuclear star-forming disk within our field-of-view, as judged from the (lack of) Br~$\gamma$ emission, the $^{12}$CO (2-0) bandhead flux and equivalent width and the continuum flux.

We therefore conclude that the spatial resolution of our observations and data analysis is sufficient to separate the nuclear from the circum-nuclear stellar populations.

\subsubsection{Telluric and flux calibration}

Telluric calibrator observations have been provided by the observatory. We used the tool {\em molecfit} \citep{smette2015,kausch2015} to model the telluric transmission based on the observation of the closest telluric calibrator. Compared to directly using the telluric calibrator spectrum, this has the advantage of producing a noise-free transmission spectrum. The strongest atmospheric spectral feature in the wavelength range of interest to this work is the atmospheric A band of molecular oxygen starting at 760 nm. Other (weaker) features include the atmospheric B band of molecular oxygen starting at 686 nm as well as two broad bands of water vapour between 715-740 and 810-838 nm. Visually, the correction even in the strongest feature of these (the A band) is very good with the molecfit model spectra and the main obvious effect is an increased noise due to the lower transmission in these features. Quantitatively, we looked at the difference between the molecfit model and the raw spectrum of the science source and found the residuals to be $\lesssim$ 10\% of the observed spectrum for the first few strong absorption peaks of the A band, but typically less than $\pm$ 5 \% for the whole A band and lower in the other bands.

For the absolute flux calibration we used the flux calibrated spectra of a small list of spectro-photometric standard stars provided by ESO \citep{moehler2014} after applying the telluric absorption correction described above. These flux standards have usually been observed within the same night, and in several cases within up to five nights before or after the science observation. Only in one case (OB id {\tt NGC5506\_2}) did we have to use a flux calibrator 13 days later. However, we looked at the count rates of the flux calibrator GD71 in five observations spread over 21 months and found the variations to be smaller than $\sim 2 \%$ in the wavelength range of interest for this work. Since the flux calibrators provided by ESO are cross-calibrated to much better than this \citep{moehler2014}, we therefore consider our flux calibration (per X-SHOOTER arm) to be accurate at the 2 \% level.

\subsubsection{Further X-SHOOTER calibration steps and tests}
Further calibration steps in the X-SHOOTER data processing included an interpolation (in wavelength direction) for bad pixels and a flux-matching between the UVB and VIS arms. We had to multiply the flux-calibrated VIS spectrum by a factor of $1.145 \pm 0.140$ (median and standard deviation, respectively) to match the UVB spectrum at the connecting point of the two spectra of 560 nm. While we do not know the origin of this flux mismatch, we can determine this ``inter-calibration'' factor between the UVB and VIS arms with high accuracy for every observation. It does not affect the fidelity of the spectra; in particular the absorption lines, which are the property of most significance to this work, are unaffected by this calibration step.

\subsubsection{X-SHOOTER calibration pipeline}
While the ESO pipeline for X-SHOOTER provides a complete set of tools for processing the raw data to the final uncalibrated spectra, higher-order calibration tasks, such as described here above, need to be performed by the user. We have therefore developed a comprehensive high-order calibration pipeline for X-SHOOTER which is implemented as a Python class {\tt Xspec} and freely available at our github project\footnote{https://github.com/astroleo/xshtools} containing various tools to reproduce the results presented in this paper.

%% file: sections/3-sps.tex
\section{Stellar Population Synthesis}
\label{sec:sps}
The goal of stellar population synthesis is to derive a number of properties of a stellar population of which only the integrated spectrum is known. This method essentially determines the best fitting linear superposition of model spectra of single stellar populations, while simultaneously fitting for extinction and kinematics. The main result of the synthesis is a ``population vector'' that contains the estimated fractions of light attributable to each of the input single stellar populations (SSP) of different ages (and metallicities and/or other parameters). A number of SSP models exist with different ingredients in terms of empirical or model spectral libraries, different techniques for estimating the impact of the post-main-sequence evolution, various approaches for determining the statistical uncertainties and a variety of other different assumptions. We refer to \citet{conroy2013} for a review of both the method of stellar population synthesis itself and its ingredients.

While the plethora of assumptions and models is somewhat perplexing, the situation is less dismal than it seems: it has been shown by \citet{baldwin2018} and \citet{dahmer_hahn2018} that population synthesis results in the optical wavelength range do not depend sensitively on the stellar population library that is used and, with a sufficiently high signal-to-noise ratio, state-of-the-art stellar population synthesis (SPS) codes such as \STARLIGHT{} \citep{cidfernandes2004,cidfernandes2005} or {\textsc pPXF} \citep{cappellari2004} are able to recover reddened input spectra with realistic noise very accurately \citep{cidfernandes2018}.

\subsection{STARLIGHT fitting}

For our stellar population synthesis we use the publicly available and well-tested tool \STARLIGHT{} in its version 4 \citep{cidfernandes2004,cidfernandes2005} in combination with the well-known single stellar population (SSP) library by \citet{bruzual2003} (hereafter ``BC03''), and using the \citet{chabrier2003} initial mass function. More modern stellar population libraries such as MILES \citep{vazdekis2016} come with higher spectral resolutions, but do not cover the youngest stellar populations that are particularly interesting for AGN feeding. Qualitatively the stellar population synthesis with these models gives similar results to the ones with the BC03 models \citep{riffel_r2021}.

In order to decrease the ambiguity in the fitting process, we use a reduced version of this library containing 15 single stellar population spectra of solar metallicity (see discussion about metallicity in Section~\ref{sec:metallicity}). Since the BC03 spectra come with a spectral resolution of 3 \AA{}, considerably lower than our X-SHOOTER spectra, we convolved our X-SHOOTER spectra to match this resolution and we further gridded our spectra in bins of constant wavelength range $\Delta \lambda = 1$ \AA{}.

In \STARLIGHT{}, we used the default normalisation range for the observed spectra (6810 -- 6870 \AA{}) and the reddening law by \citet{calzetti1994}. We fitted the wavelength range from $\lambda = 3850$ \AA{} to $\lambda = 8800$ \AA{}. The blue end is given by our self-imposed SNR limit. At 3850 \AA{} the SNR is typically only a few for the re-binned spectra. The red end was essentially set to include the \ion{Ca}{II} triplet for an accurate determination of the redshift and line-of-sight velocity dispersion from these absorption lines. Extending the wavelength coverage further into the infrared did not result in overall good \STARLIGHT{} fits. This may be due to the lack of spectral resolving power in the near-IR in the models of BC03 \citep{dahmer_hahn2018} or the fact that the near-IR light is dominated by different stellar populations (and possibly affected by a different amount of extinction) than the optical light and that a joint fit is therefore not meaningful.

\subsection{Masking}
\label{sec:masking}
Almost all of our galaxies show emission lines which need to be masked in order not to bias the fitting of the stellar absorption lines. While our philosophy for this research is to compare as equally as possible the stellar populations in AGNs and control galaxies, we cannot quite adhere to this approach for the process of masking the emission lines. If we chose a single mask that blocks all emission lines seen in any of our galaxies, essentially no data would be left to fit. We therefore choose a different approach and start the fit for each galaxy with a conservative mask including many of the frequently seen emission lines in AGNs. With this mask we perform a first SSP fit with \STARLIGHT{} and apply a smoothed version of sigma-clipping to the result. Our goal here is not to remove all outliers from the fit -- this may easily bias the result -- but to exclude strong lines reliably and completely.

To accomplish this, we define the center of new masked areas where the observed flux is more than 5 $\sigma$ larger than the flux of the first fit (with the default mask). At this stage, $\sigma$ is the uncertainty of the observed spectrum as derived by the ESO X-SHOOTER pipeline and propagated through the various subsequent processing steps (telluric calibration, re-binning). We then follow the residual spectrum bluewards and redwards until it reaches a value of $< 1 \sigma$. The interval between the so-defined limits are added to the mask. We note that it does not matter for \STARLIGHT{} if the mask definition contains overlapping regions. In order to further ensure a proper scaling of the uncertainties, we rescale the residual spectrum such that $\sim$ 68\% of the residuals actually lie within $\pm 1 \sigma$ of the distribution. We then repeat the fitting, masking and uncertainty-rescaling three times. Our tests have shown that after the third iteration, both the mask and the resulting stellar population vector have converged to a stable solution.

\subsection{Uncertainty determination}
While stellar population synthesis is a commonly used technique in astronomy, there is no simple way to estimate the uncertainty of its results. The reason for this is that the final uncertainty depends on a number of parameters. In this section we discuss and try to estimate the various sources of uncertainties involved in the fitting process.

\begin{figure}
\includegraphics[width=\columnwidth]{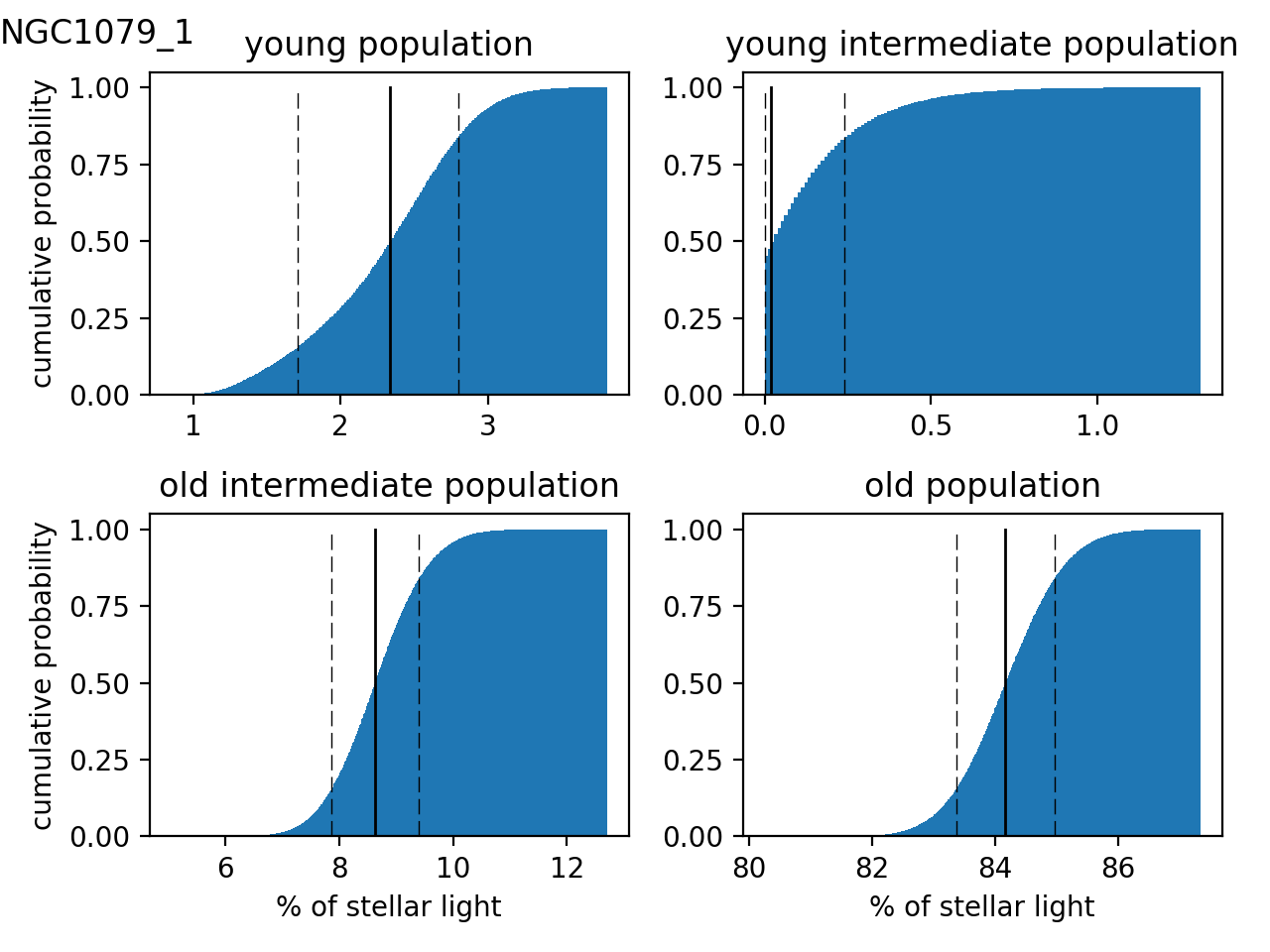}
\caption{\label{fig:starlight_uncertainties}Posterior probability distributions for the fraction of light in the four bins of stellar ages from our \STARLIGHT{} analysis of NGC~1079. We report the fractional value of stellar light associated with the 50\% value in the cumulative distribution and take as 1 $\sigma$ uncertainty the range encompassing 68.27\% of all values.}
\end{figure}

\paragraph{Statistical uncertainties of the calibration and fitting process}
The most straight-forward source of uncertainty is the noise in the spectra, the uncertainties associated with the telluric and flux calibration and possible degeneracies involved in the fitting process. These uncertainties are taken into account in the fitting process with \STARLIGHT{} internally. However, in order to actually visualise the posterior probability distributions we obtained a special version of \STARLIGHT{} (R. Cid Fernandes, priv.comm.) that saves the full Markov Chain of the Monte Carlo sampling. From these, we compute the actual statistical uncertainties involved for our spectral synthesis as shown in Fig.~\ref{fig:starlight_uncertainties}.

\begin{figure}
\includegraphics[width=\columnwidth]{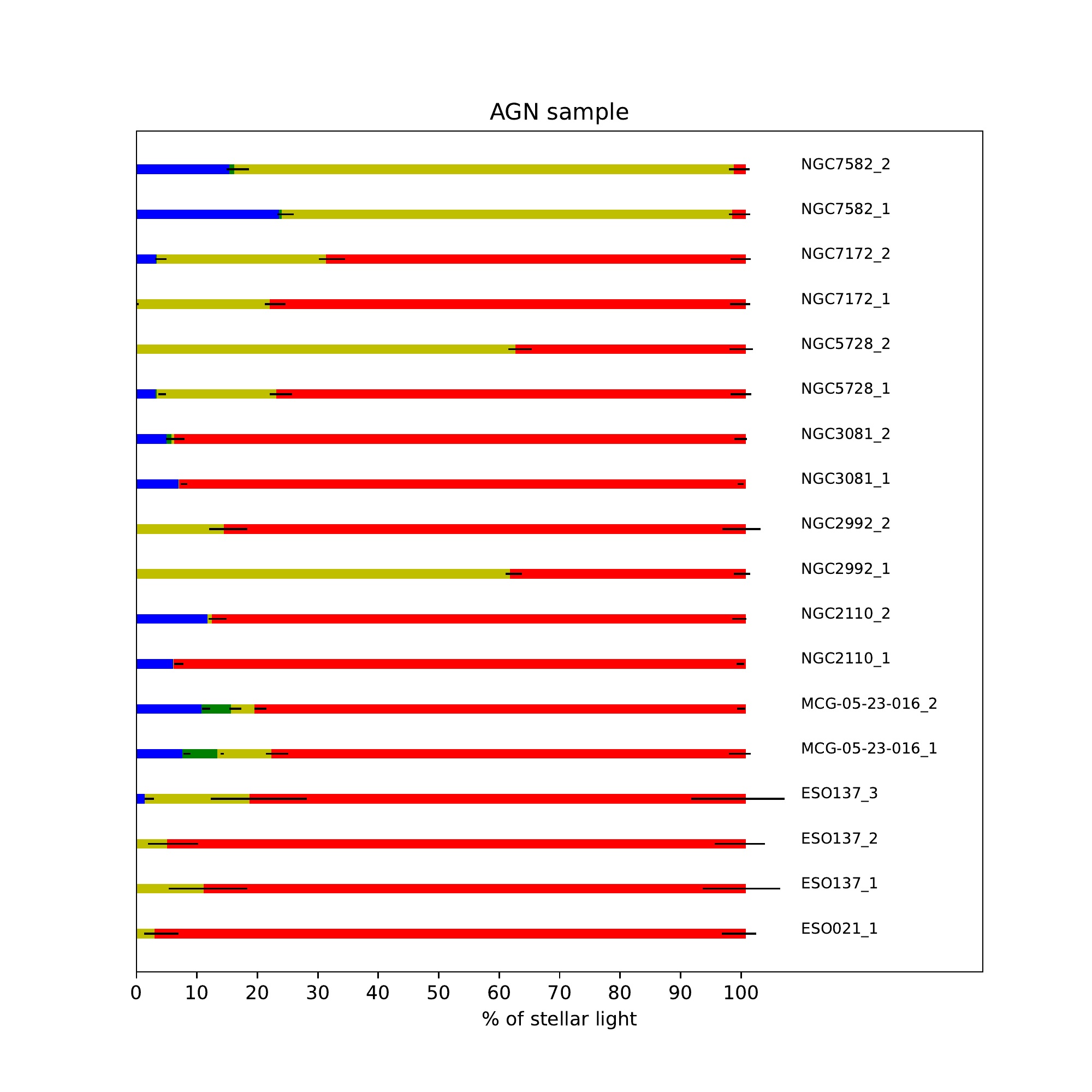}
\caption{\label{fig:STARLIGHT_OB_differences_AGN}Individual stellar population fitting results for all observations showing differences between individual observations for the same targets (the individual observations are indexed with underscores, e.g. NGC7582\_1 and NGC7582\_2). The four age bins marked in blue, green (only visible in one AGN), yellow and red correspond to log(age/yrs) $<$ 7.5 (``young''), 7.5 $<$ log(age/yrs) $<$ 8.5 (``young-intermediate''), 8.5 $<$ log(age/yrs) $<$ 9.5 (``intermediate-old'') and log(age/yrs) $>$ 9.5 (``old''), respectively. The statistical uncertainties on the fractions of stellar light for the four age bins are shown as black lines at the right edge of each of these age bins.}
\end{figure}

\begin{figure}
\includegraphics[width=\columnwidth]{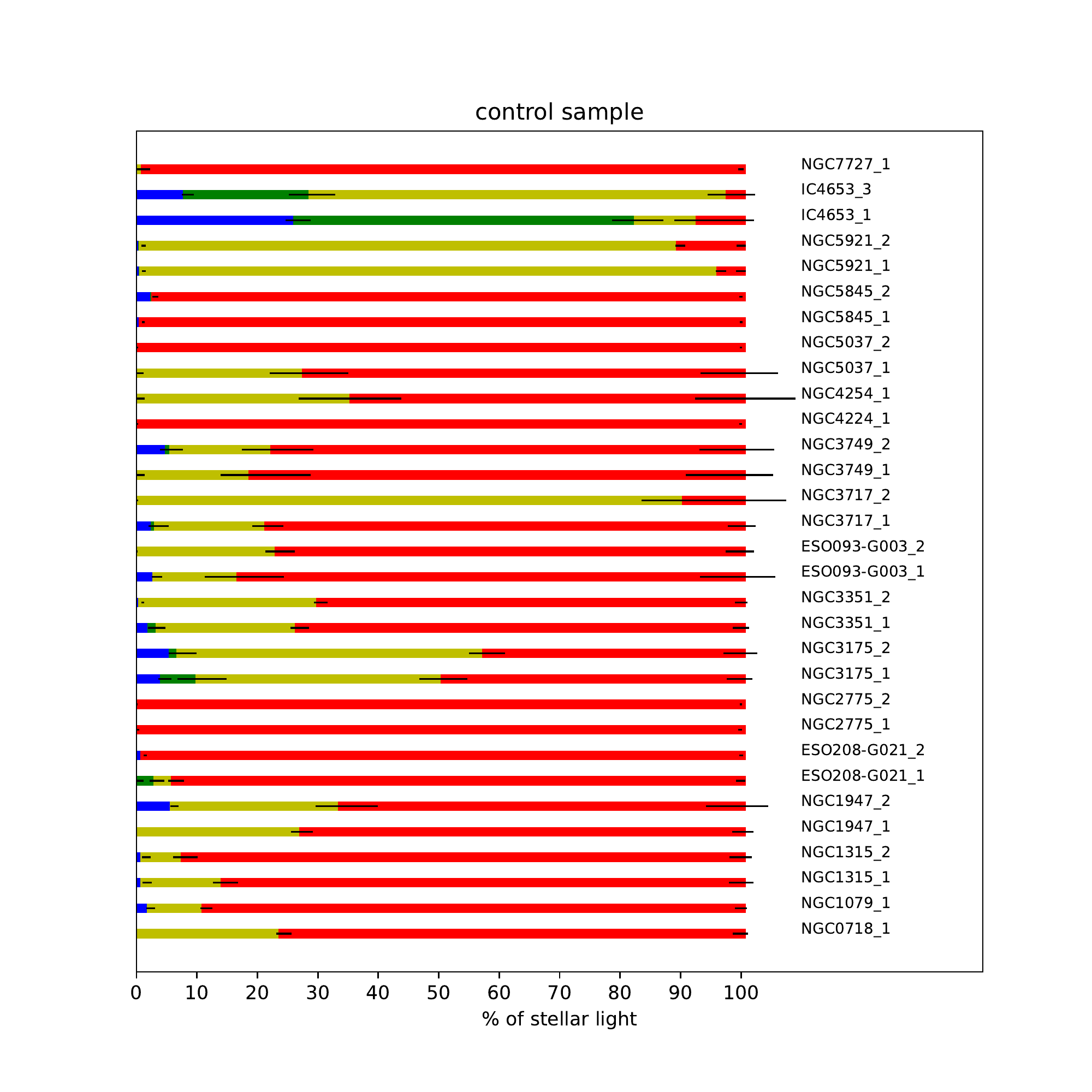}
\caption{\label{fig:STARLIGHT_OB_differences_control}Same as Fig.~\ref{fig:STARLIGHT_OB_differences_AGN}, but for the control sample.}
\end{figure}

The statistical uncertainties involved in the \STARLIGHT{} fitting process have also been analysed and discussed in \citet{cidfernandes2014} who find the uncertainties in the binned (``young'', ``intermediate'', ``old'' populations) single stellar populations to be of 3, 9 and 9\%, respectively, for random-noise induced errors at the 1-$\sigma$ level, and larger errors for shape-changing perturbations (e.g. errors in the spectral flatfield calibration). These values are considerably larger than the values we find from our own re-sampling exercise (see Figs.~\ref{fig:starlight_uncertainties}, \ref{fig:STARLIGHT_OB_differences_AGN}, \ref{fig:STARLIGHT_OB_differences_control}): The statistical uncertainties that we find are, on average, 0.34\% (0.37\%),  0.28\% (0.62\%),  2.07\% (2.37\%),  1.99\% (2.13\%) for the young, young-intermediate, intermediate-old and old populations and for the AGN (control) populations, respectively. We attribute this to the overall excellent quality and calibration of our X-SHOOTER spectra.

\paragraph{Systematic uncertainties of the observations}

Since our original signal/noise estimate required about two hours of observations per target, we received on average two one-hour observations per target. This allows us to assess the systematic uncertainties involved in the observations (e.g. pointing accuracy, slit orientation, weather conditions). We therefore show the stellar population vector for all individual observations in Figs.~\ref{fig:STARLIGHT_OB_differences_AGN} and \ref{fig:STARLIGHT_OB_differences_control} for the AGN and control sample, respectively. The statistical error bars (derived as shown in Fig.~\ref{fig:starlight_uncertainties}) are shown on top of these individual results and demonstrate that the systematic uncertainties are of the same order as the statistical uncertainties for most galaxies.

\paragraph{Other uncertainties}
Other sources of uncertainty in the derived star formation histories include metallicity (see discussion in the next Section~\ref{sec:metallicity}), the extinction correction (see discussion in Section~\ref{sec:extinction}) and of course the sample size itself which we assess with a jack-knifing test as described in Section~\ref{sec:llama_sfh}. The specific stellar population models and the SPS fitting code used (\STARLIGHT{}) do not induce further uncertainties in our estimates, as explained above and as shown by \citet{baldwin2018}. We also reiterate that the strength of our analysis lies in the comparison of the stellar populations of the AGN sample with matched analogues.

\subsection{Metallicities}
\label{sec:metallicity}
Measurements of the gas-phase metallicity in local galaxies using the direct method (i.e. using auroral instead of strong emission lines to derive electron temperature and hence metallicity, since metals are the main coolants in HII regions) show that, while the mass-metallicity relation rises steeply at low mass, it turns over at higher mass and for galaxy masses $> 10^{10} M_{\odot}$ (relevant for the LLAMA galaxies) it is fairly constant at a value consistent with solar metallicity \citep{andrews2013}.

Comparison of the mass-metallicity relation at different epochs shows that it does evolve with redshift. But while metallicity increases with cosmic time, the changes from z$\sim$2 to z$\sim$1 and from z$\sim$1 to z$=$0 are only about 0.17 dex each \citep{yuan2013}. Furthermore \citet{zahid2011} point out that this change is largely driven by lower mass galaxies, and that for galaxies with typical stellar masses of $10^{10.6} M_{\odot}$ (comparable to the LLAMA galaxies), the difference in average metallicity is much less.

The change in stellar rather than gas phase metallicity has been assessed by \citet{asari2009} by fitting stellar populations to continuum spectra tracing the stars themselves. Their conclusion is consistent with the results from studies using emission lines, that at the present day, galaxies with $M > 10^{10} M_{\odot}$ have roughly solar metallicity; and that the change with redshift or lookback time is modest for this mass range. Thus, to reduce the number of basis spectra for our fitting in order to avoid complications in our comparative analysis resulting from the age-metallicity degeneracy, we use only solar metallicity ($Z = 0.02$) spectra rather than including supersolar ($Z = 0.05$) or sub-solar ($Z = 0.008$) spectra.

The caveat is that the solar metallicity in the BC03 models is slightly higher than the most recent estimates of solar metallicity: The standard value for $Z_{\rm sun}$ was 0.02 \citep{anders1989}, and this was adopted by BC03.
But \citet{asplund2009} recommend a solar chemical composition for which $Z_{\rm sun}$ is 0.0134, substantially less. They also indicate that solar oxygen abundance $12+\log(O/H) = 8.69$.
However, in terms of the comparative analysis performed here, this difference in what is adopted as solar metallicity has negligible impact.

\subsection{Selection of best OB}
\label{sec:ob_selection}

Our following interpretation is not limited by the statistical uncertainties from the stellar population fitting (which are small, see above), but rather by the small sample size and, to a lesser extent, by the systematic uncertainties from differences between OBs. We therefore use multiple observations of a source, when available, to estimate the systematic uncertainty rather than to marginally improve the S/N at the cost of more systematic uncertainty.

Usually the \STARLIGHT{} fits to one of the OBs was noticeably better than the other one(s). We therefore selected only the best-fitting OB for the further analysis. This selection was done on the basis of (1) SNR, (2) masking as little as possible, (3) quality of the spectra\footnote{in particular, some of our X-SHOOTER spectra show a strong instrumental signature in the region where the UVB/VIS arms are overlapping. This calibration residual is possibly related to variable dichroic transmission due to humidity changes (priv. comm. J. Vernet) or from flatfielding residuals when the pipeline chooses different calibration files for the science and flux calibrators, respectively (priv. comm. A. Mehner).}, (4) goodness of residuals (we tried to avoid structure in the residuals and preferred residuals that looked like white noise), (5) if in doubt, we put more weight on the blue part of the spectrum (despite its lower SNR) since this is where more distinctive stellar population features lie.

Our selection of OBs can be seen from Tab.~\ref{tab:OBs}.

We illustrate our selection on the basis of three AGNs which show differing results between the multiple OBs:

\begin{itemize}
    \item {\bf NGC~7172} does not have a young population in NGC7172\_1, but shows one in NGC7172\_2. However the stellar population fit for the latter observation leads to a bad fit and is therefore de-selected. It is also one of the very few galaxies where the mask actually has a noticeable impact on the population vector. Using our starting mask we find a small fraction of young stellar light even for the observation NGC7172\_1. After optimising the mask iteratively based on actually present emission lines (see Section~\ref{sec:masking}), the young population disappears in this galaxy.
    \item {\bf NGC~5728} shows a young population in OB 1 and not in OB 2, but OB 2 is a worse fit than OB 1 and has a large masked area which is why we choose the observation NGC5728\_1 for the further analysis.
    \item {\bf ESO~137-G034} shows a young population in OB 3 which is the best fit. OBs 1 results in a fit where a large range between H $\alpha$ and the TiO feature at $\sim$ 7500 \AA{} is masked. OB 2 does not see a young population and is also a fairly good fit, but has a strong ``emission'' feature at $\sim$ 5500 \AA{} that is not real (see discussion above), which is why we deselect it.
\end{itemize}

%% file: sections/4-sps-results.tex
\section{Stellar Population Synthesis results}
\label{sec:sps_results}
\begin{figure*}
\includegraphics[width=\textwidth]{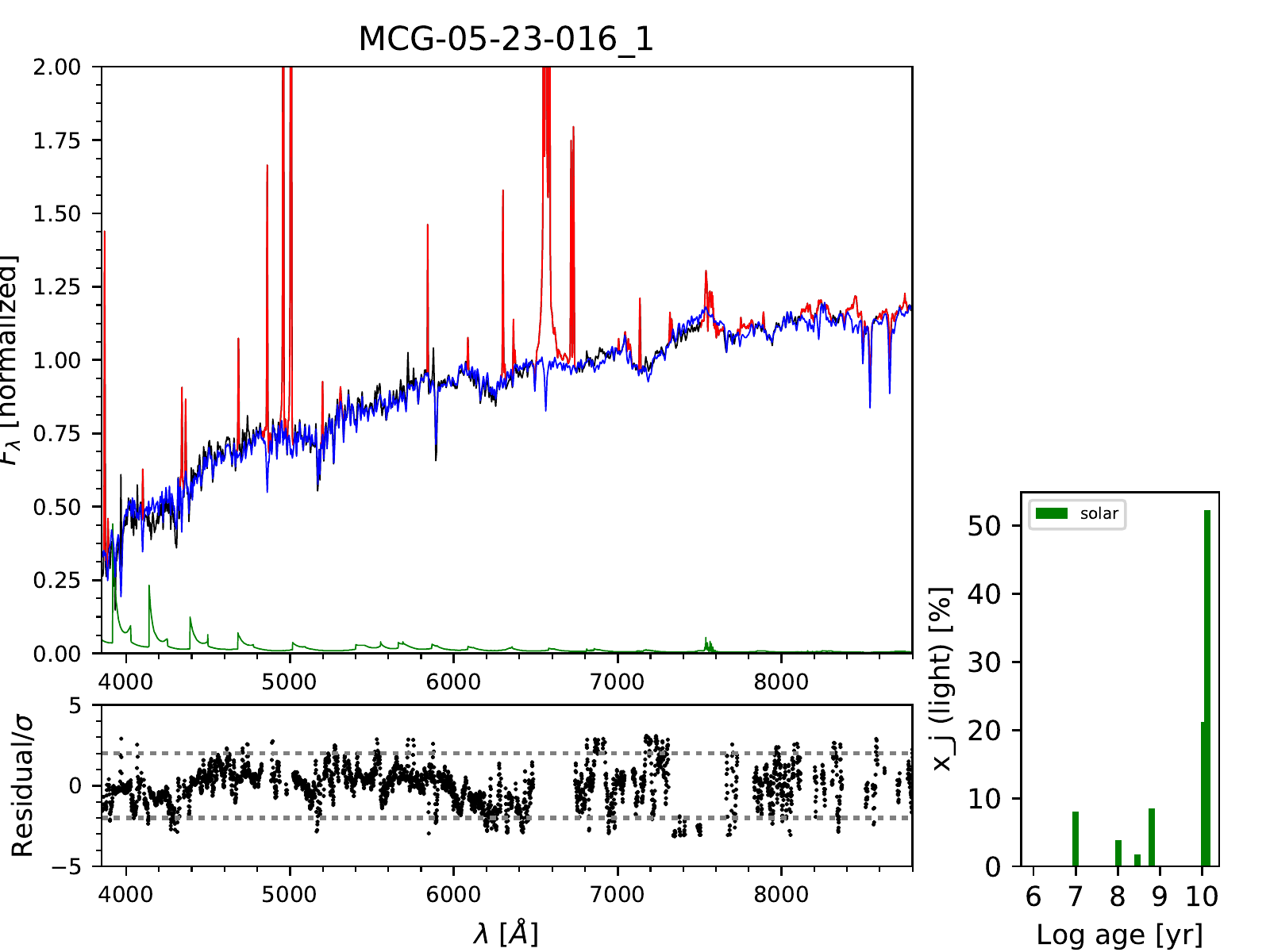}
\caption{\label{fig:SSP_MCG523_1}Example of a stellar population fit, for the AGN MCG-05-23-016. In the main panel the X-SHOOTER spectrum is displayed, down-sampled to the BC03 spectral resolution (black), the \STARLIGHT{} fit (blue) with masked areas (in red) and the noise in the data in green. Below the spectrum, the residuals are displayed in terms of the local uncertainty in the data (masked areas are not displayed). In the side panel, the light-weighted star formation history (or stellar population vector) is shown.}
\end{figure*}

\begin{figure*}
\includegraphics[width=\textwidth]{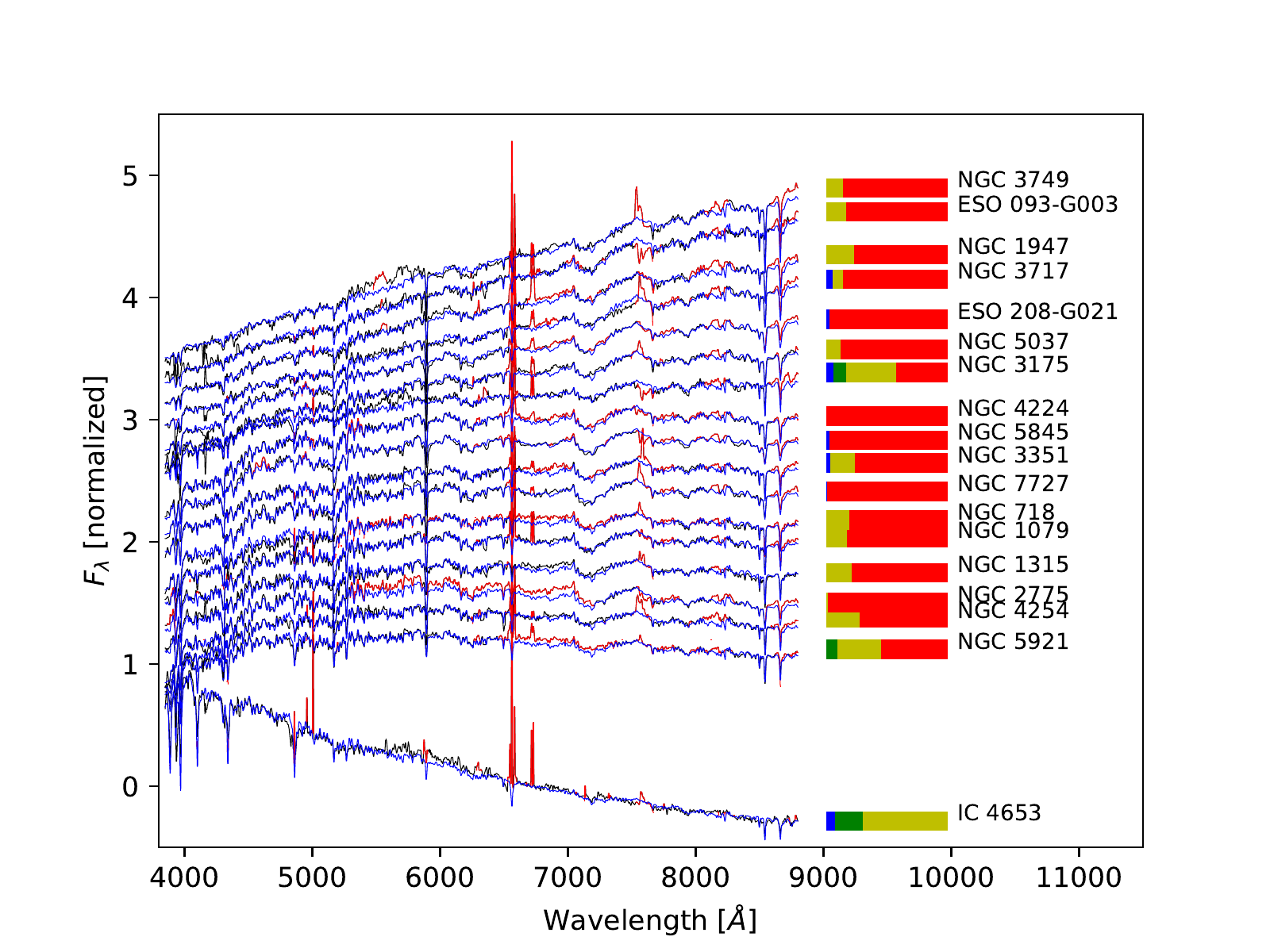}
\caption{\label{fig:sps_inactive_summary}Collage of all LLAMA non-AGN X-SHOOTER spectra with best fitting stellar population synthesis overplotted. To the right of each spectrum the summarised stellar population vector shows the best-fitting light-weighted fractions of the young (blue), young-intermediate (green), intermediate-old (yellow) and old (red) populations with boundaries log (age/years) at 7.5, 8.5, and 9.5.}
\end{figure*}

\begin{figure*}
\includegraphics[width=\textwidth]{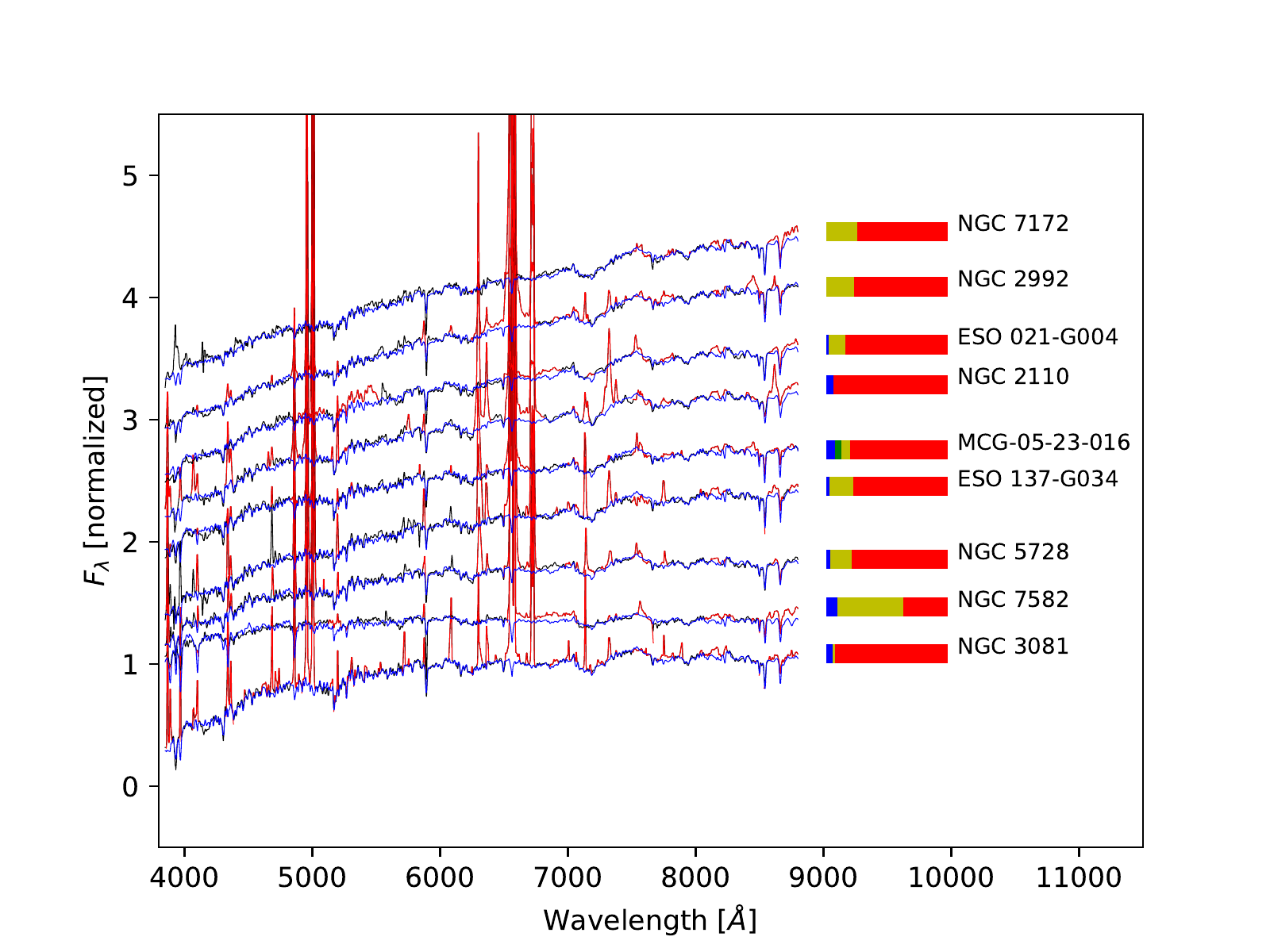}
\caption{\label{fig:sps_active_summary}Like Fig.~\ref{fig:sps_inactive_summary} but for the active galaxies in this paper.}
\end{figure*}

The result of our stellar population synthesis is a so called stellar population vector that gives the fraction of light (or mass) attributed to each single stellar population spectrum. We show the resulting fit, the (down-sampled and re-binned) data, the masked emission lines, the residuals and the stellar population vector for one galaxy in Fig.~\ref{fig:SSP_MCG523_1} and for all galaxies in Appendix~\ref{sec:appendix:fits}. We also show a summary plot of all spectral syntheses in Figs.~\ref{fig:sps_active_summary} and \ref{fig:sps_inactive_summary} for the AGNs and inactive galaxies, respectively.

The resulting fit parameters for both light and mass fractions are displayed for all galaxies in Tab.~\ref{tab:starlight}. In this table (and for the further analysis) we bin the stellar population vector into four age groups: log(age/yrs) $<$ 7.5 (``young''), 7.5 $<$ log(age/yrs) $<$ 8.5 (``young-intermediate''), 8.5 $<$ log(age/yrs) $<$ 9.5 (``intermediate-old'') and log(age/yrs) $>$ 9.5 (``old''). The choice of these age groups is driven by one of the original aims of this project, namely to discriminate between the youngest stars and young-intermediate stars. The table further contains the signal-to-noise-ratio (SNR) in the fiducial wavelength range 605-608 nm, the velocity offset of this galaxy relative to the assumed redshift (see Tab.~\ref{tab:sample})\footnote{For NGC~5037 we use the redshift reported by Simbad since the redshift from NED (0.009893) is wrong based on our spectral fitting. Also for IC~4653 the initial recession velocity assumed for the fit had been taken from NED \citep[][ 1890 km/s]{corwin1994}, which is wrong. We thus use and confirm the updated redshift as published previously by \citet{wegner2003}. See \citet{davies2020} for more discussion about the systemic velocities of these two galaxies.}, the measured velocity dispersion, dust extinction and reduced $\chi^2$ of the best fitting combination of single stellar population spectra. Slightly negative values for $A_{\rm V}$ arise from (small) uncertainties in the spectral flat field and flux calibrations and give an indication of the accuracy of our extinction values ($\lesssim 0.2$ mag). Finally we quote the total stellar mass within our aperture $\log M_{\star}/M_{\odot}$ as fitted by STARLIGHT and using the distance quoted in Tab.~\ref{tab:sample}.

We will now discuss in this Section the accuracy of our fits and the emission line diagnostics, both of which are required in order to reliably analyse and interpret the stellar population results in Section~\ref{sec:discussion}.

\begin{table*}[ht]
\caption{Results from the STARLIGHT fitting}
\label{tab:starlight}
\centering
\begin{tabular}{ccccccccccccccc}
\hline\hline
Galaxy name & SNR & $v_0$ & $\sigma$ & $A_V$ & $\chi^2_{\rm red}$ & $x_y$ & $x_{yi}$ & $x_{io}$ & $x_o$ & $\mu_y$ & $\mu_{yi}$ & $\mu_{io}$ & $\mu_o$ & $\log M_{\star}$\\
&&[km/s]&[km/s]&&&[\%]&[\%]&[\%]&[\%]&[\%]&[\%]&[\%]&[\%]&[$M_{\odot}$]\\
\hline
NGC 2110 & 62 & 9 & 227 & 1.18 & 1 & 7.0 & 0.0 & 0.0 & 93.0 & 0.06 & 0.00 & 0.00 & 99.9 & 9.00 \\
NGC 2992 & 104 & -10 & 164 & 1.87 & 1 & 0.0 & 0.0 & 26.6 & 73.4 & 0.00 & 0.00 & 7.42 & 92.6 & 8.80 \\
MCG-05-23-016 & 64 & -68 & 128 & 1.24 & 1 & 8.3 & 5.8 & 8.9 & 76.9 & 0.09 & 0.38 & 1.26 & 98.3 & 8.85 \\
NGC 3081 & 97 & 3 & 131 & 0.51 & 1 & 5.7 & 0.0 & 2.5 & 91.8 & 0.05 & 0.00 & 0.50 & 99.5 & 8.97 \\
ESO 021-G004 & 67 & -102 & 178 & 1.40 & 1 & 2.3 & 0.0 & 15.7 & 82.0 & 0.02 & 0.00 & 5.04 & 94.9 & 9.15 \\
NGC 5728 & 58 & -19 & 174 & 0.83 & 1 & 4.0 & 0.0 & 20.2 & 75.8 & 0.05 & 0.00 & 4.69 & 95.3 & 9.13 \\
ESO 137-G034 & 26 & 14 & 134 & 1.32 & 1 & 2.6 & 0.0 & 23.7 & 73.7 & 0.03 & 0.00 & 10.19 & 89.8 & 8.99 \\
NGC 7172 & 57 & -49 & 167 & 1.79 & 1 & 0.0 & 0.0 & 30.0 & 70.0 & 0.00 & 0.00 & 7.33 & 92.7 & 9.17 \\
NGC 7582 & 91 & 32 & 137 & 1.18 & 1 & 10.9 & 0.0 & 64.3 & 24.8 & 0.23 & 0.00 & 25.80 & 74.0 & 8.73 \\
\hline{}
NGC 718 & 58 & 22 & 88 & 0.39 & 2 & 0.0 & 0.0 & 22.1 & 77.9 & 0.00 & 0.00 & 9.07 & 90.9 & 8.89 \\
NGC 1079 & 52 & 17 & 102 & 0.28 & 1 & 0.0 & 0.0 & 19.7 & 80.3 & 0.00 & 0.00 & 4.23 & 95.8 & 8.49 \\
NGC 1315 & 45 & 18 & 81 & 0.23 & 2 & 0.0 & 0.0 & 24.8 & 75.2 & 0.00 & 0.00 & 5.61 & 94.4 & 8.27 \\
NGC 1947 & 72 & 95 & 132 & 1.88 & 1 & 0.0 & 0.0 & 26.8 & 73.2 & 0.00 & 0.00 & 7.59 & 92.4 & 8.72 \\
ESO 208-G021 & 83 & -25 & 211 & 1.16 & 2 & 3.0 & 0.0 & 0.0 & 97.0 & 0.02 & 0.00 & 0.00 & 100.0 & 8.71 \\
NGC 2775 & 4 & -23 & 163 & -0.16 & 2 & 0.0 & 0.0 & 1.2 & 98.8 & 0.00 & 0.00 & 0.13 & 99.9 & 7.87 \\
NGC 3175 & 41 & 3 & 74 & 1.40 & 1 & 7.0 & 11.7 & 49.3 & 31.9 & 0.15 & 2.05 & 21.50 & 76.3 & 7.90 \\
NGC 3351 & 63 & 3 & 84 & 0.54 & 1 & 3.9 & 0.0 & 23.8 & 72.4 & 0.04 & 0.00 & 5.77 & 94.2 & 7.38 \\
ESO 093-G003 & 45 & -4 & 78 & 2.07 & 1 & 0.0 & 0.0 & 18.8 & 81.2 & 0.00 & 0.00 & 4.31 & 95.7 & 9.01 \\
NGC 3717 & 68 & 11 & 127 & 1.70 & 1 & 6.0 & 0.0 & 9.9 & 84.1 & 0.05 & 0.00 & 2.09 & 97.9 & 8.71 \\
NGC 3749 & 96 & 43 & 155 & 2.34 & 1 & 0.0 & 0.0 & 16.0 & 84.0 & 0.00 & 0.00 & 5.11 & 94.9 & 9.37 \\
NGC 4224 & 65 & 26 & 141 & 0.29 & 1 & 0.0 & 0.0 & 0.0 & 100.0 & 0.00 & 0.00 & 0.00 & 100.0 & 9.33 \\
NGC 4254 & 47 & 37 & 68 & 0.26 & 1 & 0.0 & 0.0 & 32.1 & 67.9 & 0.00 & 0.00 & 14.41 & 85.6 & 7.63 \\
NGC 5037 & 86 & 20 & 162 & 1.06 & 1 & 0.0 & 0.0 & 13.9 & 86.1 & 0.00 & 0.00 & 2.95 & 97.1 & 9.02 \\
NGC 5845 & 102 & -8 & 255 & 0.29 & 2 & 3.1 & 0.0 & 0.0 & 96.9 & 0.03 & 0.00 & 0.00 & 100.0 & 9.10 \\
NGC 5921 & 60 & 15 & 67 & 0.31 & 1 & 0.0 & 10.3 & 43.1 & 46.6 & 0.00 & 1.53 & 17.20 & 81.3 & 8.67 \\
IC 4653 & 58 & -12 & 69 & -0.03 & 1 & 8.2 & 27.4 & 64.4 & 0.0 & 0.33 & 13.57 & 86.10 & 0.0 & 7.27 \\
NGC 7727 & 73 & -44 & 188 & 0.15 & 1 & 0.3 & 0.0 & 0.0 & 99.7 & 0.00 & 0.00 & 0.00 & 100.0 & 8.95 \\
\hline
\end{tabular}
\tablefoot{
For each galaxy (top: AGNs, bottom: inactive galaxies) we give the SNR in the range 605-608 nm, the velocity offset from the adopted redshift given in Tab.~\ref{tab:sample} (without any corrections applied), the velocity dispersion, the extinction $A_V$ in magnitudes, the reduced $\chi^2$ and the summarised star-formation histories. They are given as fractional contributions to the light (mass) as normalised at 684 nm in the four bins designated young $x_y$ ($\mu_y$), young-intermediate $x_{yi}$ ($\mu_{yi}$), intermediate-old $x_{io}$ ($\mu_{io}$) and old $x_o$ ($\mu_o$). They are defined by their boundaries of log(age/yr) = 7.5, 8.5, 9.5. Total stellar mass within our aperture: $\log M_{\star}/M_{\odot}$
}
\end{table*}

\subsection{How robustly can we determine the young stellar population in AGNs?}
\begin{figure}
\includegraphics[width=\columnwidth]{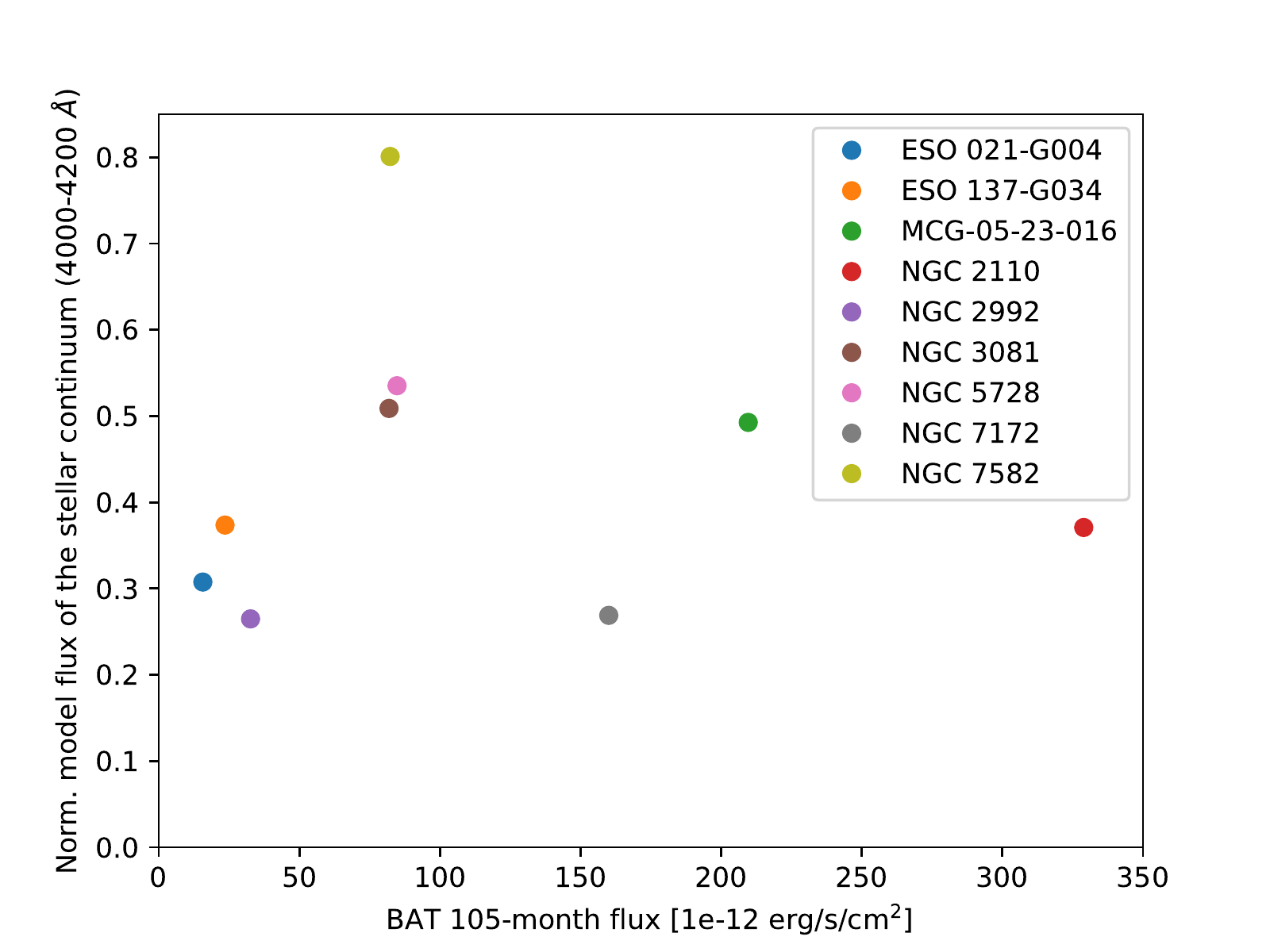}
\caption{\label{fig:young_vs_fx}Relation of the blue continuum and the ultra-hard X-ray flux. The (noise-free) model flux at the blue end of our spectrum is plotted against the BAT 105-month ultra-hard X-ray (14-195 keV) flux. No relation between the two fluxes is seen providing further evidence that the blue continuum in our spectra is not dominated by AGN light. The uncertainties of the BAT fluxes are within the symbol size.}
\end{figure}

It has been known for some time that even obscured (Seyfert~2) AGNs show a significant fraction of blue continuum in their nuclei. It is sometimes called ``featureless continuum'' although spectral features of massive stars have been found early on \citep{gonzalez_delgado1998}. It also shows relatively low levels of polarisation \citep{tran1995}. According to AGN unification, it should therefore not originate from the central engine, as any direct light from there would be blocked and only reflected, and thus polarised, light should be observable in type~2 AGNs \citep{cidfernandes1995}.

Nevertheless, its origin has remained contentious. For example, \citet{colina1997} showed, based on HST data, that most of the circum-nuclear light originates in young star-forming regions, but could not say whether the nuclear blue light is from nuclear star formation or from an AGN-type power-law. \citet{sarzi2007b}, on the other hand, find a good correlation between a tracer of the AGN continuum ([Ne V]) and the blue light in their stellar population synthesis and deduce that the blue light they find originates most likely from the AGN itself rather than from a young stellar population. However, not enough detail is given in \citet{sarzi2007b} to fully understand these results, e.g. we do not know about the strength of the [NeV] line in the Seyfert~2~AGNs for which they do not find a blue continuum.

\begin{figure}
\includegraphics[width=\columnwidth]{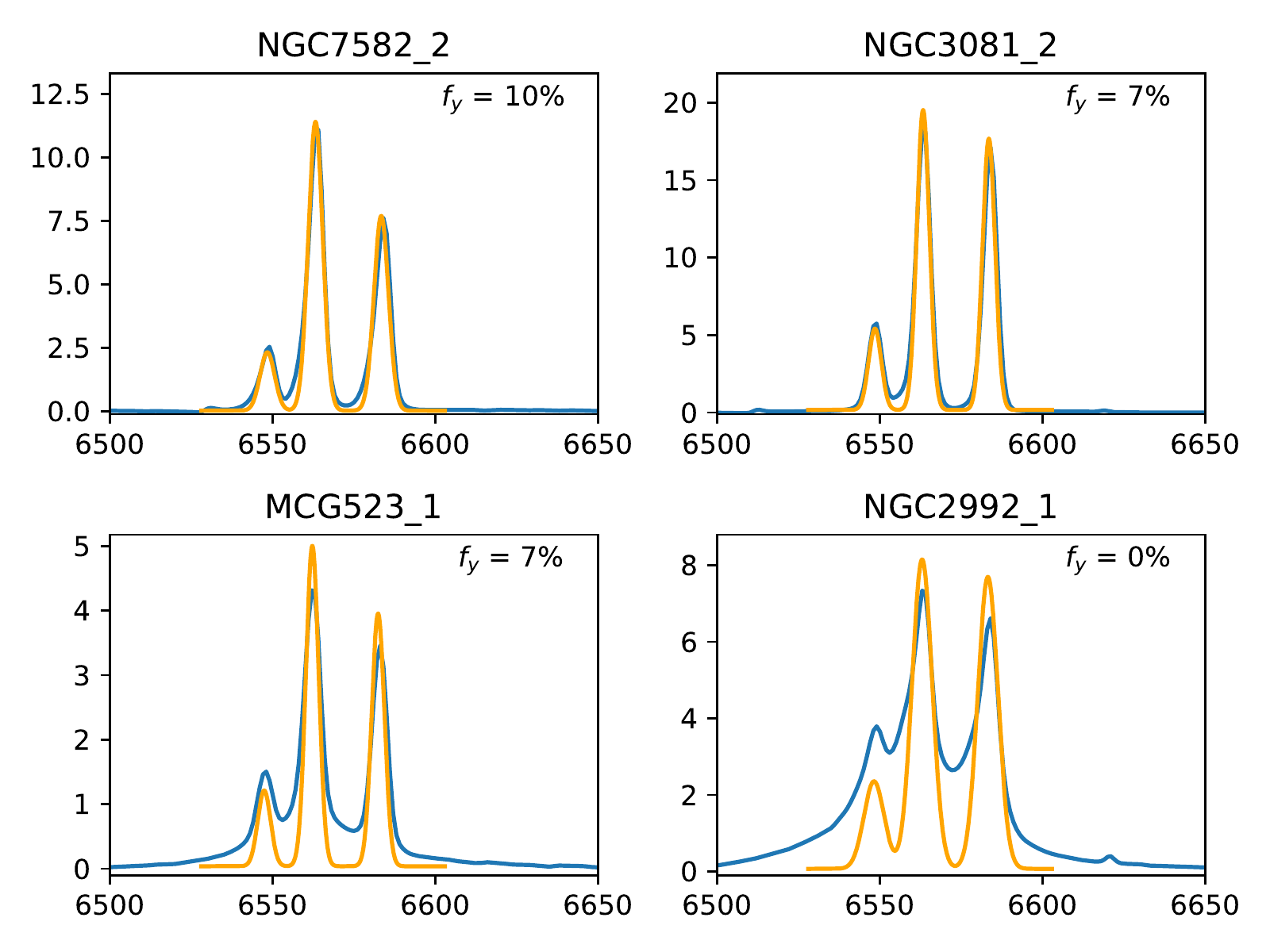}
\caption{\label{fig:broad_ha_vs_young}Presence (or not) of a broad component of the H$\alpha$ line in four AGNs with different fractions of light in the young population ($f_y$) according to the \STARLIGHT{} fit: NGC~7582 and NGC~3081 do not show a broad component to the H$\alpha$ line, and yet STARLIGHT finds a young stellar population in these two galaxies. And, vice versa, in NGC~2992, there is a very strong broad component to the H$\alpha$ line, but no young stellar population is found. Only in MCG-05-23-016 do we find both a young population and a broad component. Overall this confirms the robustness of the detection of young stellar light, and that it is likely not confused with the non-stellar continuum from the AGN.}
\end{figure}

Both a (very) young stellar population and a partly hidden AGN produce a blue and nearly featureless continuum. We therefore need to carefully control whether we can determine the fraction of light in the youngest population robustly in our AGN sub-sample or whether we mis-interpret scattered AGN light as young stars. 

We test this possible ambiguity in two ways:
\begin{enumerate}
	\item We compare the flux in the blue continuum (at 4000 -- 4200 \AA{}) with the hard X-ray flux from the BAT-105 month survey \citep{oh2018}. Figure \ref{fig:young_vs_fx} clearly shows that there is no correlation between these two fluxes, i.e. that there is no direct relation between the hard X-ray flux and the flux in the blue continuum. Pearson's $R$ coefficient for this relation is -0.0064.
	\item We further take a closer look at the type 1.9 AGNs in our sample where we know that direct AGN light is included in our observed spectra since a broad component is observed in the H$\alpha$ line. Figure \ref{fig:broad_ha_vs_young} shows the H$\alpha$ / [N II] complex for two ``regular'' type 2 AGNs with a significant young population and the two type 1.8/1.9 AGNs (NGC~2992 and MCG~-05-23-016) in our sample, one with and one without a young population, according to our \STARLIGHT{} fits. This shows both that a young stellar population is found in spectra {\em without} broad line wings and also conversely that in one spectrum {\em with} broad lines no young stellar population is found. The presence of a young population is therefore not obviously related to the presence of broad lines. This is consistent with the results from \citet{cidfernandes1995} who found a broad component in H $\beta$ only when the scattered AGN light contributes $\gtrsim$ 20 \% of the optical continuum.
\end{enumerate}

We conclude that the fraction of light in the young stellar population is not polluted by AGN light and that, therefore, we can determine the presence of a young stellar population accurately also in the AGN sub-sample. We therefore do not add a power-law component to account for the AGN featureless continuum in our comparative study in order not to confuse our fits \citep[e.g.][]{cidfernandes2004} and to not bias the AGN fit against that of the inactive galaxies.

\subsection{Lower bound on the young stellar populations in AGNs}
\label{sec:lower_age_bound}

We can use the equivalent width of H$\alpha$ ($EW_{\rm H\alpha}$, see Table~\ref{tab:bpt_ratios}) to put additional constraints on the recent star formation history of the AGNs. To do so, we first need to estimate the largest fraction of the H$\alpha$ luminosity that could arise from star formation and for this we use the group of AGNs with the highest $EW_{\rm H\alpha}$. There are a number of reasons why AGNs could have lower $EW_{\rm H\alpha}$, but the maximum values will only be found when both AGN and star-formation contribution are maximal. The group of six AGNs with the highest $EW_{\rm H\alpha}$ consists of ESO 137-G034, NGC~3081, NGC~2110, NGC~7582, NGC~2992, and NGC~5728, and their median $EW_{\rm H\alpha}$ is 70\AA. These also tend to have the highest log [OIII]/H$\beta$, with a median ratio of 1.07. Similarly, their median log [NII]/H$\alpha$ is 0.01. To assess the star formation contribution to the H$\alpha$ flux, we consider lines from the most extreme AGN photoionisation models of \citet{groves2004b}, to the location of solar metallicity star forming galaxies \citep[see also][]{davies_rl2014b}. Since the AGNs are very close to the highest [OIII]/H$\beta$ ratios that the models can produce, we find that at most 10\% of their H$\alpha$ flux could be due to star formation. Applying this correction reduces $EW_{\rm H\alpha}$ that is associated with star formation to $<$ 7\AA.
We also need to make a correction for the old stellar population, since $EW_{\rm H\alpha}$ in models compares the line flux to the continuum due to stars associated with the line emission. We have found that 5\% of the continuum is associated with young stars. We therefore estimate that the maximum $EW_{\rm H\alpha}$ associated with the most recent star forming episode is $<$ 7\AA / 5\%, i.e. $<$ 140\AA.
Comparing this to Starburst99 models (specifically Fig.~83 and 84 in \citet{leitherer1999}) rules out continuous star formation since it would imply a timescale exceeding several Gyr, inconsistent with the stellar population synthesis results.
Instead it means that the recent star formation must have ceased. Based on the rate at which the $EW_{\rm H\alpha}$ falls after star formation stops, we estimate that the end of star formation must have typically been at least 6 Myr previously.

In addition, we can look into the youngest bin of the individual stellar population fits (Appendix~\ref{sec:appendix:fits}) and find that six out of the seven AGNs in which we find such a population show this young population in the $\sim 10^7$ year single stellar population model (and only one in the youngest, $10^6$ year, population), further confirming the interpretation that the youngest stellar population in AGNs has ceased forming stars since at least 6 Myr.

\subsection{Stellar masses}
The stellar masses quoted as the last column of Tab.~\ref{tab:starlight} are derived from the STARLIGHT fitting itself\footnote{Using Eq. (2), page 22, of the \STARLIGHT{} manual, v0.4, but with the correct number for $M_{\odot}$}, i.e. using the decomposition into various stellar populations with its mass/light ratios. The normalisation from measured flux to stellar masses includes the uncertainties of the absolute spectrophotometric flux calibration (better than $\sim 2 \%$, see Section~\ref{subsec:data_reduction}) and the uncertainty in the distance (see \citet{davies2015} for a discussion of the distances used in the LLAMA sample).

The median mass that STARLIGHT finds within our aperture is log $M_{\star}/M_{\odot}$ = 8.8, while the median total mass that we have previously \citep{davies2015} inferred from H band luminosity is log $M_{\star}/M_{odot}$ = 10.35, i.e. a factor of 35 lower (or $\sim 3 \%$). This is roughly consistent with the fact that we cover (in the median galaxy) 15 \% of the mass of the bulge (see Section~\ref{sec:discuss:nonpara}). Since our galaxies have a median bulge/total ratio of 0.25 \citep{lin2018}, this would lead to roughly the $\sim$ 3\% number found above. The masses obtained by STARLIGHT are therefore consistent with the masses obtained from H band luminosity for the entire galaxy and scaled to the fraction of the galaxy that we include in our aperture for this work.

The total mass in the aperture can be used to convert the fraction of mass in a certain age bin to the star-formation {\em rate} on this time scale, e.g. the star-formation rate within the last $10^{7.5}$ years can be computed by evaluating $\mu_y \times M_{\star}/10^{7.5}$ years, where $\mu_y$ is the mass fraction in the young population, given in Tab.~\ref{tab:starlight}.

%% file: sections/5-emission-lines.tex
\section{Emission line diagnostics}
\label{sec:bpt_analysis}

\subsection{Emission line fitting}

In order to allow for a number of diagnostic plots, we determined the flux in the classical BPT and VO07 emission lines \citep{baldwin1981,veilleux1987}, after subtracting the stellar population fit from our spectra. Our fit strategy is as follows: For the [O III] / H$\beta$ region, we first fit a Gaussian and an offset to the [O III] $\lambda$ 5007 line and bind both the mean ($\pm$ 0.5 \AA) and the standard deviation ($\sigma_{\mathrm H \beta} = [0.9 \ldots 1.1] \cdot \sigma_{\mathrm [O III]}$) of the H$\beta$ line to the [O III] fit.

For the region encompassing the H$\alpha$ / [N II] complex as well as the [O I] $\lambda$ 6300 and the [S II] $\lambda$ (6716, 6731) doublet, our fit strategy was similar: In general, we first fitted the [S II] doublet with a constant offset and a Gaussian each, and tied the mean and standard deviation of that fit to the H$\alpha$/[N II] complex as described above. Lastly we fitted [O I] with its mean and standard deviation tied to [S II] as well.

This fitting strategy returned reliable fit results with the {\tt LevMarLSQFitter} of {\tt astropy}'s {\tt modeling} package \citep{astropy2013,astropy2018}, but for several cases we had to adjust the binding conditions manually and/or fix the continuum level in order to get a good fit of the emission line. The results of the emission line fitting are presented in Tab.~\ref{tab:bpt_ratios}.

We also show the extinction $A_{\rm V}$ from the Balmer decrement (see discussion in Section~\ref{sec:extinction}) and the equivalent width of H$\alpha$. For the emission lines used in the BPT and VO87 diagnostic diagrams, we give the logarithmic flux ratios rather than the individual fluxes. For the actual fluxes, we refer to \citet{davies2020} who have measured the fluxes of the main emission lines using higher spectral resolution stellar templates compared to the lower spectral resolution single stellar population models, used in this paper. \citet{davies2020} also show that emissions lines in AGNs are generally dominated by outflows, although few have as complex profiles as the LLAMA galaxy NGC~5728 \citep{shimizu2019}. For the purpose of determining the dominant ionisation mechanism, integrating the entire line flux as we do in this paper is, however, still applicable.

\begin{table*}
\caption{Results of the emission line fitting}
\label{tab:bpt_ratios}
\centering
\begin{tabular}{ccccccc}
\hline\hline
Galaxy name & $A_V$ & EW (Ha) & log([OIII]/H$\beta$) & log([NII]/H$\alpha$) & log([SII]/H$\alpha$) & log([OI]/H$\alpha$]) \\
\hline
NGC 2110 & 2.2 & 75.3 & 0.71 & 0.11 & 0.05 & -0.4 \\
NGC 2992 & 4.51 & 62.6 & 1.05 & -0.03 & -0.31 & -0.96 \\
MCG-05-23-016 & 1.72 & 24.4 & 1.08 & -0.1 & -0.46 & -0.9 \\
NGC 3081 & 0.88 & 94.0 & 1.14 & -0.04 & -0.3 & -0.99 \\
ESO 021-G004 & 2.35 & 2.5 & 0.77 & 0.3 & 0.17 & -0.47 \\
NGC 5728 & 1.63 & 53.1 & 1.09 & 0.12 & -0.16 & -0.91 \\
ESO 137-G034 & 2.24 & 116.9 & 1.09 & 0.04 & -0.13 & -0.83 \\
NGC 7172 & 1.89 & 7.1 & 0.57 & 0.04 & -0.34 & -1.02 \\
NGC 7582 & 2.44 & 65.5 & 0.39 & -0.17 & -0.54 & -1.64 \\
\hline{}
NGC 718 & 0.41 & 0.9 & 0.44 & 0.4 & 0.2 & -0.63 \\
NGC 1079 & 1.08 & 19.2 & -0.13 & -0.33 & -0.51 & -1.91 \\
NGC 1947 & 4.37 & 4.8 & 0.35 & -0.06 & 0.12 & -0.52 \\
ESO 208-G021 & 1.35 & 2.3 & 0.1 & 0.18 & 0.14 & -0.67 \\
NGC 3175 & 3.26 & 26.4 & -0.76 & -0.4 & -0.64 & -1.74 \\
NGC 3351 & 0.85 & 6.6 & -0.33 & -0.2 & -0.59 & -1.46 \\
ESO 093-G003 & 4.96 & 17.6 & -0.2 & -0.34 & -0.42 & -1.46 \\
NGC 3717 & 2.76 & 8.3 & -0.4 & -0.26 & -0.51 & -1.38 \\
NGC 3749 & 1.76 & 2.4 & -0.04 & -0.05 & -0.3 & -0.77 \\
NGC 4224 & 3.25 & 1.9 & 0.45 & 0.16 & 0.02 & -0.34 \\
NGC 4254 & 1.35 & 6.3 & 0.59 & -0.12 & -0.61 & -1.18 \\
NGC 5037 & 2.95 & 3.0 & 0.57 & 0.32 & 0.06 & -1.0 \\
NGC 5921 & 2.01 & 6.2 & -0.0 & -0.02 & -0.17 & -0.86 \\
IC 4653 & 1.16 & 62.7 & -0.01 & -0.56 & -0.47 & -1.66 \\
NGC 7727 & 0.07 & 1.5 & 0.16 & 0.2 & 0.06 & -0.73 \\
\hline
\end{tabular}
\tablefoot{
$A_{\rm V}$ in magnitudes from the Balmer decrement, following the prescription in \citet{dominguez2013}, equivalent width of H$\alpha$ in \AA; logarithmic flux ratios for the BPT and VO87 diagnostic diagrams.
}
\end{table*}

\subsection{Emission line diagnostic diagrams}

\begin{figure}
\includegraphics[width=\columnwidth]{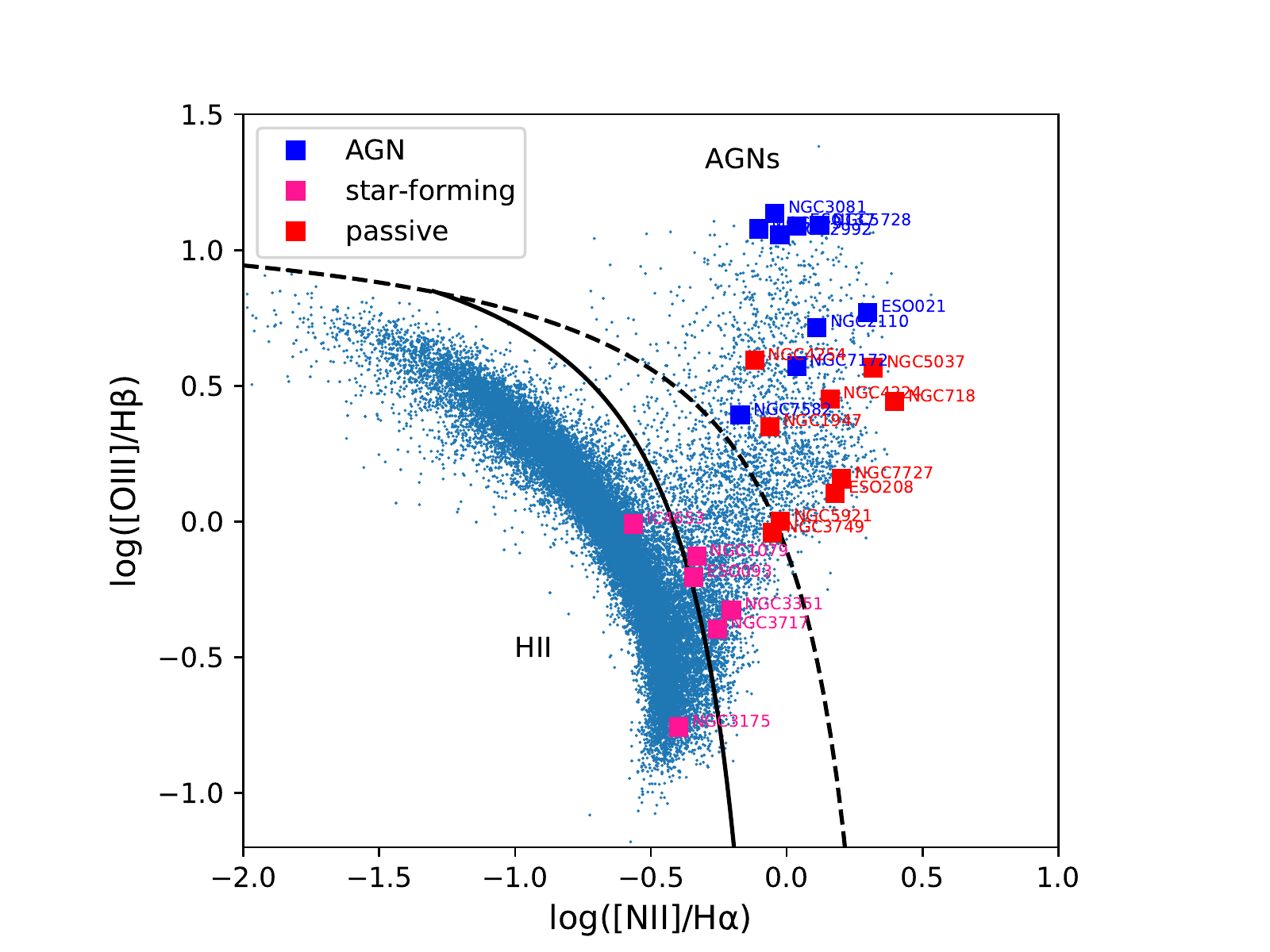}
\caption{\label{fig:BPT_N2}BPT plot with [N II]/H$\alpha$ vs. [O III]/H$\beta$. Galaxies from the AGN as well as the control sample of star-forming and passive galaxies are shown in blue, deep pink, and red, respectively. The empirical \citep{kauffmann2003c} and theoretical \citep{kewley2001} boundaries between starburst and AGN galaxies are marked with black straight and dashed lines, respectively. The underlying blue cloud of points is from the MPA-JHU emission line analysis of SDSS DR7 galaxies \citep{brinchmann2004}.}
\end{figure}

\begin{figure}
\includegraphics[width=\columnwidth]{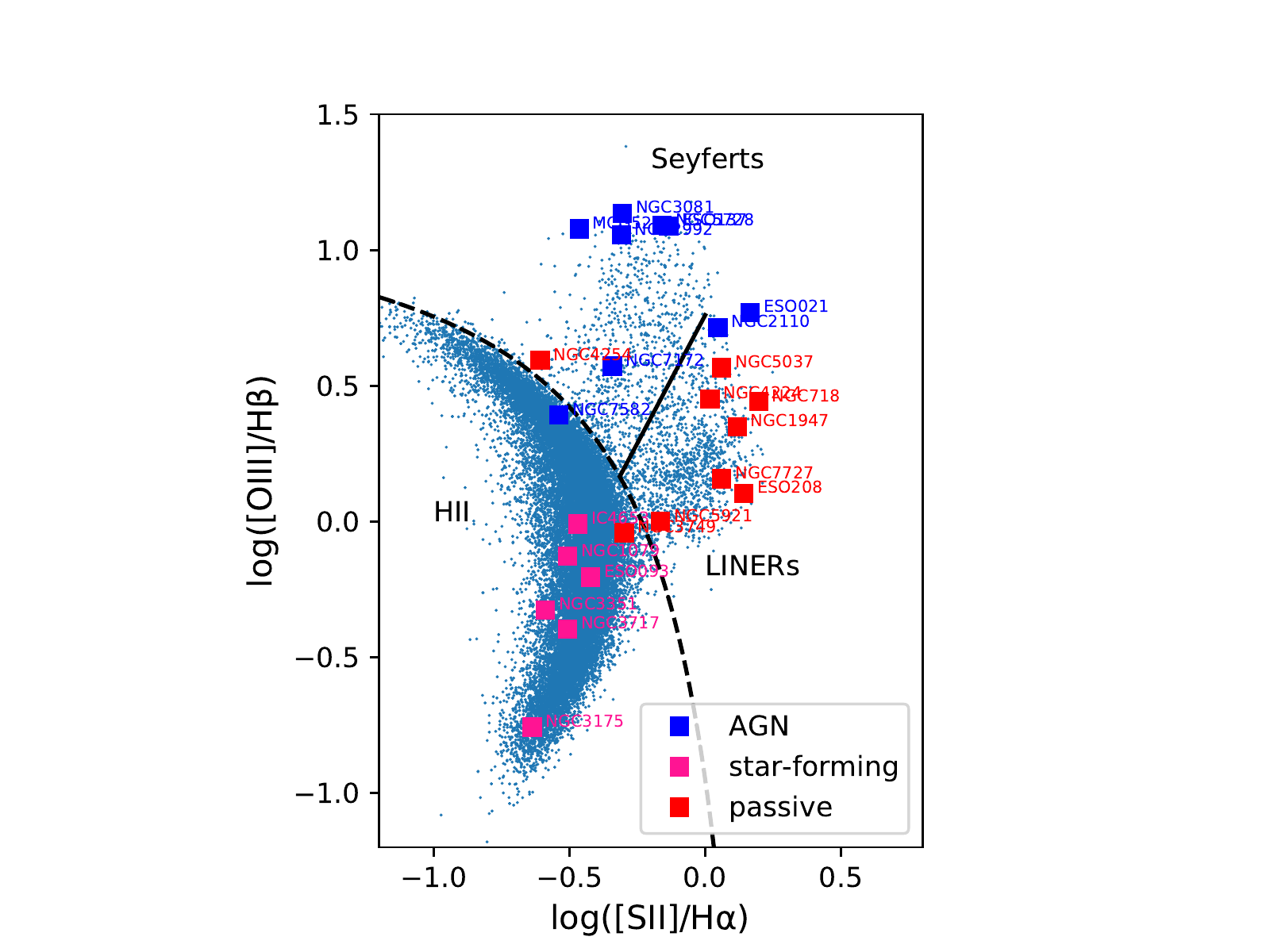}
\caption{\label{fig:BPT_S2}Same as Fig. \ref{fig:BPT_N2}, but here for [S II]/H$\alpha$ vs. [O III]/H$\beta$ (``VO87 plot'').}
\end{figure}

\begin{figure}
\includegraphics[width=\columnwidth]{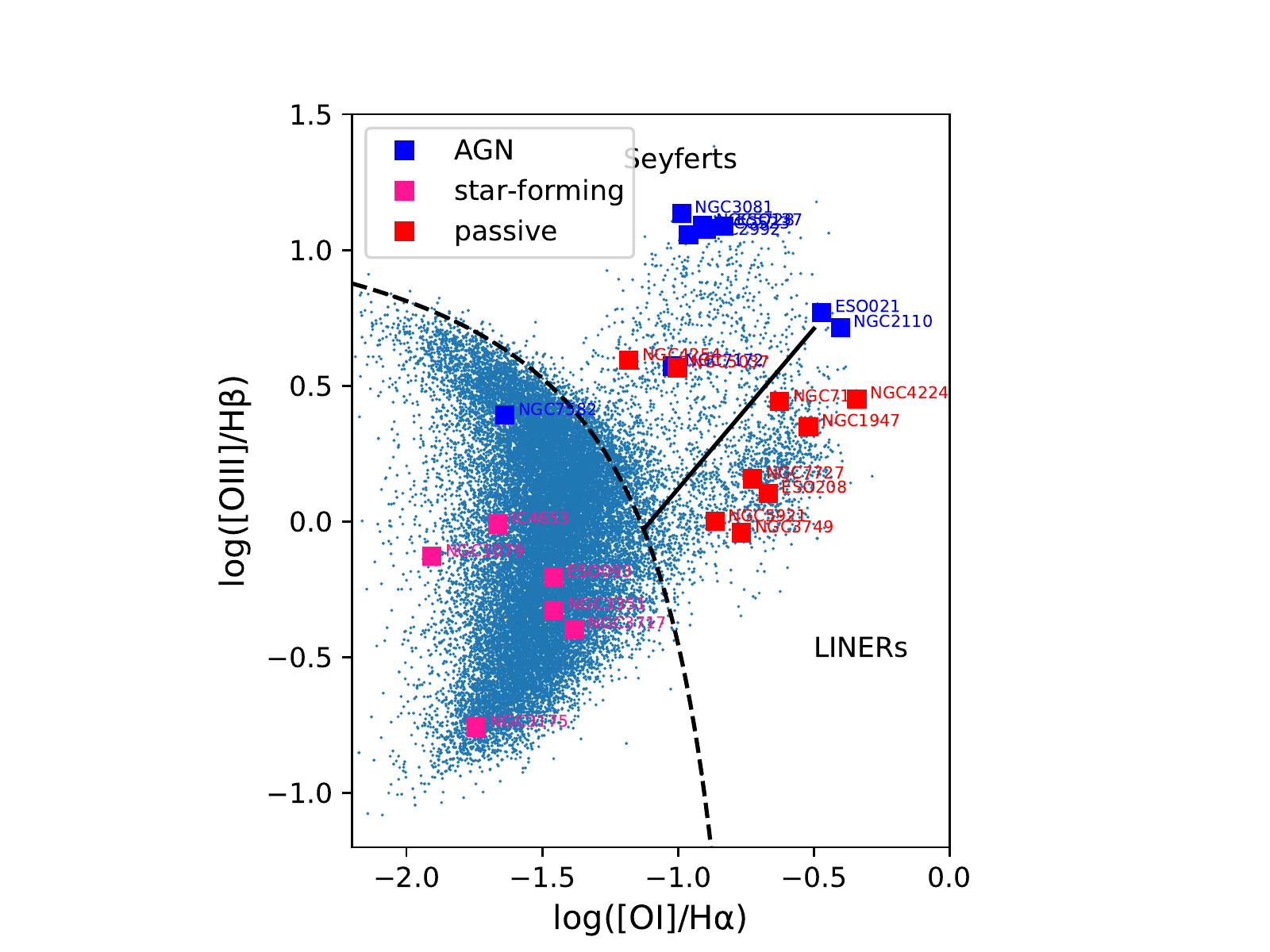}
\caption{\label{fig:BPT_OI}Same as Fig. \ref{fig:BPT_N2}, but here for [O I]/H$\alpha$ vs. [O III]/H$\beta$}
\end{figure}

\begin{figure}
\includegraphics[width=\columnwidth]{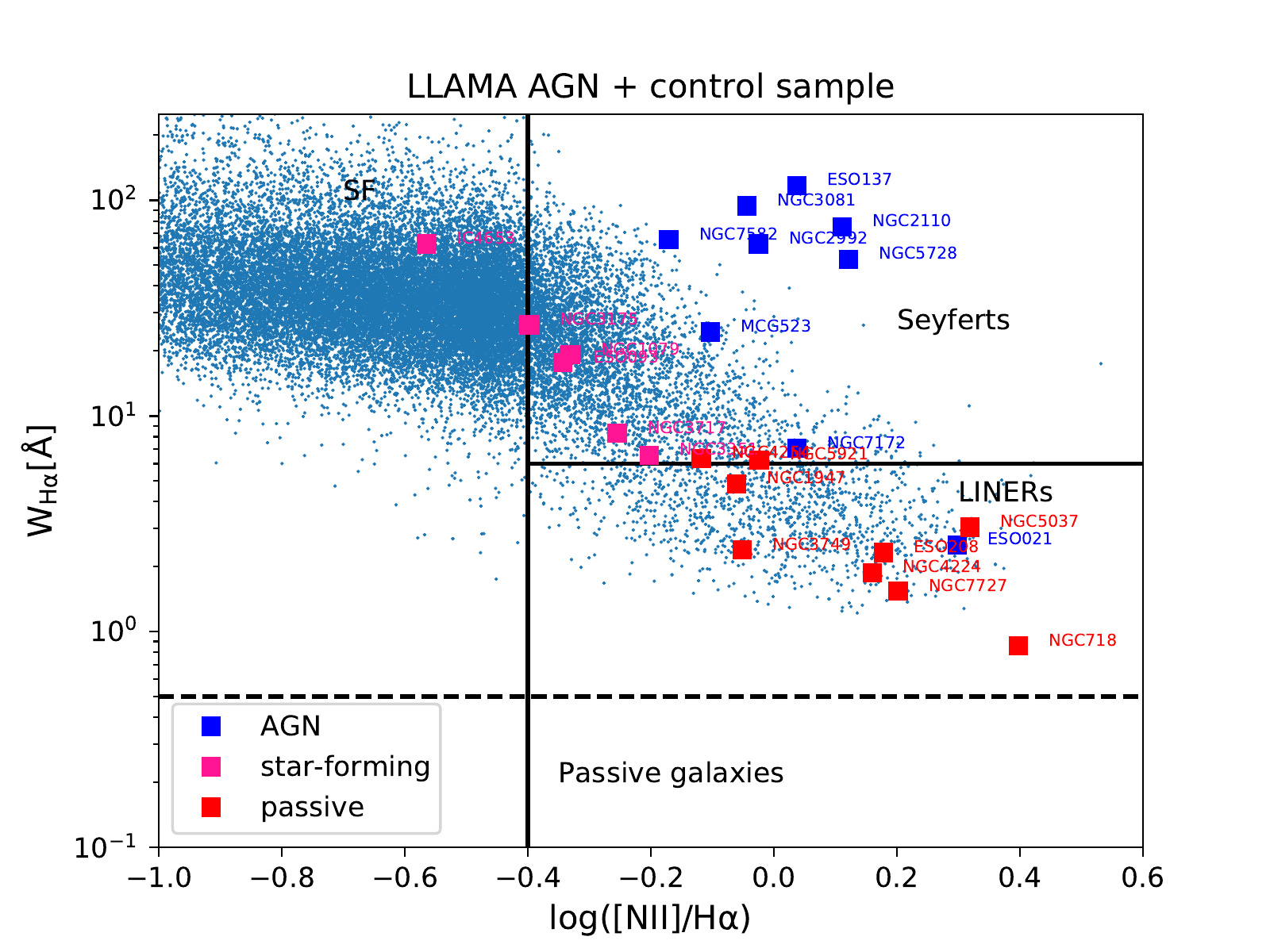}
\caption{\label{fig:whan}``WHaN plot'' showing the equivalent width of H$\alpha$ on the y axis vs. the BPT flux ratio [N II]/H$\alpha$. Galaxies from the AGN as well as the control sample of star-forming and passive galaxies are shown in blue, deep pink, and red, respectively. This plot is useful to show the relative strength of emission line features compared to the continuum. The demarcation lines are from \citet{cidfernandes2011}, the blue cloud of points in the background is again from \citet{brinchmann2004} as in Figs. \ref{fig:BPT_N2} -- \ref{fig:BPT_OI}.}
\end{figure}

In Figs.~\ref{fig:BPT_N2}--\ref{fig:BPT_OI} we show the location of our sample on the classical ``BPT'' plots including the extension to [S II] $\lambda$ (6716, 6731) by \citet{veilleux1987}. We further show the (empirically derived) boundary between starburst galaxies and AGNs from \citet{kauffmann2003c} on the [N II]/H$\alpha$ vs. [O III]/H$\beta$ plot as well as the (theoretically found) ``maximum starburst line'' from \citet{kewley2001} on all emission line diagrams. For reference, we also show the corresponding line ratios from the MPA-JHU emission line analysis of the SDSS DR 7 galaxies \citep{brinchmann2004} in our redshift range where {\em all} the six relevant lines (required to produce the four ratios shown here) have been detected at $> 5 \sigma$.

It is noteworthy to mention that all nine AGNs and 14 out of 17 control galaxies are shown on these plots, i.e. our spectra have sufficient SNR to identify all six BPT/VO87 lines in all objects except three non-star-forming (passive/retired) control galaxies. These passive galaxies, where no single BPT line is robustly detected, are NGC~1315, NGC~2775, and NGC~5845. It is furthermore useful to also show our objects on the ``WHaN'' (equivalent width of H$\alpha$ vs. $\log$ [N II]/H$\alpha$) plot, Fig.~\ref{fig:whan}, which allows to discriminate better between bona-fide AGN and retired galaxies with low H$\alpha$ equivalent widths \citep{cidfernandes2010,cidfernandes2011}. Indeed, our AGNs are better separated from the control galaxies in the WHaN plot compared to the BPT/VO87 plots. 

In the BPT/VO87 plots, our AGN population is split into a group of four galaxies that are in the ``extreme'' Seyfert regime, another one (NGC 7172) that would also be clearly classified as an AGN and three AGNs (NGC~2110, ESO~021-G004, and NGC~7582) that would not be classified as an AGN based on their BPT/VO87 optical emission line ratios; NGC~2110 would, however, be classified as an AGN in the WHaN diagram. This is similar to the results for a larger sample of ultra-hard X-ray selected sources from the {\em BAT AGN Spectroscopic Survey} \citep[BASS; ][]{koss2017} where the authors find that only half of their sources would be selected as AGNs based on optical emission lines.

A similarly split view is seen in our control population. Six of our control galaxies are clearly on the sequence of star-forming galaxies as seen from the [N II]/H$\alpha$ vs. [O III]/H$\beta$ plot, and we will refer to them as ``star-forming control galaxies'' in the further analysis. But what is the nature of the other eight galaxies (+3 that are not detected in H$\alpha$)? The clearest separation is seen in the [O I]/H$\alpha$ plot, Fig.~\ref{fig:BPT_OI}, from which one would deduce they are in the ``LINER'' regime which in this case is most likely an indication for an old stellar population since we tested the control sample rigorously for the existence of any nuclear (AGN) activity. The odd-one out is NGC~4254, an inactive control galaxy according to our initial classification, which is located in the Seyfert region (see below for a discussion on this). We refer to this latter population as ``non-star forming control galaxies''.

Since galaxies are more likely identified as AGN using nuclear spectra (such as ours) rather than integrated spectra \citep[e.g.][]{gavazzi2018}, the fact that none (except perhaps one, see below) of our control galaxies are in the AGN regime can be considered as additional evidence showing the purity of our control sample (we know that our AGNs are bona-fide AGNs due to the ultra-hard X-ray selection).

\subsection{Metallicities from the BPT diagrams}
\citet{kewley2006} point out that star-forming galaxies form a tight sequence from low metallicities (low [NII]/\halpha{}, high [O III]/\hbeta{}) to high metallicities (high [N II]/\halpha{}, low [OIII]/\hbeta{}). This can be seen in the photoionisation calculations presented in \citet{kewley2001}. While \citet{kewley2002} caution that [NII]/\halpha{} alone is not a good metallicity indicator because of its dependence on ionisation parameter, they also note that if it can be confirmed that ionisation ratios are due to star formation rather than AGNs or shocks -- for example by location in diagnostic diagrams -- it can be used to provide a zeroth order estimate of metallicity. Our inactive galaxies form a sequence that extends to the star-forming region of the BPT corresponding to approximately solar metallicity abundances \citep[see e.g. Fig 9 of][]{kewley2001}. And we take that as confirmation that we can adopt solar abundance for our stellar population fitting as described in Section~\ref{sec:metallicity}.

\subsection{The ``misfit'' in the BPT relation: NGC~4254}
The nuclear emission line ratios in NGC~4254 place this object on the Seyfert side of the extreme starburst line in diagnostic plots. If one did an optical emission line selection, this object would be included as an AGN. However, both the available X-ray and mid-infrared observations indicate that there is no significant AGN in NGC~4254.
The 0.3-8~keV nuclear luminosity \citep{grier2011} is $\log L$/erg/s $<$ 38.2, adjusted to our adopted distance of 15~Mpc.
This is well below that seen at 2-10~keV for the low luminosity AGNs in \citet{asmus2011}, and even below the typical range of star forming galaxies. Similarly, the 3$\sigma$ limit on the 10$\mu$m luminosity in a 6\arcsec{} aperture of $\log L$/erg/s $<$ 40.6 \citep{scoville1983} is well below that typically seen in Seyferts and LINERs.

Can the emission line ratios in NGC~4254 be due to shocks?
While \citet{allen2008} show that models of shocks with speeds of $\sim$ 400km/s with a precursor (i.e. an H II region in front of the shock caused by ionising photons from the shock front) can produce high [OIII]/H$\beta$ ratios, these are typically associated with lower [NII]/H$\alpha$, and higher [SII]/H$\alpha$ and [OI]/H$\alpha$ ratios than measured in NGC~4254.
However, we can consider whether the addition of shock excitation to stellar photoionisation (by either young stars or post-AGB stars) can move the location of a galaxy from where we find the other inactive galaxies to where we find NGC~4254. This could indeed reproduce [OIII]/H$\beta$ and [NII]/H$\alpha$, but not the other ratios. This is because it produces too much [SII] and [OI] to be able to match the ratios involving H$\alpha$.

While one can find a particular metallicity and ionisation parameter that roughly matches the location of NGC~4254 \citep{stasinska2008}, it is hard to understand why this object should be so different from the other inactive galaxies.

We are left with speculating that NGC~4254 may be a recently turned off AGN which would explain its lack of X-ray and mid-IR emission (that are generated on scales of a few pc). The larger narrow-line region could then be the afterglow of such a recently turned off AGN.

Alternatively, it could indeed be a very low luminosity AGN. The typical 12$\mu$m luminosity of the AGNs in our sample is $\log L_{12 \mu{}m}$/(erg/s) = 42.8 and the typical observed 2-10~keV luminosity is $\log L_{\rm 2-10 keV}$/(erg/s) = 42.2. Comparing these to the luminosities above (and not worrying that the bands are slightly different) suggests that if it is an AGN, NGC~4254 is at least two orders of magnitude lower luminosity than our sample, and that in addition there is likely to be a high X-ray absorbing column. If all the H$\alpha$ flux is due to the AGN, one would expect the equivalent width of H$\alpha$ to be proportionally lower than in the other AGNs. However it is only a factor 10, rather than 100, less than the majority of the AGNs in our sample. As such, much of the line flux should originate in other processes. But, as for the shocks, considerations of mixing also do not explain the low [SII]/H$\alpha$ ratio.

\subsection{Extinction estimates: Comparison between SPS fitting and Balmer decrement analysis}
\label{sec:extinction}

\begin{figure}
\includegraphics[width=\columnwidth]{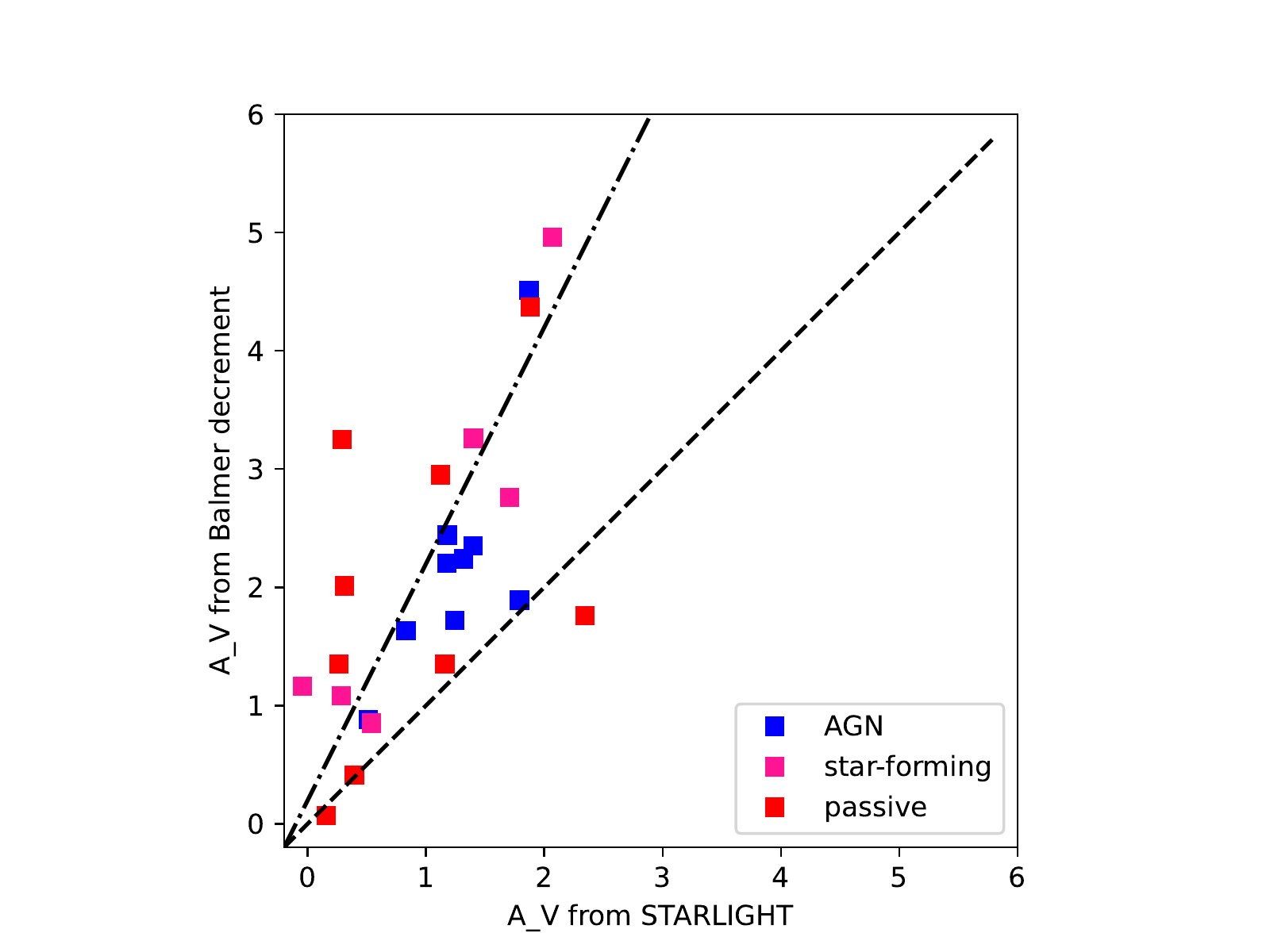}
\caption{\label{fig:A_V}Comparison of the extinction estimate from emission line fitting vs. absorption-line fitting (\STARLIGHT{}). The dashed line corresponds to a 1:1 ratio, the dash-dotted line is a ratio of 2:1. Statistical uncertainties in both methods are minute ($\lesssim$ 0.2 mag) and the systematic difference between the two estimates is discussed in the main text. Galaxies from the AGN as well as the control sample of star-forming and passive galaxies are shown in blue, deep pink, and red, respectively.}
\end{figure}

In Fig.~\ref{fig:A_V} we show the comparison of the extinction estimate from our absorption-line fitting (\STARLIGHT{}) and from the emission line fitting discussed above. In the latter, we convert the Balmer decrement to an extinction $A_V$ using the Eqs. 4 and 7 from \citet{dominguez2013} who in turn use the ``Calzetti reddening law'' \citep{calzetti1994,calzetti2000}. They assume an intrinsic H$\alpha$/H$\beta$ ratio of 2.86, which is applicable for star forming galaxies, but not strictly for AGNs, where a better value to use would be 3.1 \citep{wild2011}. However, since we take a logarithm of that number to derive the extinction in magnitudes, the impact on $A_{\rm V}$ is in fact only $\sim 0.03$, i.e. insignificant. We therefore opt for just using one reddening law for the all galaxies.

Rather than following a 1:1 line (dashed line in Fig.~\ref{fig:A_V}), the $A_V$ as derived from the emission line fitting is {\em higher} than the extinction derived from the absorption line fitting by roughly a factor of two (dash-dotted line in Fig.~\ref{fig:A_V}). This is to be expected since the Balmer decrement preferably probes regions of active star formation, typically embedded in dust, while the extinction from \STARLIGHT{} is dominated by light from old stars within an essentially dust-free environment. The reddening law \citep{calzetti1994,calzetti1997}, as stated more simply at the bottom of page 1394 of \citet{foersterschreiber2009}, gives $A_{\rm V}$(nebular) = 2.27 $A_{\rm V}$(stars), which is consistent with the relation we find, too. A similar result has recently been shown with the larger sample of MaNGA galaxies by \citet{riffel_r2021}.

%% file: sections/6-sample-discussion.tex
\section{Population analysis}
\label{sec:discussion}

\subsection{Non-parametric star-formation histories}
\label{sec:discuss:nonpara}

The result of the stellar population synthesis is a so called stellar population vector that contains the fractional contribution of light that the fitting tool ascribes to the included base spectra. In this so-called non-parametric description of the star formation history (SFH) of a galaxy, stellar ages are described in discrete age bins. It is good to keep in mind how some simple star-formation histories would look like in this description. For example, if the star formation history is composed of a number of bursts, we expect to find a depression of light ascribed to an intermediate-age stellar population. This is a combined effect of mass/light ratio (favouring the very young populations) and logarithmic binning on the age axis (favouring the oldest populations) which results in a ``bias'' against an intermediate age population.

We can in addition ask observationally, how large the fraction of light of the oldest population should be. Photometric decomposition of many of the LLAMA galaxies have shown a large variety of bulge sizes \citep{lin2018} with a median effective radius $r_e = 4\farcs5 \pm 3\farcs8.$ Compared to our fixed aperture of $1\farcs8$ (side of a square extraction region), and using de Vaucouleur's law \citep{devaucouleur1948}, this means that we cover $\sim 15\%$ of the {\em mass} of the bulge in our aperture.

Our observations are further designed such that only the light from the nuclear stellar population is seen (apart from the ``foreground'' bulge light) and the light from the circum-nuclear stellar ring is excluded (see Sec.~\ref{sec:aperture}). The mass in the nuclear stellar population is about ten times larger than the mass of the super-massive black hole \citep[e.g.][]{davies2007,schartmann2009}, and the mass of the bulge is $\sim$ 400-700 times the mass of the black hole\footnote{\citet{haering2004} give a ratio of $\sim$ 700, \citet{mcconnell2013} look at this ratio for early-type galaxies and give a ratio of about 400 for our typical black holes mass of $\sim 3\times10^7 M_{\odot}$ \citep{caglar2020}. \citet{davis2019} (Eq. 11, noting that $\nu$=1 from their Eq. 10) give a ratio of $\sim$ 500 for our typical black hole mass.}. From this we can infer that the bulge should contribute $\sim 15\% \times 500/10 \sim 7.5 \times$ more mass than, or about 90\% of the mass of the nuclear stellar populations. This is consistent with the $\sim$ 90\% mass fraction we find from the \STARLIGHT{} fitting (see Table~\ref{tab:starlight_summary}). This simple calculation is not supposed to lead to an accurate stellar light fraction of the bulge, but should clarify that we expect to see a significant fraction of light in the oldest stellar population. Additional uncertainties include the large variance of apparent bulge sizes in our sample \citep{lin2018} as well as the likely pseudo-bulge nature of many of our galaxies \citep{caglar2020}.

\subsection{Summarised star formation histories}
\label{sec:llama_sfh}

\begin{figure}
\includegraphics[height=0.85\textheight]{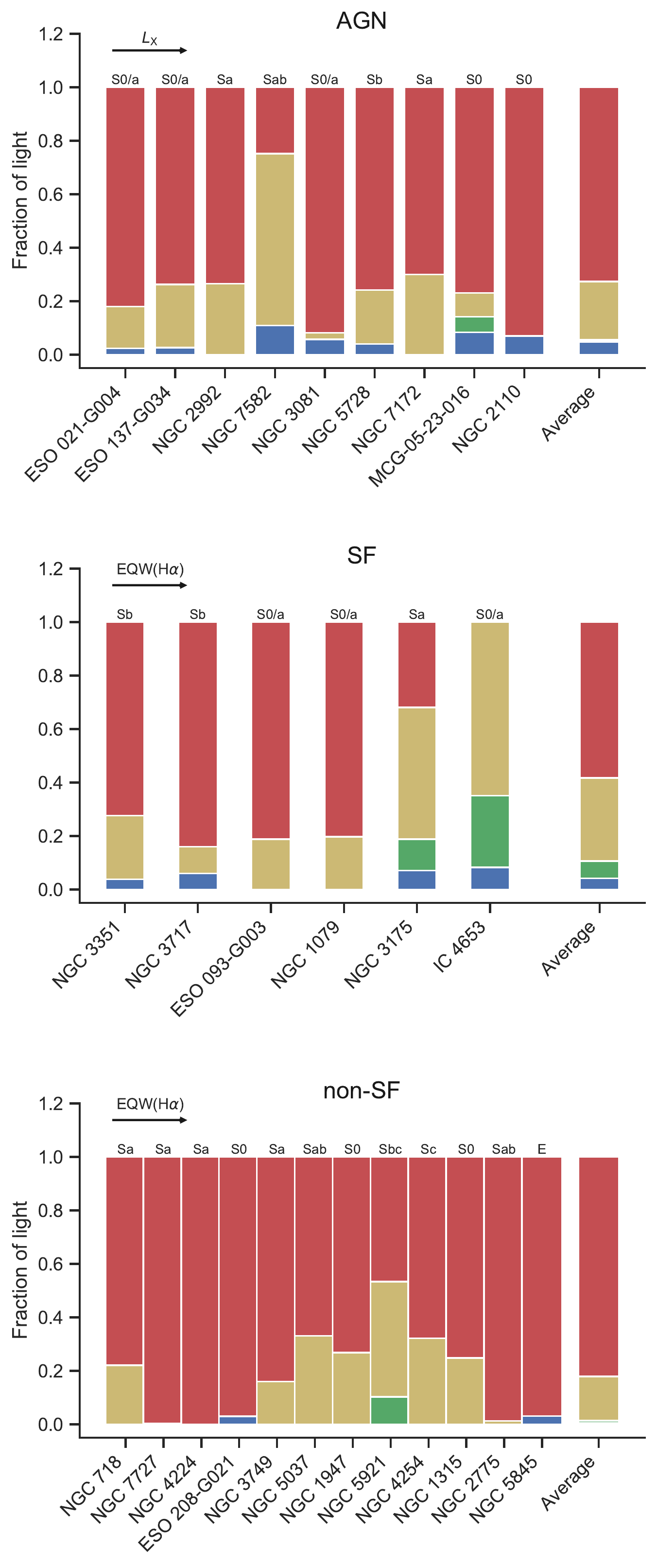}
\caption{\label{fig:sfh_result_sample}Comparison of stellar populations between the AGN and inactive sample where the inactive sample is split by location on the BPT diagram into a star-forming (SF) sub-population and a non-star-forming (non-SF) sub-population. The stellar populations are split into the same four age bins as in the previous section, i.e. young (blue), young-intermediate (green), intermediate-old (yellow), old (red) with boundaries at $\log(age/yr)$ = 7.5, 8.5, 9.5. The AGN hosting galaxies are sorted by increasing X-ray luminosity $L_X$ and the inactive control population is sorted by increasing equivalent width of H$\alpha$ (EQW(H$\alpha$)).}
\end{figure}

\begin{figure}
\includegraphics[width=\columnwidth]{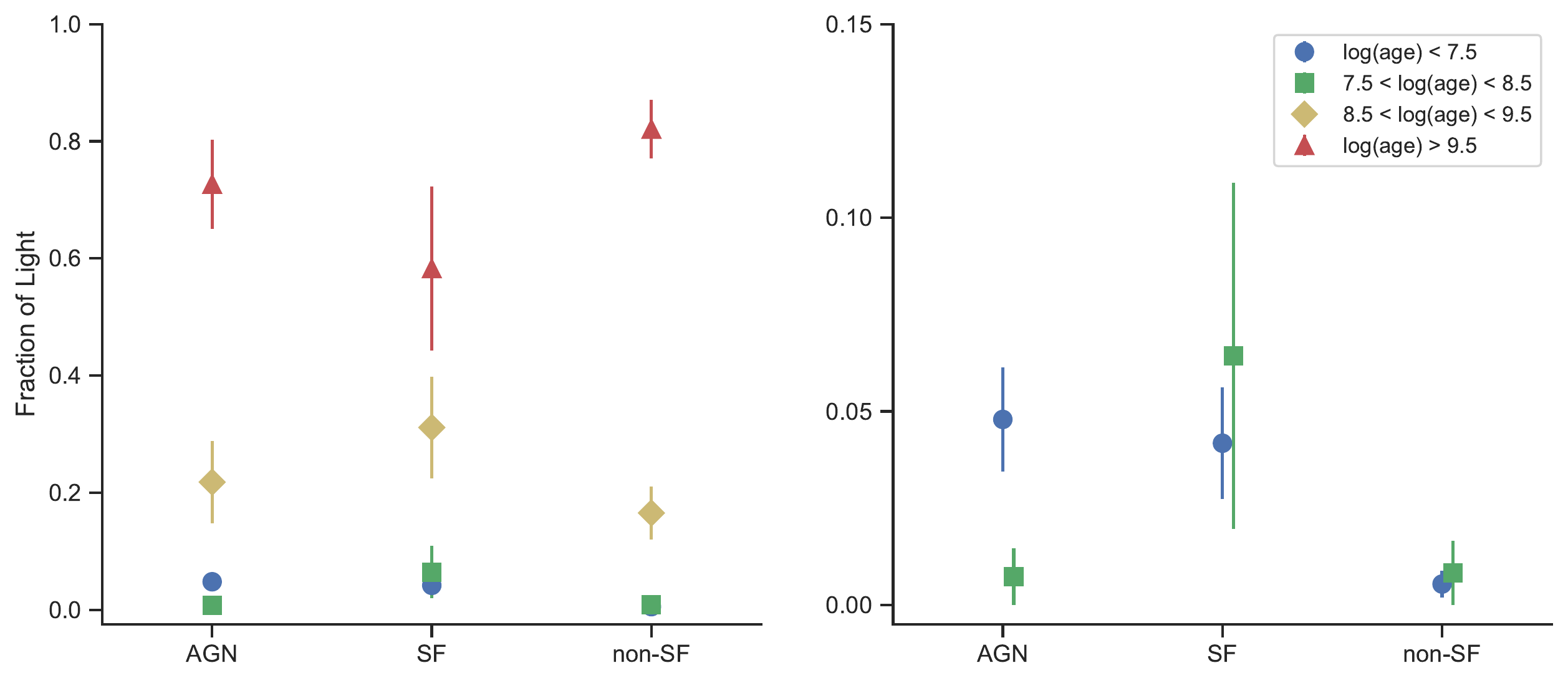}
\caption{\label{fig:sfh_result_sample_detail}Average star-formation histories of AGNs, star-forming and non-star-forming control galaxies as in Fig.~\ref{fig:sfh_result_sample} (left panel) and zooming into the youngest stellar populations (right panel). The uncertainties are derived through jack-knifing, see text for further discussion.}
\end{figure}

\begin{table*}
\caption{Summary of stellar populations by galaxy class with uncertainties from jackknifing as described in the text.}
\label{tab:starlight_summary}
\centering
\begin{tabular}{cccccccccc}
\hline\hline
class & $x_y$ & $x_{yi}$ & $x_{io}$ & $x_o$ & $\mu_y$ & $\mu_{yi}$ & $\mu_{io}$ & $\mu_o$\\
&[\%]&[\%]&[\%]&[\%]&[\%]&[\%]&[\%]&[\%]\\
\hline
AGN&
$ 4.53 \pm  1.24$&
$ 0.64 \pm  0.64$&
$21.32 \pm  6.37$&
$73.49 \pm  6.62$&
$ 0.07 \pm  0.02$&
$ 0.04 \pm  0.04$&
$ 6.91 \pm  2.61$&
$93.00 \pm  2.62$&
\\
SF&
$ 4.18 \pm  1.43$&
$ 6.52 \pm  4.57$&
$30.98 \pm  8.56$&
$58.32 \pm 14.03$&
$ 0.08 \pm  0.05$&
$ 2.62 \pm  2.21$&
$20.67 \pm 13.33$&
$76.65 \pm 15.58$&
\\
non-SF&
$ 0.53 \pm  0.34$&
$ 0.86 \pm  0.85$&
$15.00 \pm  4.29$&
$83.61 \pm  4.70$&
$ 0.00 \pm  0.00$&
$ 0.13 \pm  0.12$&
$ 5.17 \pm  1.71$&
$94.71 \pm  1.79$&
\\
\hline
\end{tabular}
\tablefoot{
Quoted are the fractional contributions to the light (mass) in percent in the four bins designated young $x_y$ ($\mu_y$), young-intermediate $x_{yi}$ ($\mu_{yi}$), intermediate-old $x_{io}$ ($\mu_{io}$) and old $x_o$ ($\mu_o$). They are defined by their boundaries of log (age/yr) = 7.5, 8.5, 9.5.
}
\end{table*}

Looking at the summarised star formation histories it is useful to split our sample not only in AGN and control galaxies, but to further discriminate between star-forming and non-star-forming control galaxies, as defined in Sec.~\ref{sec:bpt_analysis}. In Figs.~\ref{fig:sfh_result_sample} and \ref{fig:sfh_result_sample_detail} we show the summarised stellar population vectors for all galaxies with a successful stellar population fit and binned into the four age ranges.

Apart from the large fraction of light from old stars that is expected (see above), the most obvious observation is that seven out of nine AGNs show a measurable fraction of young stellar light. This is consistent with previous observations that (Seyfert-type) AGN activity is often associated with young nuclear stellar populations \citep[e.g.][]{cidfernandes2004}. We further find a similar occurrence rate of young stars in star-forming control galaxies (four out of six galaxies). This may sound trivial, but it is actually a solid cross-check between the absorption and emission line analysis since our SFHs are only derived from the absorption line fitting while our classification is solely based on emission line diagnostics. The SFH for non-star-forming galaxies, on the other hand, is dominated by old stars.

We evaluate the significance of this difference from a jackknife resampling test\footnote{We use \texttt{jackknife\_stats} from \texttt{astropy.stats} with a $1 \sigma$ confidence level.} which indicates the uncertainty due to our relatively small sample size, the dominant uncertainty in this analysis. The average and $1 \sigma$ uncertainties determined in this way are given in Tab.~\ref{tab:starlight_summary} and are shown in Fig.~\ref{fig:sfh_result_sample_detail}. Both the AGN and the star-forming inactive control galaxies show a significant fraction of light in our youngest age bin. The two sub-samples are indistinguishable within the uncertainties in terms of their nuclear young stellar population but are both very different, at the $\sim 3 \sigma$ and $\sim 2.5 \sigma$ levels respectively, to the non-star-forming control population. In terms of the young-intermediate population, however, AGNs appear identical to the non-star-forming control population and the fraction of light in this age bin for the star-forming galaxies is dominated by a single very young galaxy, IC~4653. Without this galaxy, the fraction of light in the intermediate-young population for the star-forming sub-sample would be 2.27 $\pm$ 2.25 \%, i.e. consistent with the AGN and non-star-forming sub-samples.

This finding is consistent with the idea that any galaxy that has a sufficient amount of cold, molecular gas close to its nucleus will turn ``on'' as an AGN eventually \citep[e.g.][]{davies2012} and that accretion onto black holes is a highly variable process \citep[e.g.][]{novak2011} that quickly turns AGNs (hosting nuclear starbursts) into inactive nuclear star-forming galaxies \citep[e.g.][]{hickox2014}. Regarding their star-formation histories, the LLAMA star-forming control galaxies could indeed be the parent population of the LLAMA AGNs.

\subsection{Number densities of AGNs and star-forming galaxies and an indication for the AGN lifetime}
If we identify the LLAMA star-forming control galaxies with the parent population of LLAMA AGNs, we can infer the occurrence rate of AGNs -- at the given luminosity threshold and at the stellar masses probed in LLAMA -- thanks to the volume-completeness of our sample. For this test, we select all star-forming galaxies from the MPA-JHU catalogue of SDSS DR7 galaxies\footnote{These galaxies have been identified in the SDSS DR7 spectroscopic sample using BPT line ratios. While the SDSS fibre aperture is almost twice as large (3\farcs0) as our extraction aperture (1\farcs8), we do not expect that this leads to a significant bias when comparing our sample to the SDSS sample since most of our (powerful) AGNs are clearly detected as AGNs in the emission-line diagnostic diagrams.} \citep{brinchmann2004} within our redshift limits of $0.005 < z < 0.01$ and above our lower stellar mass limit of $\log M_{\star}/\log M_{\odot} \gtrsim$ 9.5.\footnote{We note that the stellar mass distribution of so-selected SDSS star forming galaxies is different from the stellar mass distribution of our (LLAMA) AGNs. This is expected since BAT-selected AGNs are known to be found preferentially in the most massive host galaxies \citep{koss2011}. Our sample is too small to differentiate the AGN incidence per stellar mass bin and the analysis presented here should only be interpreted as an average number for massive galaxies.}

We apply a correction for the respective areas probed: LLAMA applied a cut for $\delta < 15 \deg$, i.e. it covered $1/2 + \int_0^{15 \deg}\cos\delta {\rm d}\delta \approx 3/4$ of the sky while SDSS-II DR7 (Legacy survey/spectroscopy) covered about 8000 deg$^2$. Since the whole sky is $4 \pi (180/\pi)^2 \approx 41000$ deg$^2$ the area correction factor is 3.8 (this is the factor by which the LLAMA area is larger than the SDSS area). We tested that in the LLAMA volume, SDSS is complete for star-forming galaxies of the given stellar mass. In this volume there are 127 star-forming galaxies in SDSS which, scaled to the LLAMA volume would give 476 star-forming galaxies. Further assuming that all AGNs of this luminosity show similar SFHs (not just the nine that we could probe, but also the other eleven for which we could not perform the stellar population synthesis since they have broad AGN lines or which were not observed) and correcting for the (small) number of Compton-thick AGNs that the LLAMA selection missed due to their high absorbing column \citep[5 sources, ][]{ricci2015a}, we contrast this number with in total 25 AGNs of $\log L_{\rm 14-195 keV}$/erg/s$ > 42.5$ and Declination $\delta < 15 \deg$. In other words, in this stellar mass range, we expect to find one AGN (defined by the X-ray luminosity cutoff of our sample) among twenty star-forming galaxies.

We can compare this to an upper bound for this ratio that we derive from the flickering time scale of $10^5$ years that \cite{schawinski2015} reported for AGNs. Compared to the youngest age bin of our SFH analysis (30 Myr), a total of 300 ``flickering events'' might have happened within this time if AGN activity and nuclear star formation are totally uncorrelated, i.e. one might expect to find star-forming galaxies to be up to 300 times more frequent than AGNs (if only one AGN flickering event occurs per nuclear starburst). This upper limit is consistent with the number derived above. We conclude that our analysis supports the notion that a (luminous, $\log L_X^{14-195 {\rm keV}} / {\rm erg/s} > 42.5$) AGN phase occurs in about 5 \% of all star-forming galaxies with the given mass limits in the local universe. For comparison, the prevalence of any optically selected AGNs (without luminosity cut) is about 4\% of all galaxies as found recently in the CALIFA survey \citep{lacerda2020}. This is also consistent with larger X-ray selected AGN samples. For example \citet{xue2010} find a median AGN fraction in their most local redshift bin ($0 \le z \le 1$) between 1.5 and 6 \% dependent on luminosity. Our luminosity selection is in between their lower and upper limits and the numbers are therefore quite consistent.

\citet{goto2006}, on the other hand, identified 840 galaxies in SDSS that show both a post-starburst signature and an AGN; they make up 4.2 \% of all galaxies in their volume limited sample. Since these authors were looking for Balmer absorption signatures, indicative of A-type stars, their sample includes starbursts with ages up to 1 Gyr. Our analysis is consistent with this, but we show in addition that the ``intermediate-young'' population is insignificant in our nine X-ray selected AGNs and that post-starburst features are dominated by a population that is between 6 and 30 Myr old (see Section~\ref{sec:lower_age_bound}). We conclude from our analysis that the AGN is indeed not ``on'' during the circum-nuclear starburst, but shortly after it, consistent with e.g. \citet{davies2007}.

\subsection{Comparison with molecular gas masses}
As part of the LLAMA project, we have also obtained molecular gas estimates using APEX observations to search for molecular gas in the central 1-3 kpc \citep{rosario2018}. Despite the large differences in aperture, we still find some relation between the APEX results and our analysis here. Of the 26 galaxies in common to both studies, molecular gas was not detected by APEX in only four. Three of these galaxies, i.e. NGC~2775, NGC~5845 and ESO~208-G021, are inactive galaxies characterised by a dominant very old population of stars and the 0.3 - 3 Gyr population is invisible in these (two of them do show a small fraction of young stellar light, though).  Looking at this the other way, for three of the non-star-forming control galaxies in our sample, we have only upper limits for gas masses from the APEX observation, while none of the star-forming control galaxies and only one AGN show that behaviour. That AGN is MCG-05-23-016, which shows a substantial fraction of both young and young-intermediate stars in the analysis presented in this article. The APEX limit was not as deep as for the inactive galaxies, and comparable to some of the detections in other galaxies. Interferometric data from ALMA show that there is indeed molecular gas in the central arcsec, from which stars could form and which could feed the AGN, but very little gas in the surrounding 5--10\arcsec. This highlights the importance of high spatial-resolution studies of gas in AGNs to study the relation between gas reservoirs, stellar populations and AGN activity.

Other recent studies largely corroborate the picture that AGN activity is related to gas mass and star formation, although the observations tend to cover scales of kiloparsecs or more and so are more related galaxy-wide than nuclear properties. 
\citet{jarvis2020} find high molecular gas masses in $z\sim0.1$ powerful quasars, most of which also had high star formation rates typical of starbursts. \citet{husemann2017} had also found some starbursts in their $z<0.2$ sample, but also highlighted a relation between gas content and AGN accretion rate for the disk-dominated (but not bulge-dominated) QSOs in their sample. Similarly, for a sample of QSOs at $z<0.3$, \citet{shangguan2020} found that AGN luminosity correlates with both CO luminosity and black hole mass, and that the molecular gas content of the AGN was similar to star forming galaxies of comparable stellar mass. Using the amount of reddening as a proxy for cold molecular gas mass in order to increase the sample size by an order of magnitude, \citet{zhuang2020} found a similar gas content in AGNs and star-forming galaxies. \cite{yesuf2020} also used a proxy for molecular gas content -- in this case the dust absorption derived from the H$\alpha$/H$\beta$ ratio -- in order to study a large sample of galaxies. Based on their analysis, these authors proposed an evolutionary life cycle in which the gas content mediates the star formation rate and AGN activity. 
Using direct observations of the CO(2-1) line of more than 200 nearby ($z<0.05$) hard X-ray selected galaxies, \cite{koss2021} found that AGN hosts had a higher gas content than inactive galaxies matched by stellar mass but a similar gas content to inactive galaxies matched by star formation rate.
They argued this was because the AGN sample is largely missing a population of quenched passive galaxies that are found among the inactive population.
This may be because passive galaxies do not have their own large gas supply that can feed a central black hole and, as suggested by \cite{davies2014} and \cite{davies2017}, to become a radiatively efficient AGN they may have to acquire gas from the environment instead.
Together, these studies show that, when assessing the links between gas content, star formation and AGN activity, one needs to take into account the global properties of host galaxy (such as the stellar mass, host morphology, etc) since these can affect the nature of that relation. Our expectation is that, by switching the focus from large to small scales, high resolution measurements of the nuclear gas content (Rosario et al. in prep) will lead to a better understanding of those relations.

\subsection{Comparison with previous estimates of the nuclear star formation rate and history}

When comparing the central star forming properties of AGNs, it is important to bear in mind how the AGNs were selected, the size of the region being considered, and the way the star formation rate or history was derived.
These all have an impact on how the various results may fit together.
The AGNs presented here were selected solely by their very hard X-ray luminosity (and would not all be classified as AGN based on their optical line ratios and H$\alpha$ equivalent widths); and their proximity means that our aperture extends to radial scales of typically ~$\sim150$\,pc.
The AGNs from MaNGA presented by \citet{rembold2017}, \citet{mallman_2018}, and \citet{bing2019} were selected based on BPT ratios and the equivalent width of H$\alpha$. As such, \citet{rembold2017} and \citet{mallman_2018} contain a large fraction of lower luminosity AGNs. To avoid a possible bias due to this, \citet{bing2019} also imposed an [O\,III] luminosity constraint.
In addition, the distance to the AGNs in MaNGA means that the 2\arcsec\ central fibre typically covers radii out to 1\,kpc or more. In many cases this is large enough to include a circumnuclear ring if one is present.

Finally, we note that because \citet{bing2019} focussed on secular evolution in gaseous disks, they imposed an additional constraint that the host galaxy of the AGN should be star forming.
The impact of this selection criterion is not obvious: our Figs.~\ref{fig:BPT_N2},~\ref{fig:whan}, and ~\ref{fig:sfh_result_sample} (also Fig.~2 in \citealt{rembold2017}) show that among the matched inactive galaxies in our sample one finds central regions that are star forming and also those that are not; but there appears to be little relation between large scale host galaxy classification (spiral versus lenticular) and whether there has been recent nuclear star formation (as defined by emission line ratios or by stellar population fitting).
Given all these caveats, it can be difficult to compare published results directly with ours. 
Nevertheless, here we summarize some of those results and attempt an assessment of whether there is a consistent picture emerging.

The underlying question is whether accretion onto the central black hole is associated with star formation in the nuclear or circumnuclear region; and if so whether the AGN is synchronous with star formation or occurs after it in a post-starburst phase. Answers to these questions can indicate whether nuclear star formation and AGN activity are only related because both require inflowing gas, whether the young stars themselves help regulate the inflow, or whether the AGN regulates the star formation.
Following the early work of \citet{norman1988}, a variety of theoretical scenarios have been proposed that support all these options.
The analytical model of \citet{kawakatu2008} and detailed hydrodynamical simulations of \citet{wada2009} suggest that inflow and accretion onto the black hole should increase with turbulence in the ISM, and occur simultaneously with star formation.
Alternatively, \citet{hobbs2011} suggested that supernovae may ballistically drive enough of the ISM inwards to fuel an AGN.
\citet{schartmann2009,schartmann2010} used hydrodynamical simulations to model a scenario where, once star formation had ceased, slow stellar winds from AGB stars would flow inwards and fuel the AGN.
\citet{vollmer2008} proposed an analytical model in which inflow to the circumnuclear region triggers star formation, but soon after, it is quenched by the subsequent supernovae. \citet{vollmer2013} argued that turbulence generated by the inflowing gas may also quench the star formation.
\citet{hopkins2012} argued that one might simply expect a delay between star formation and AGN activity due to the time it takes the gas to flow in from 10--100\,pc scales.

Observationally, analyses of the nuclear and circumnuclear region based on stellar population synthesis consistently provide evidence that AGNs are generally characterised by young stellar populations \citep{storchibergmann2000,kauffmann2003c,cidfernandes2004,sarzi2007,riffel_r2009}. But these populations do not always seem to be present, and it is not always easy to separate the contribution of a very young stellar population from a featureless continuum associated with the AGN accretion disk.
More recent studies have focussed on SDSS-III or SDSS-IV MaNGA data.
Using 62 AGNs together with a larger number of matched inactive galaxies, \citet{rembold2017} found a relation between the contribution from young ($\sim 40$\,Myr) stars and the [O\,III] luminosity, supporting the notion that massive starbursts fuel powerful AGNs.
Similarly, and using the same active and inactive samples, \citet{mallman_2018} looked at the spatially resolved stellar properties and found that in the inner 1$R_e$ the more luminous AGNs do present larger fractions of young ($t \leq 40$\,Myr) stars than the control galaxies.
In contrast, and perhaps associated to their selection that focuses on galaxies with star forming disks and their use of H$\alpha$ to measure star formation rate, \citet{bing2019} reported that AGNs may have slightly suppressed ongoing nuclear star formation rates compared to matched star-forming inactive galaxies.

For the nuclear regions where one may expect star formation to occur in short bursts rather than continuously, the ionized emission lines are a crucial probe of the current star formation rate.
These were a key part of the analysis of \cite{davies2007} who combined them with mass-to-light ratios and an estimate of the supernova rate from radio continuum imaging, to conclude that there was recent, but no longer on-going, star formation.
\cite{hicks2009,hicks2013} and \cite{davies2014} highlighted the low equivalent widths of the Br$\gamma$ line in the near-infrared, where extinction has less impact than the optical.
After taking into account the contribution to the line from the AGN (e.g. via kinematics) and also the contribution to the continuum from hot dust (e.g. via the slope and CO2-0 bandhead depth), their conclusion was that there could not be on-going star formation.
This is consistent with the stellar synthesis analyses in the literature as well as our conclusion from this paper: it allows that star formation has occurred recently, but requires that it must have also ceased. It argues for repeated short-lived bursts of star formation.

A third approach to assessing the central star formation is based on the presence of poly-aromatic hydrocarbons (PAHs) which are known to trace young stars, but over a wider range of ages ($\sim100$\,Myr) than the ionised line emission.
However, there is a debate about whether the PAHs are excited by stars or by the AGN itself, or whether small PAHs are destroyed by the hard radiation from an AGN \citep{siebenmorgen2004,smith2007,sales2010}.
This may depend on which PAH feature one considers: \citet{diamondstanic2010} found that in AGN on kpc scales the 6.2, 7.7, and 8.6\,$\mu$m features were suppressed while the 11.3\,$\mu$m feature was not.
Partly for this reason, recent studies of PAHs close around AGN have focussed on the feature at 11.3\,$\mu$m.
\citet{esquej2014} detected PAHs on subarcsec scales in about half of their sample of 29 nearby Seyfert galaxies. They argued that the high column densities in the torus around the AGN would provide shielding that enables PAHs to survive. And the implied central star formation rate density was much higher than in the circumnuclear region.
A similar conclusion about PAH survival was reached by \citet{esparzaarredondo2018} for their sample of 19 AGN, based on the lack of relation between the X-ray luminosity and a central PAH deficit -- while PAHs were detected in most AGN, the equivalent width was lower in the centre. They attributed this to low star formation rates in that region.
On the other hand, while \citet{jensen2017} also detected PAHs in a sample of 13 AGNs, they argued that the similarity and slope of the radial profiles from tens to hundreds of parsecs, point towards an origin in a single compact source of excitation.
In order to shed more light on the excitation of PAHs in this context, \cite{alonsoherrero2020} compared PAH line ratios in 22 AGNs to models of PAH excitation, and analysed this in the context of the measured molecular gas content. They concluded that PAHs can be shielded from the hard AGN radiation.

Of the nine AGNs analysed in this paper, six appear in the PAH papers above. 
Four of those have PAH detections in the Spitzer aperture (3.7\arcsec\ slit width), but only one has a PAH detection from ground-based data in a $\la 1$\arcsec\ slit.
This detection is NGC\,7582, which has the highest young stellar population fraction of our AGN sample.
But three of the AGN with PAH non-detections also have significant young stellar populations (NGC\,2110, NGC\,3081, and MCG-05-23-016).
The other two non-detections are for NGC\,2992 and NGC\,7172, for which we did not find evidence of a young (or young-intermediate) stellar population.
Thus it seems that while in principle PAHs can be used to trace star formation close around AGN, the assessment of the implied star formation rates is still open -- although it does point in the same direction as the other studies that there is recent star formation close around many, although perhaps not all, AGN.

%% file: sections/7-conclusions.tex
\section{Conclusions and outlook}
\label{sec:conclusions}

Using VLT/X-SHOOTER spectra and \citet{bruzual2003} single-stellar population models together with the STARLIGHT fitting code, we fit the integrated light in an aperture corresponding to a physical radius of $\approx$ 150 pc. We perform the analysis on a sample of nine ``type 2'' AGNs that are part of the volume-complete sample of hard X-ray selected, powerful, local AGNs presented by \citet{davies2015}. We conduct an identical analysis on a sample of 18 inactive control galaxies that were matched in stellar mass, distance and Hubble-type. Our main findings can be summarised as follows:

\begin{itemize}
    \item We can robustly determine the fraction of young stellar light in type 2 AGNs and we are not, or only insignificantly, affected by scattered AGN continuum.
    \item We find young stellar populations in seven of the nine AGNs and constrain their age to be between 6 and 30 Myr from their location on the diagnostic emission line ratio diagrams from BPT/VO87 and from the stellar population fitting. On average, 4.5 $\pm$ 1.2 \% of the optical light of our AGNs stems from this population.
    \item The analysis of the nuclear stellar populations of a sample of inactive galaxies matched on large scales shows that they fall into two distinct groups: those with and without recent star formation according to their BPT/VO87 emission line ratios.
    \item When separated according to their BPT/VO87 emission line ratios, the star-forming inactive control galaxies show very similar fractions (4.2 $\pm$ 1.4 \%) of young stellar light and are clearly different from the non-star-forming control galaxies that only show 0.5 $\pm$ 0.3 \% of young stellar light.
    \item Interpreting the star-forming control galaxies as the parent population of our AGNs, we find that a luminous ($\log L_X^{14-195 {\rm keV}}$/erg/s $>$ 42.5) AGN phase occurs in $\sim$ 5\% of all star-forming galaxies within the mass limits of our sample in the local universe, consistent with other estimates.
    \item That the matched inactive galaxies form two sub-groups, with the star-forming inactive galaxies identified as the AGN parent population (and for which the timescale to switch between active and inactive is rather short), points also to long timescales of inactivity. This is associated with the inactive galaxies with no recent star formation. And is an indication that in those galaxies there is insufficient gas to trigger either nuclear star formation or an AGN.
    \item This supports the notion that the relevant criterion for both AGN activity and circum-nuclear star formation is the presence of gas in the central 100 parsecs \citep[e.g.][]{sabater2015}.
\end{itemize}

In the future, we plan to enhance sample size, while keeping the volume-complete nature of the selection. We note that the high spatial resolution afforded by the VLT Unit Telescopes is actually not required in the optical since we are limited by extinction, which confuses the definition of the nucleus on sub-arcsecond scales in the optical. In order to separate the $<$ 100 pc nuclear stellar populations from the nuclear rings at $>$ 100 pc, a resolution of $\approx$ 2\farcs0 is sufficient for galaxies within our redshift limit of $z<0.01$, suggesting that a larger sample could be obtained with a significantly smaller telescope. The SOXS instrument \citep{schipani2020} seems particularly suited for a follow-up of this kind of research.

In addition to expanding the sample size, we also plan to continue investigating nuclear stellar populations in the (near-)infrared both to further zoom into the highly obscured nucleus and to get the ``full picture'' regarding young populations \citep{riffel_r2015}.

%% file: sections/a1-ssp_fits_appendix.tex
\section{Individual Stellar Population Fits}
\label{sec:appendix:fits}
\subsection{Control galaxies}
\clearpage

\begin{figure*}
\includegraphics[width=\textwidth]{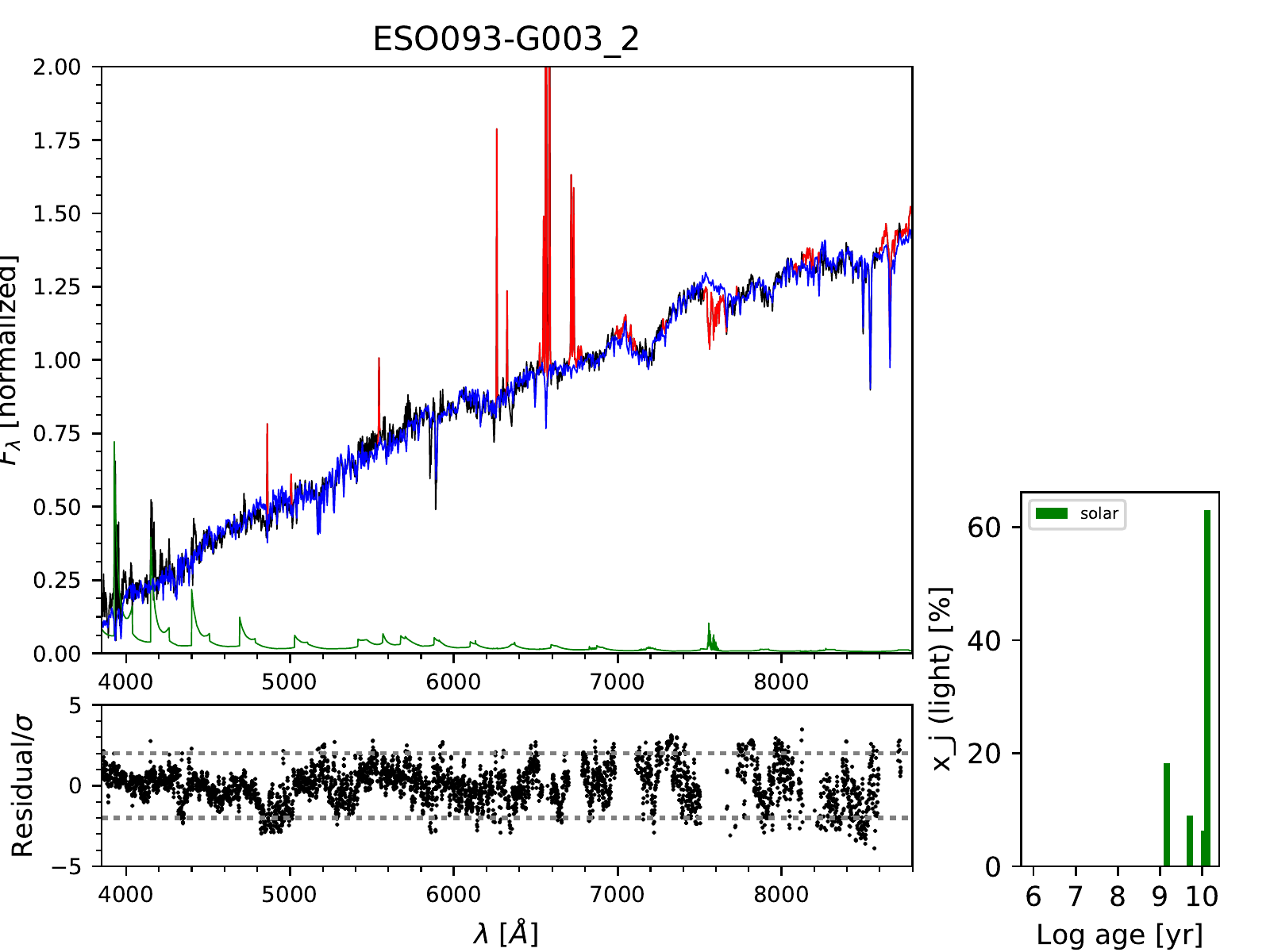}
\caption{\label{fig:SSP_ESO093_3}As Fig.~\ref{fig:SSP_MCG523_1}, but for ESO 093-G003}
\end{figure*}
\begin{figure*}
\includegraphics[width=\textwidth]{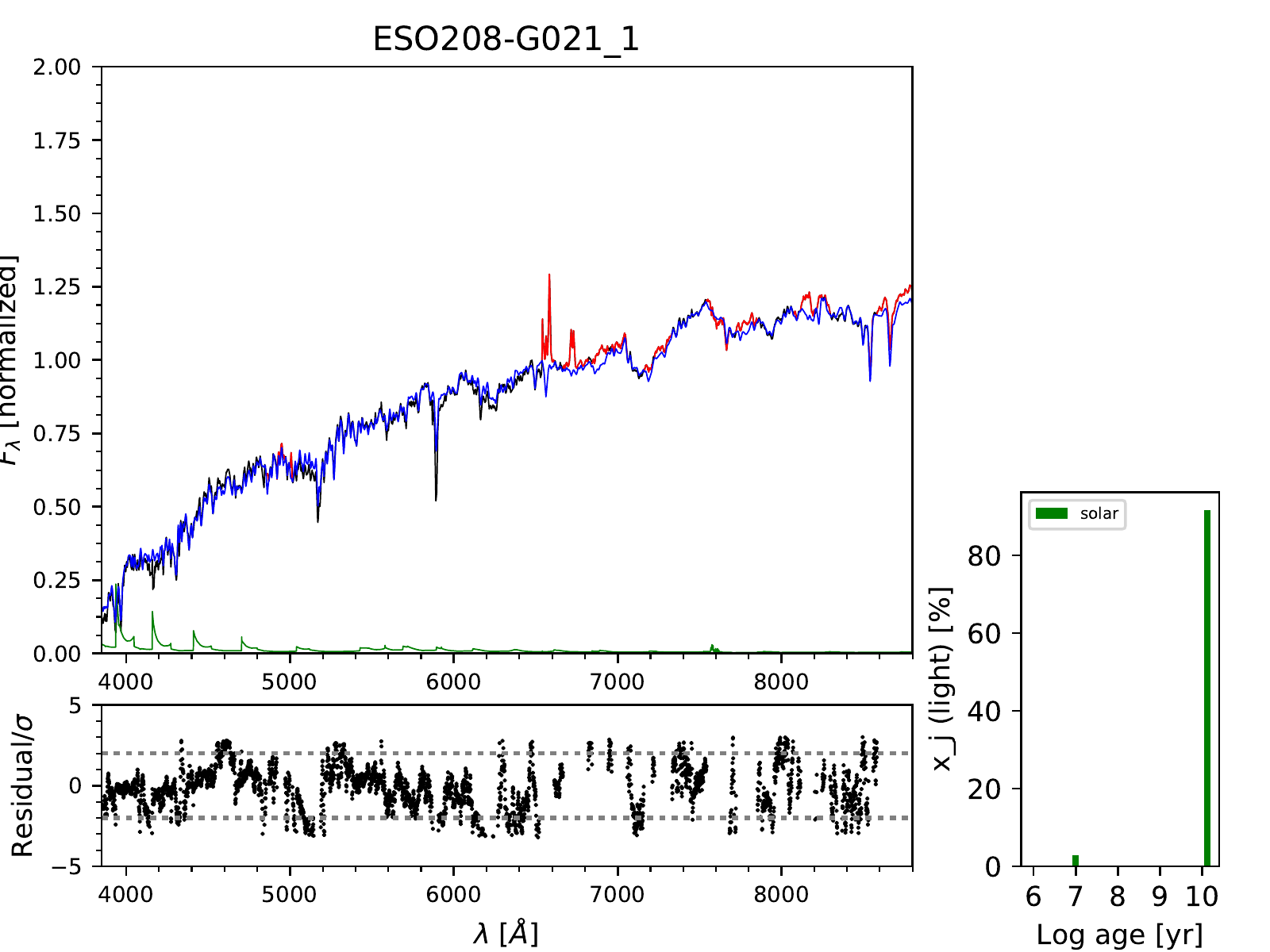}
\caption{\label{fig:SSP_ESO208_1}As Fig.~\ref{fig:SSP_MCG523_1}, but for ESO 208-G021}
\end{figure*}
\begin{figure*}
\includegraphics[width=\textwidth]{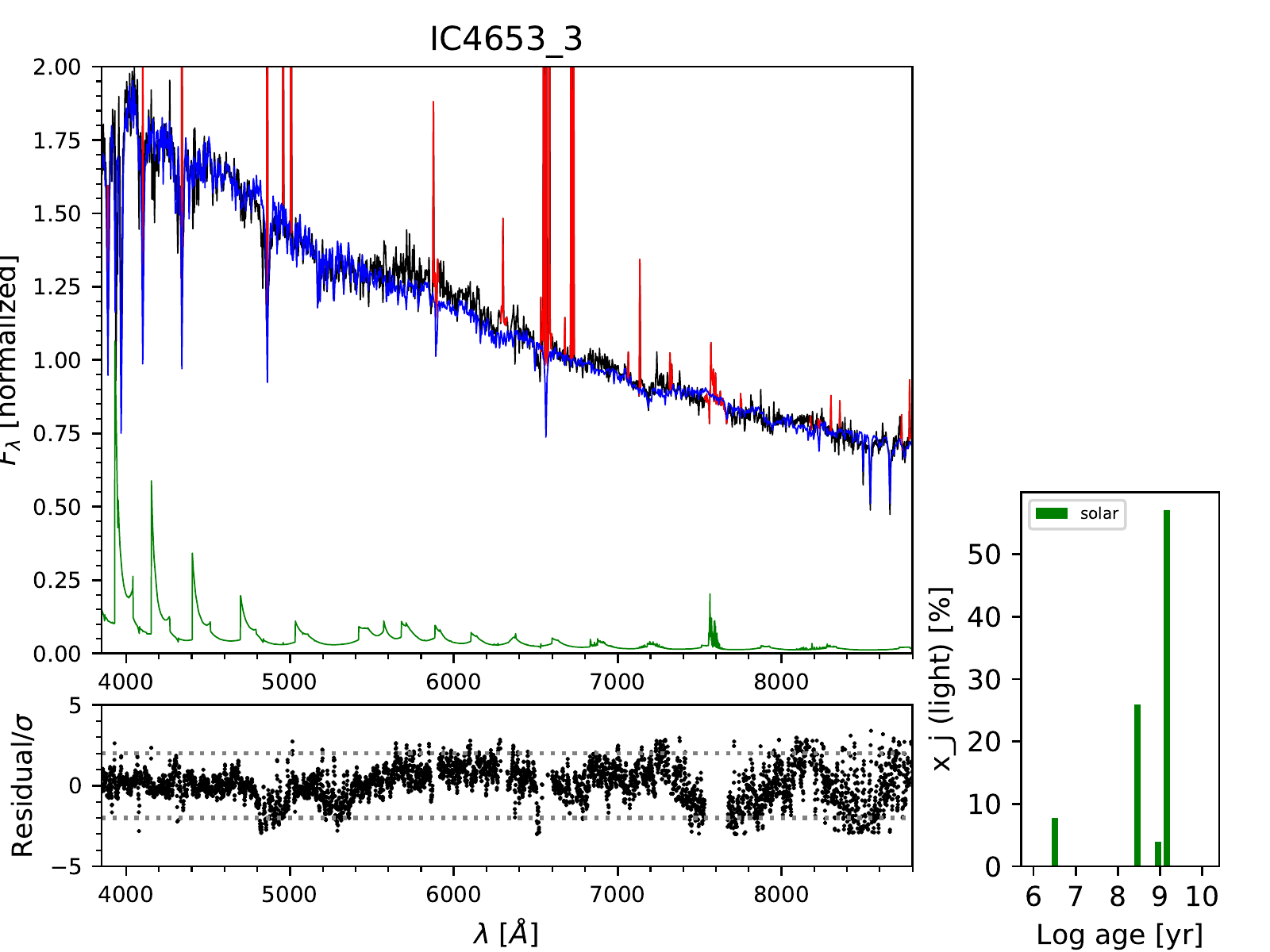}
\caption{\label{fig:SSP_IC4653_3}As Fig.~\ref{fig:SSP_MCG523_1}, but for IC 4653}
\end{figure*}
\begin{figure*}
\includegraphics[width=\textwidth]{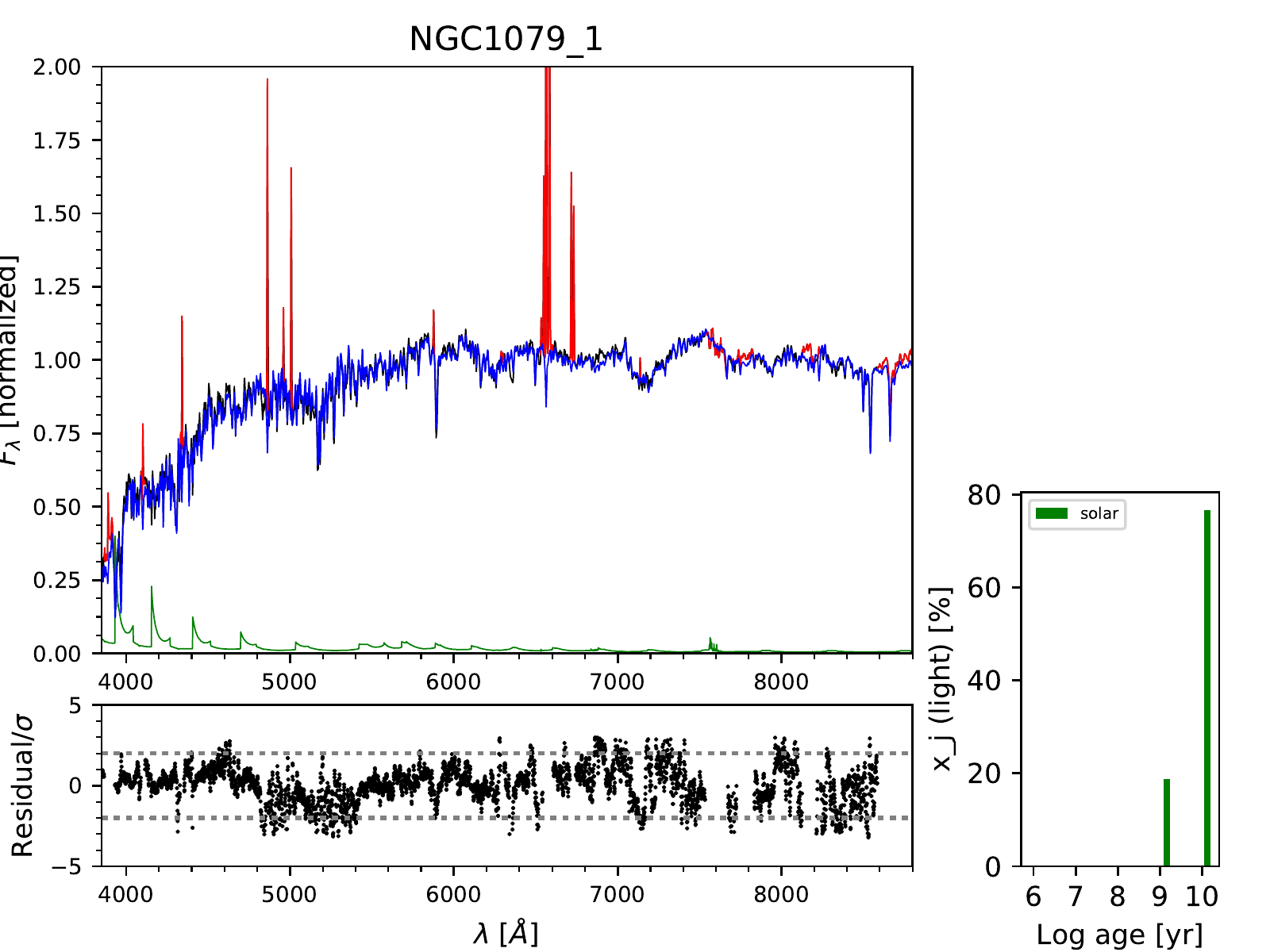}
\caption{\label{fig:SSP_NGC1079_1}As Fig.~\ref{fig:SSP_MCG523_1}, but for NGC 1079}
\end{figure*}
\begin{figure*}
\includegraphics[width=\textwidth]{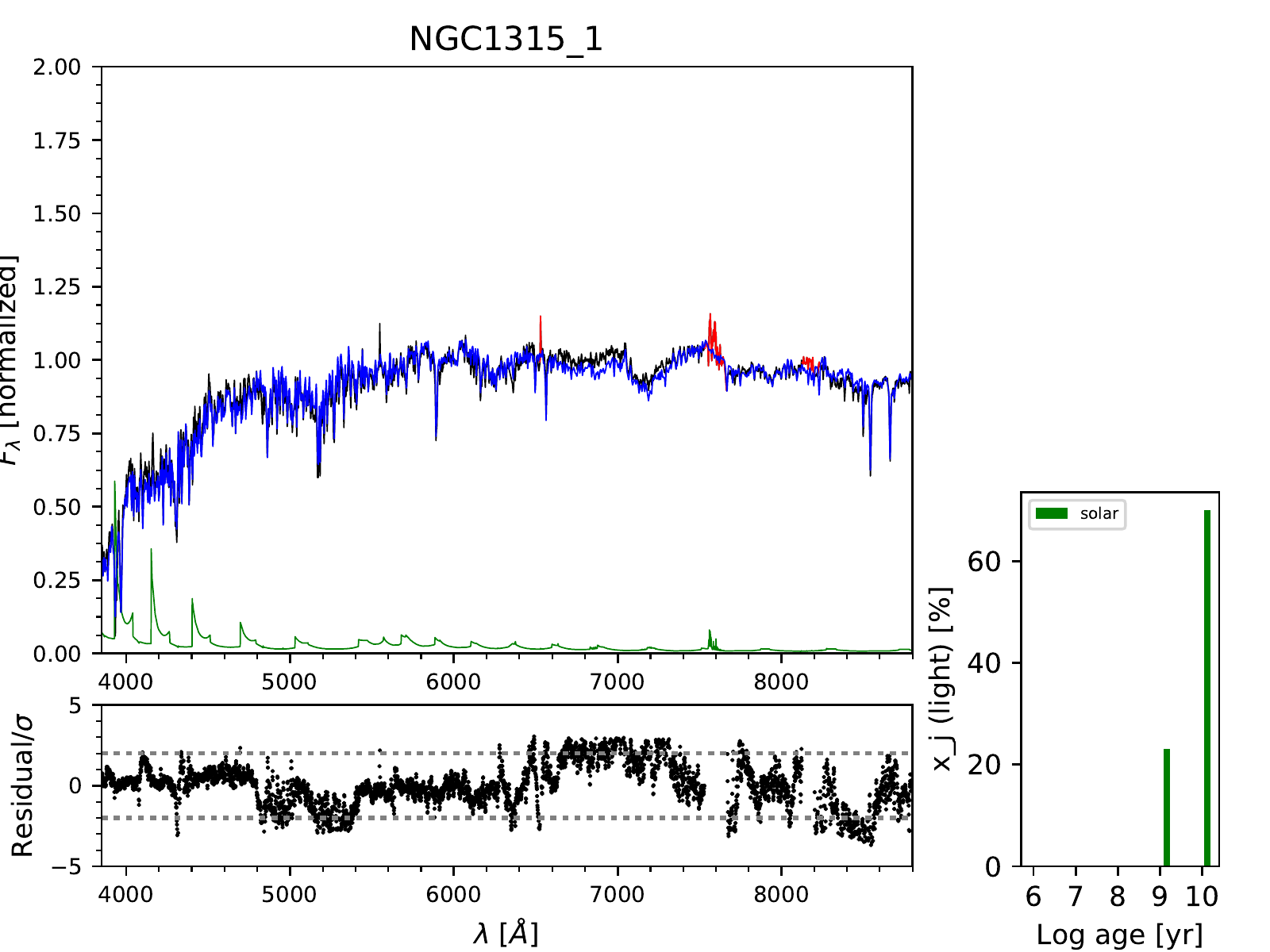}
\caption{\label{fig:SSP_NGC1315_1}As Fig.~\ref{fig:SSP_MCG523_1}, but for NGC 1315}
\end{figure*}
\begin{figure*}
\includegraphics[width=\textwidth]{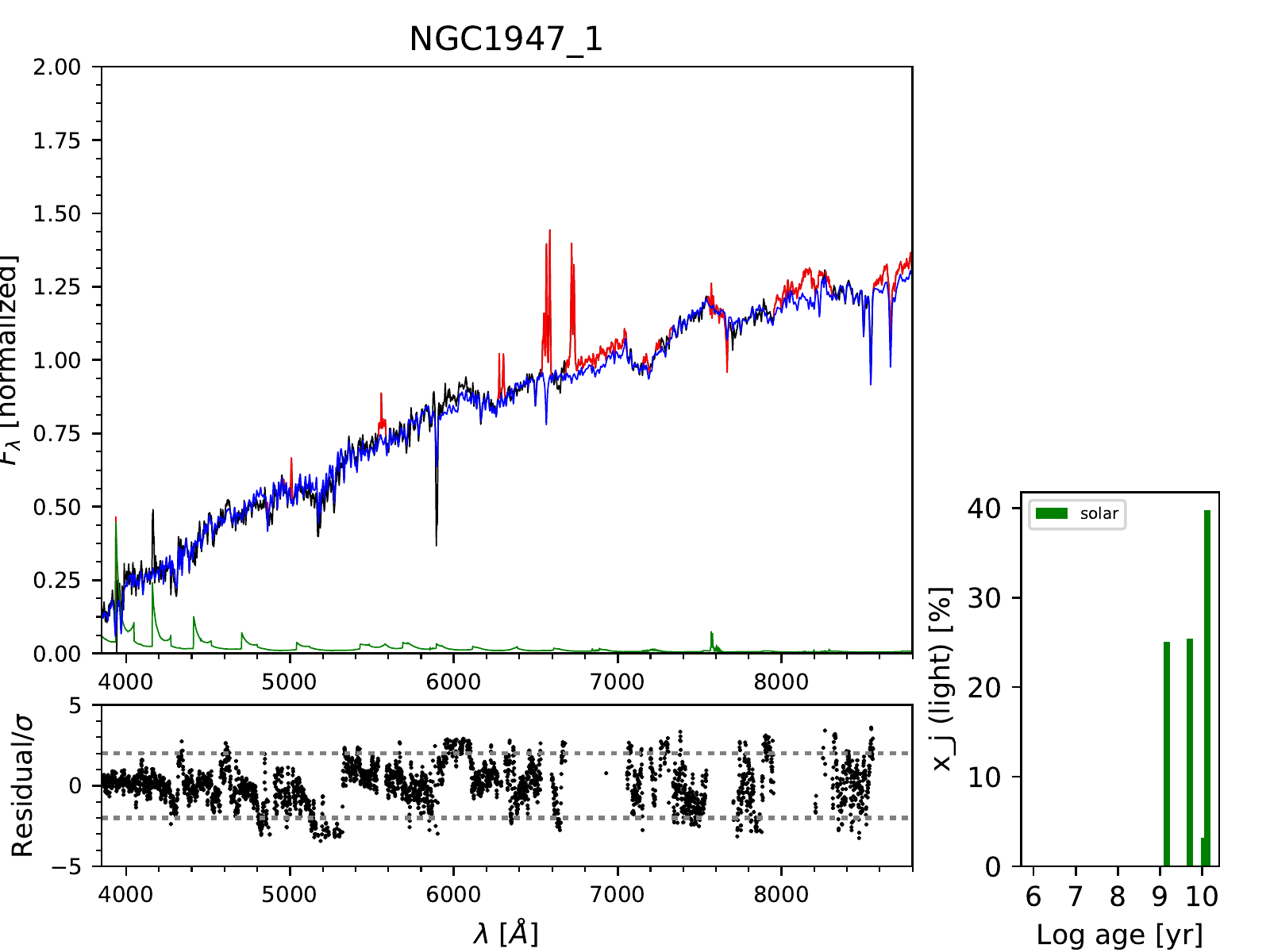}
\caption{\label{fig:SSP_NGC1947_1}As Fig.~\ref{fig:SSP_MCG523_1}, but for NGC 1947}
\end{figure*}
\begin{figure*}
\includegraphics[width=\textwidth]{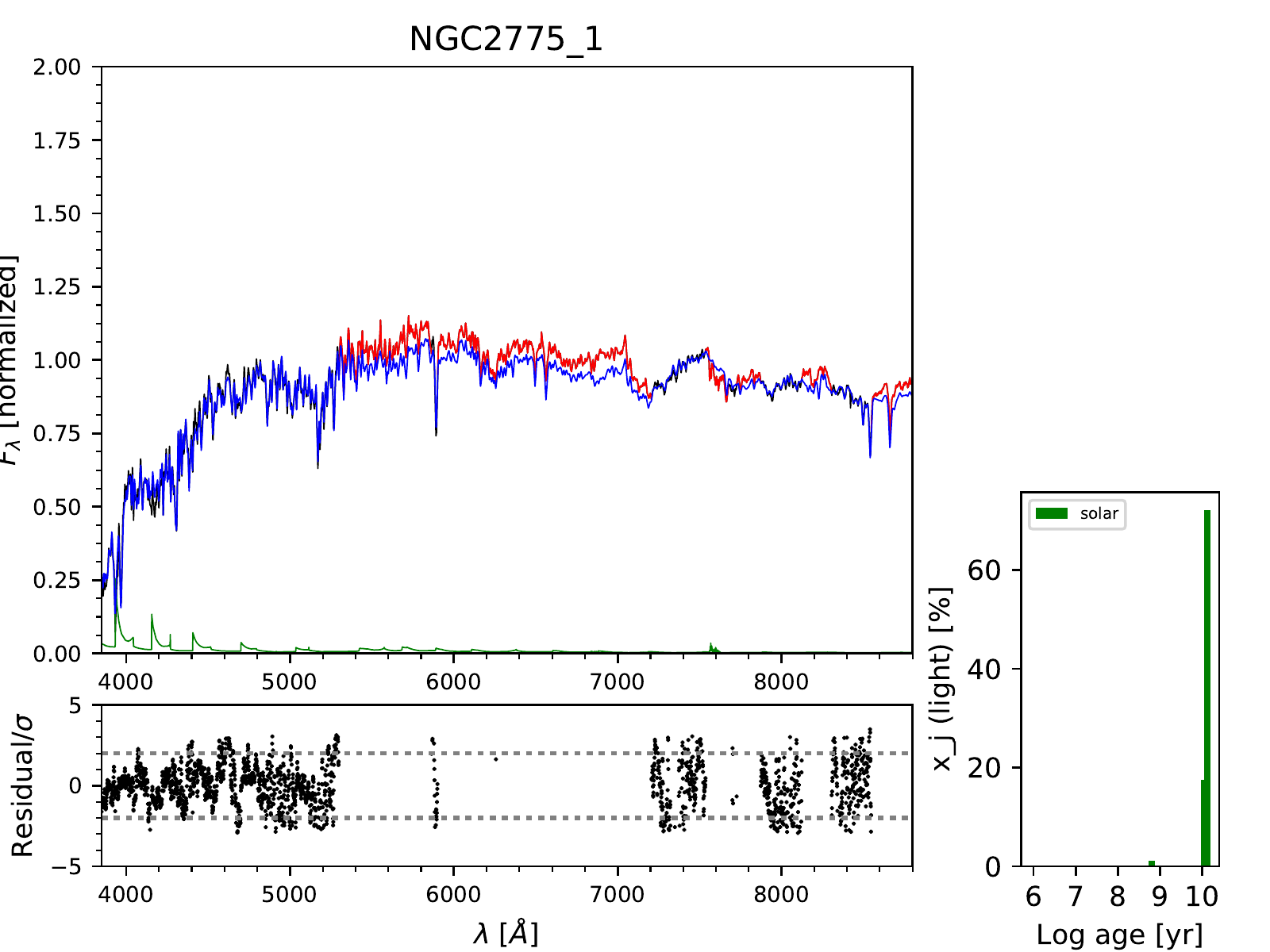}
\caption{\label{fig:SSP_NGC2775_1}As Fig.~\ref{fig:SSP_MCG523_1}, but for NGC 2775}
\end{figure*}
\begin{figure*}
\includegraphics[width=\textwidth]{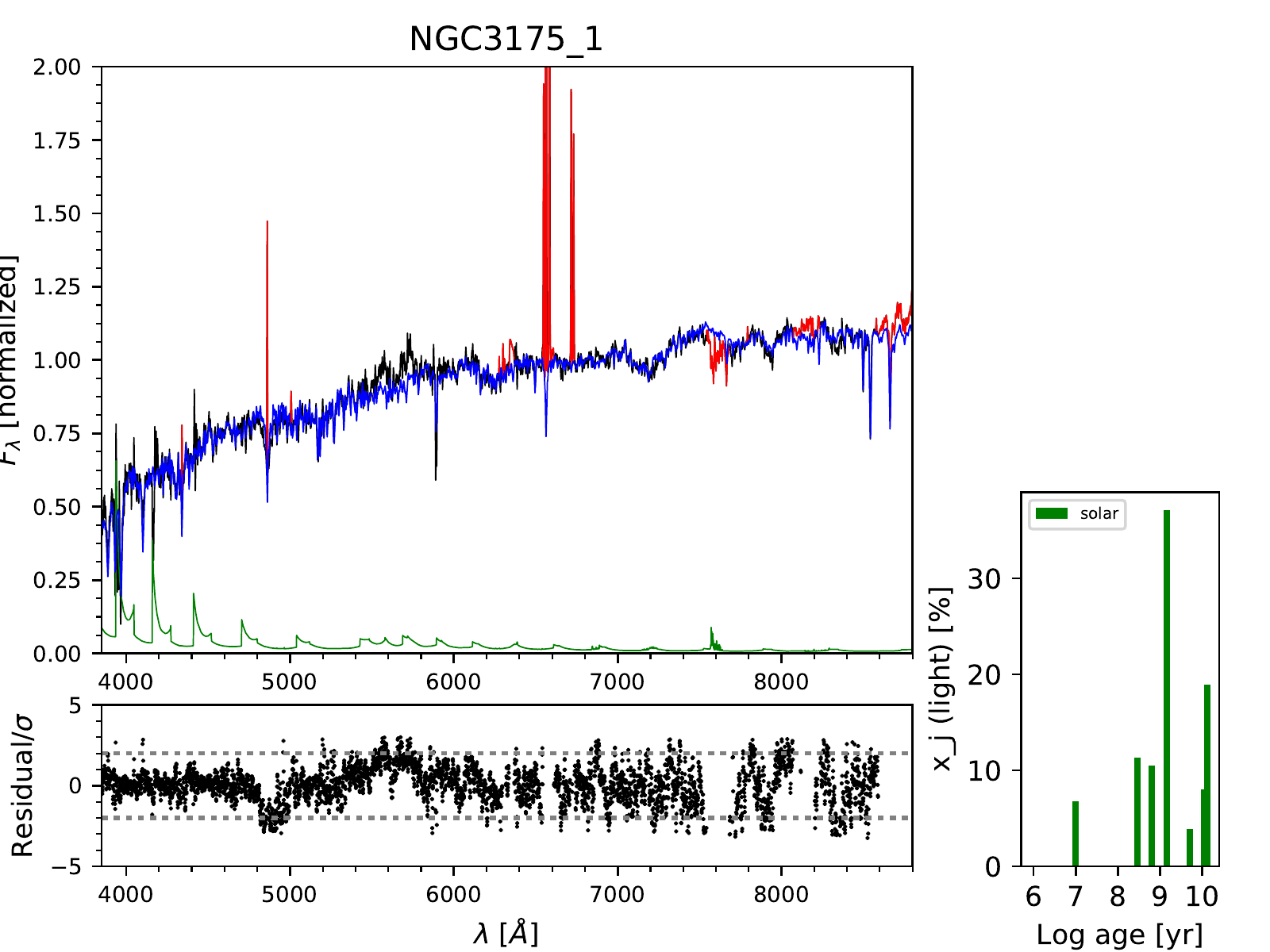}
\caption{\label{fig:SSP_NGC3175_1}As Fig.~\ref{fig:SSP_MCG523_1}, but for NGC 3175}
\end{figure*}
\begin{figure*}
\includegraphics[width=\textwidth]{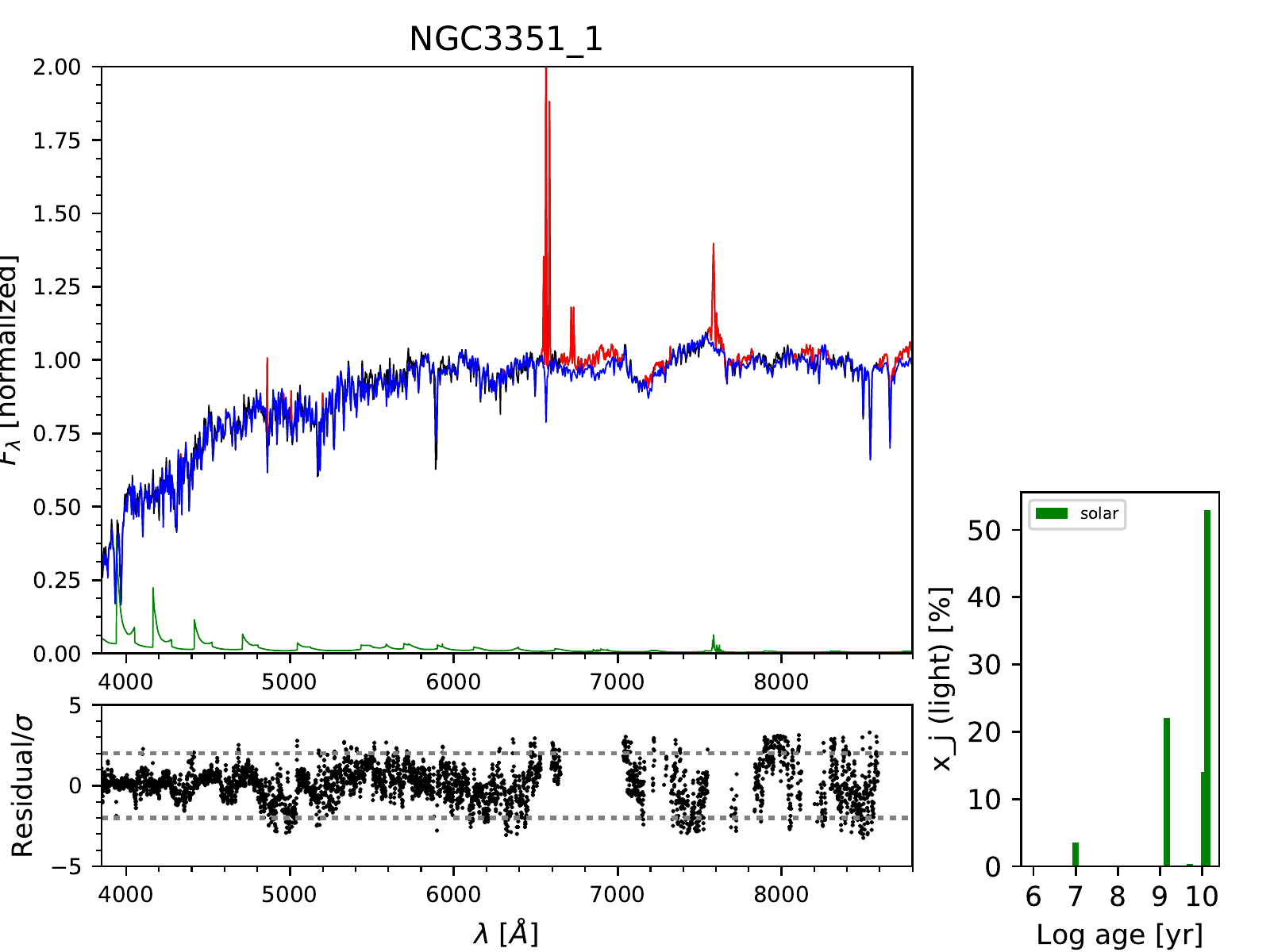}
\caption{\label{fig:SSP_NGC3351_1}As Fig.~\ref{fig:SSP_MCG523_1}, but for NGC 3351}
\end{figure*}
\begin{figure*}
\includegraphics[width=\textwidth]{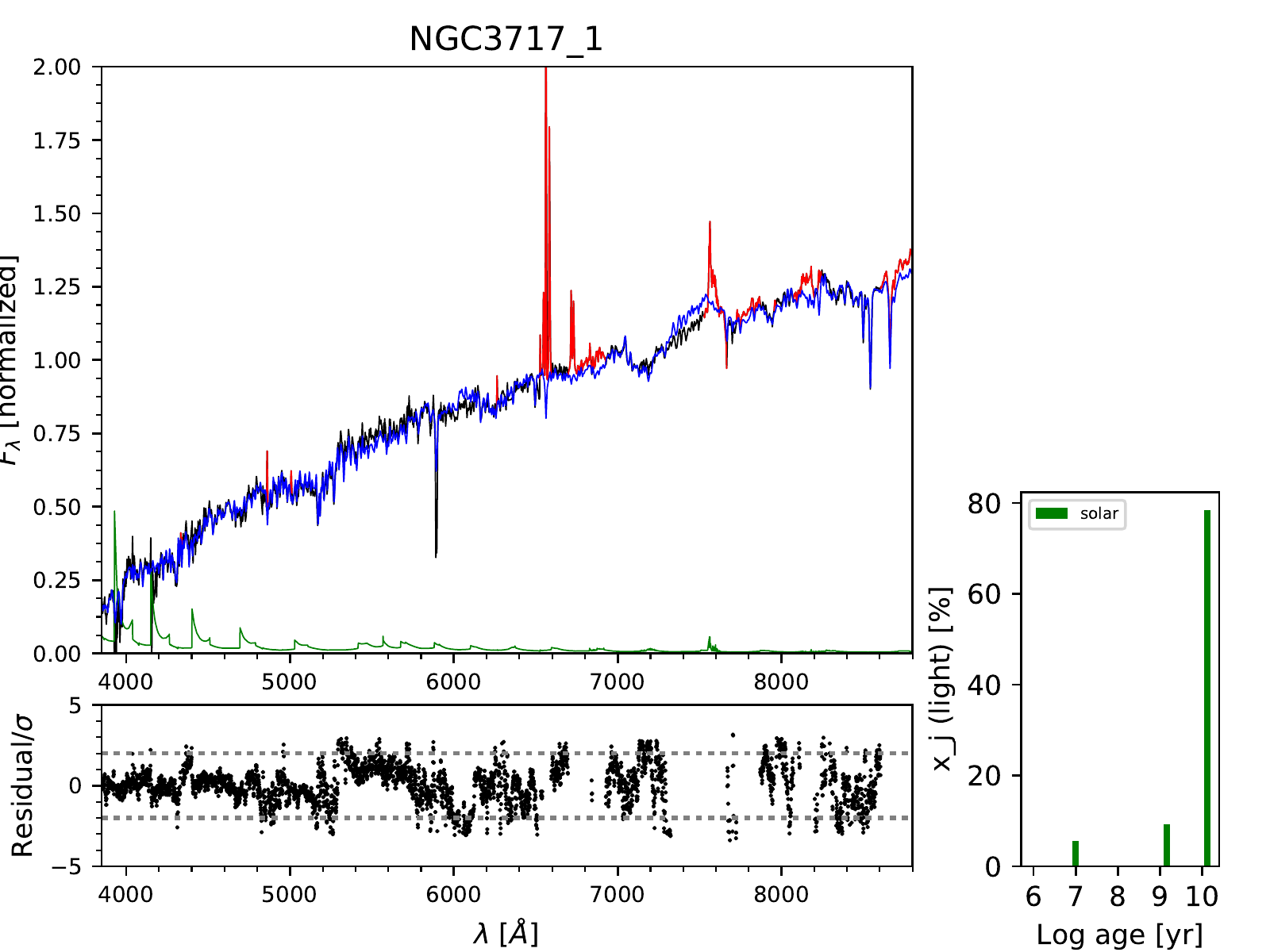}
\caption{\label{fig:SSP_NGC3717_1}As Fig.~\ref{fig:SSP_MCG523_1}, but for NGC 3717}
\end{figure*}
\begin{figure*}
\includegraphics[width=\textwidth]{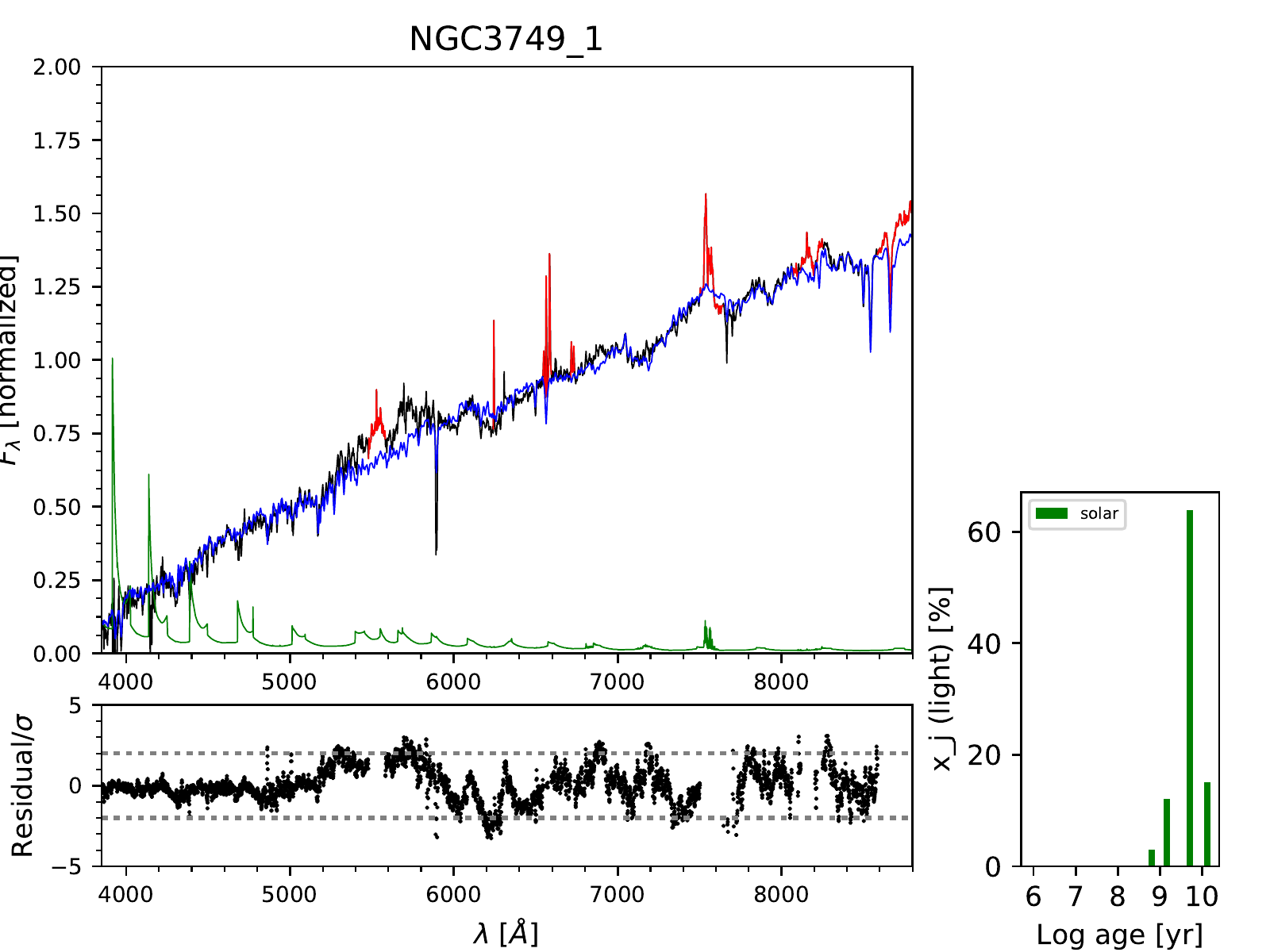}
\caption{\label{fig:SSP_NGC3749_1}As Fig.~\ref{fig:SSP_MCG523_1}, but for NGC 3749}
\end{figure*}
\begin{figure*}
\includegraphics[width=\textwidth]{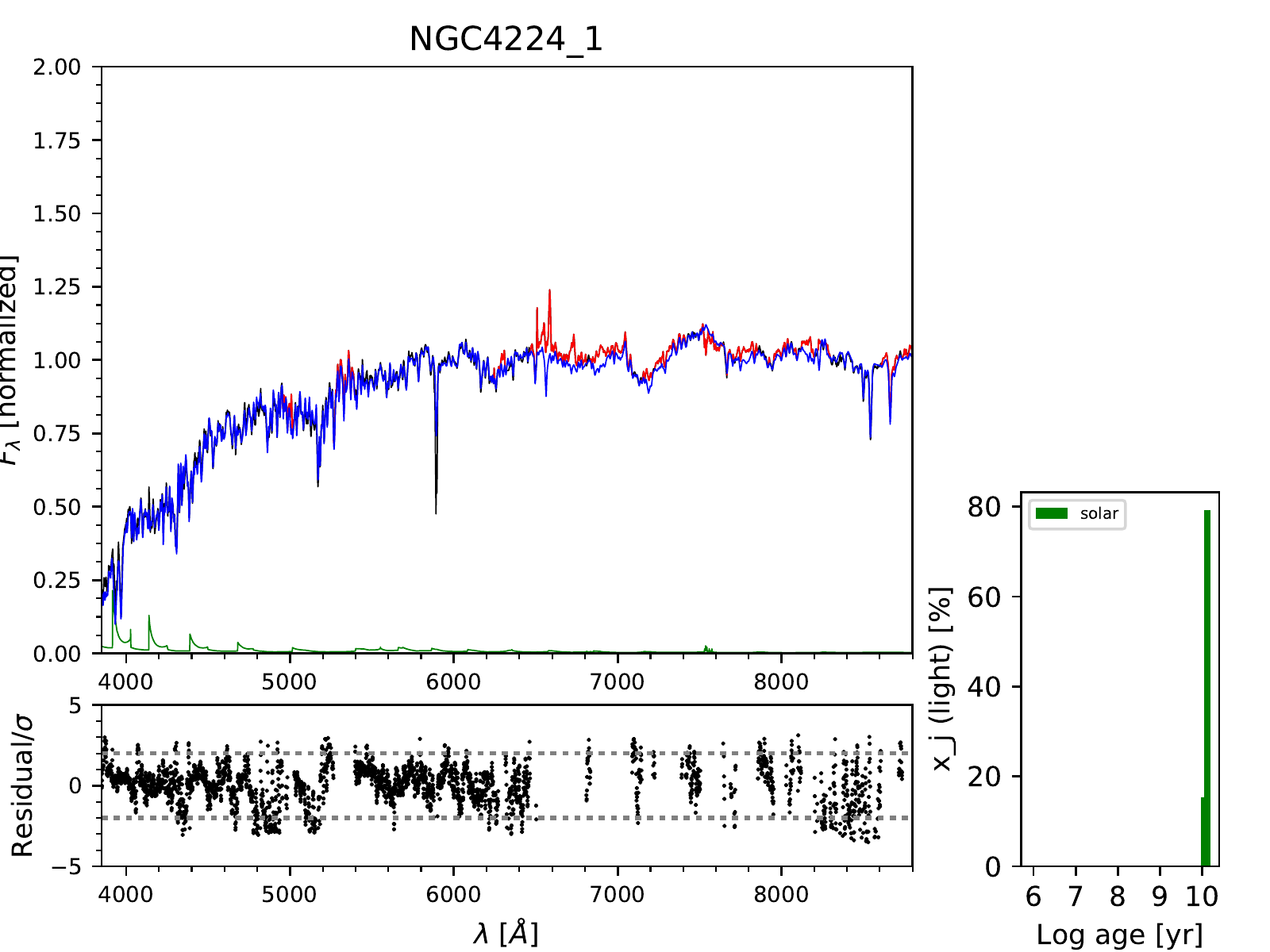}
\caption{\label{fig:SSP_NGC4224_1}As Fig.~\ref{fig:SSP_MCG523_1}, but for NGC 4224}
\end{figure*}
\begin{figure*}
\includegraphics[width=\textwidth]{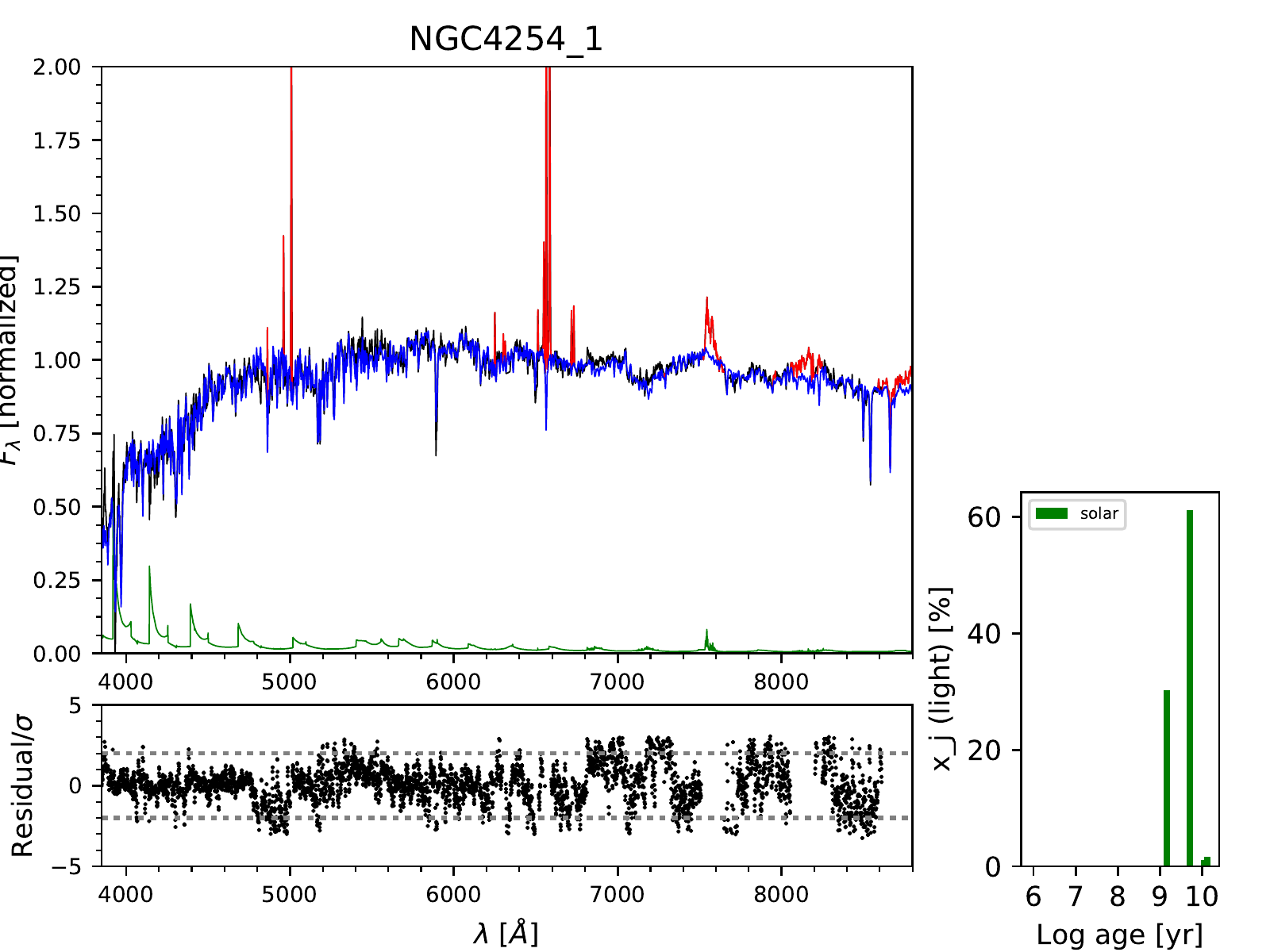}
\caption{\label{fig:SSP_NGC4254_1}As Fig.~\ref{fig:SSP_MCG523_1}, but for NGC 4254}
\end{figure*}
\begin{figure*}
\includegraphics[width=\textwidth]{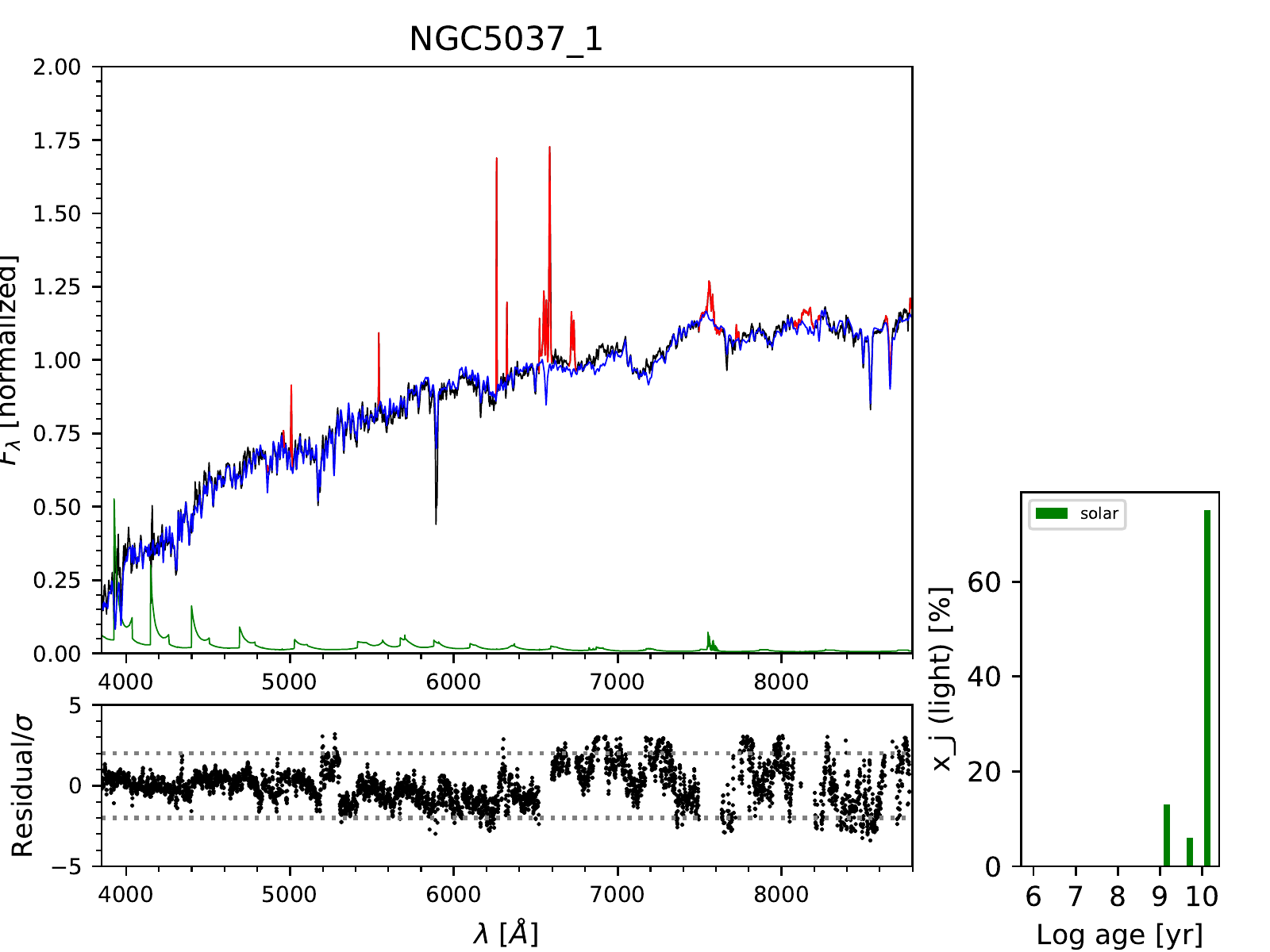}
\caption{\label{fig:SSP_NGC5037_1}As Fig.~\ref{fig:SSP_MCG523_1}, but for NGC 5037}
\end{figure*}
\begin{figure*}
\includegraphics[width=\textwidth]{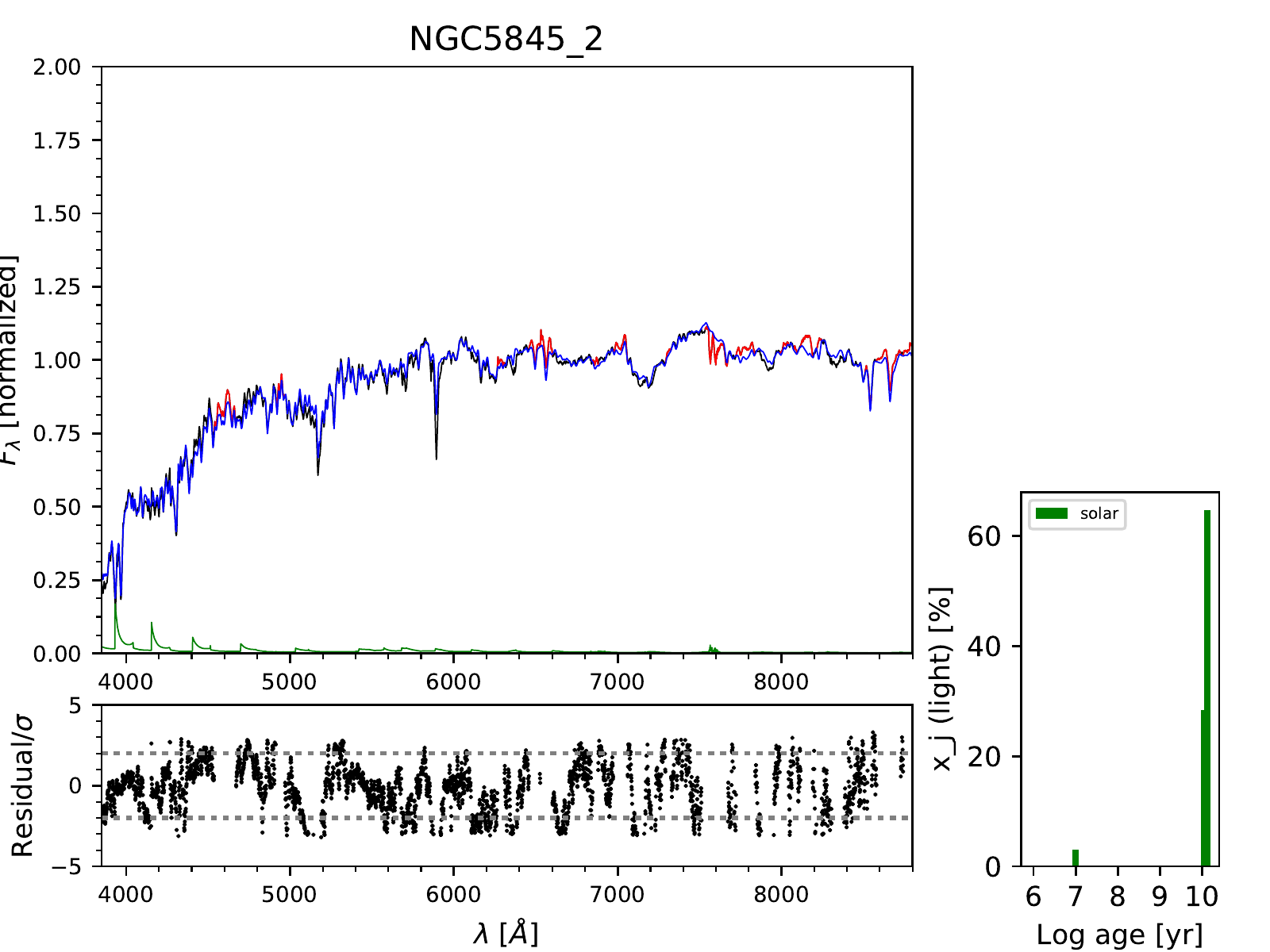}
\caption{\label{fig:SSP_NGC5845_2}As Fig.~\ref{fig:SSP_MCG523_1}, but for NGC 5845}
\end{figure*}
\begin{figure*}
\includegraphics[width=\textwidth]{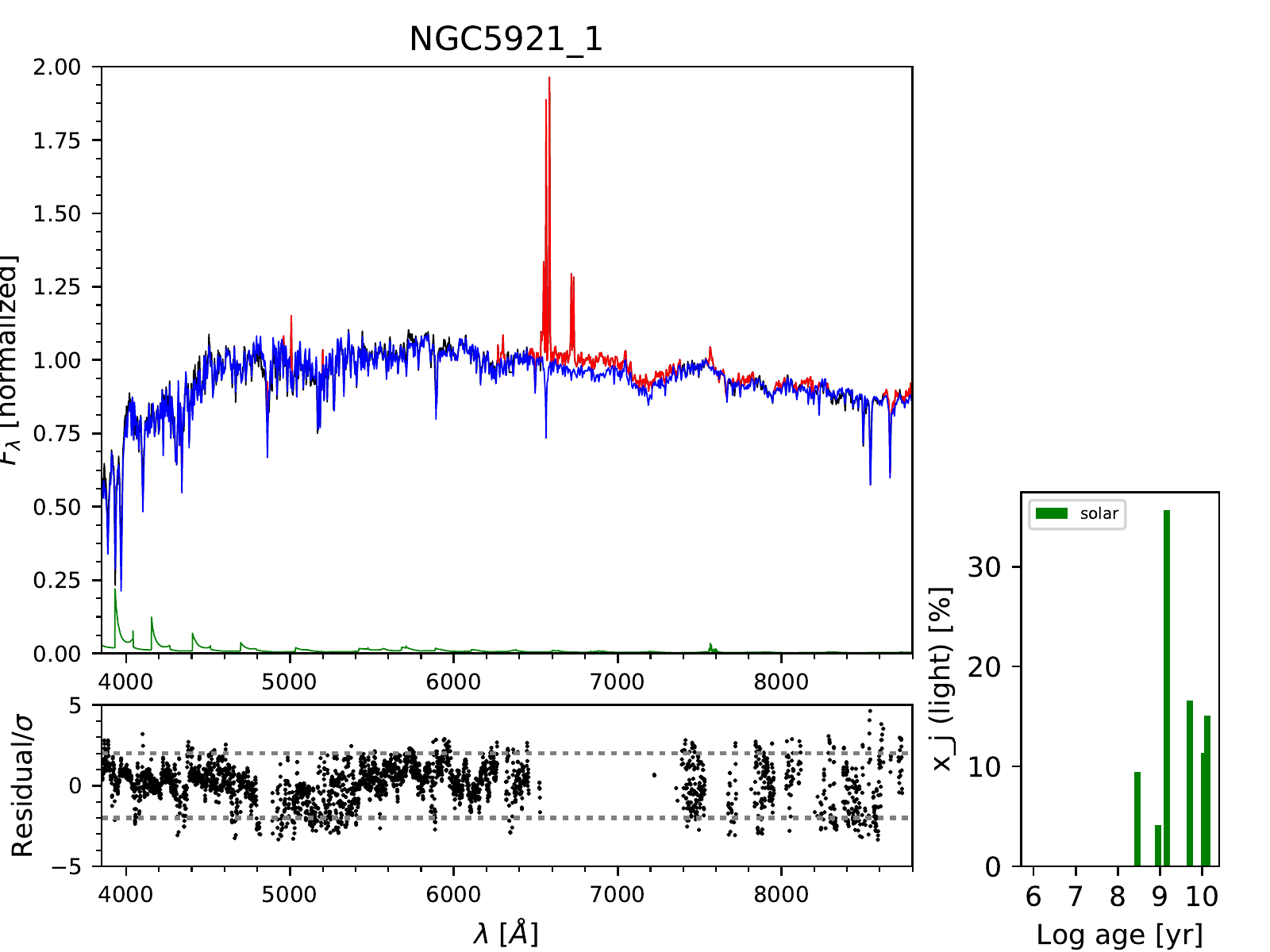}
\caption{\label{fig:SSP_NGC5921_1}As Fig.~\ref{fig:SSP_MCG523_1}, but for NGC 5921}
\end{figure*}
\begin{figure*}
\includegraphics[width=\textwidth]{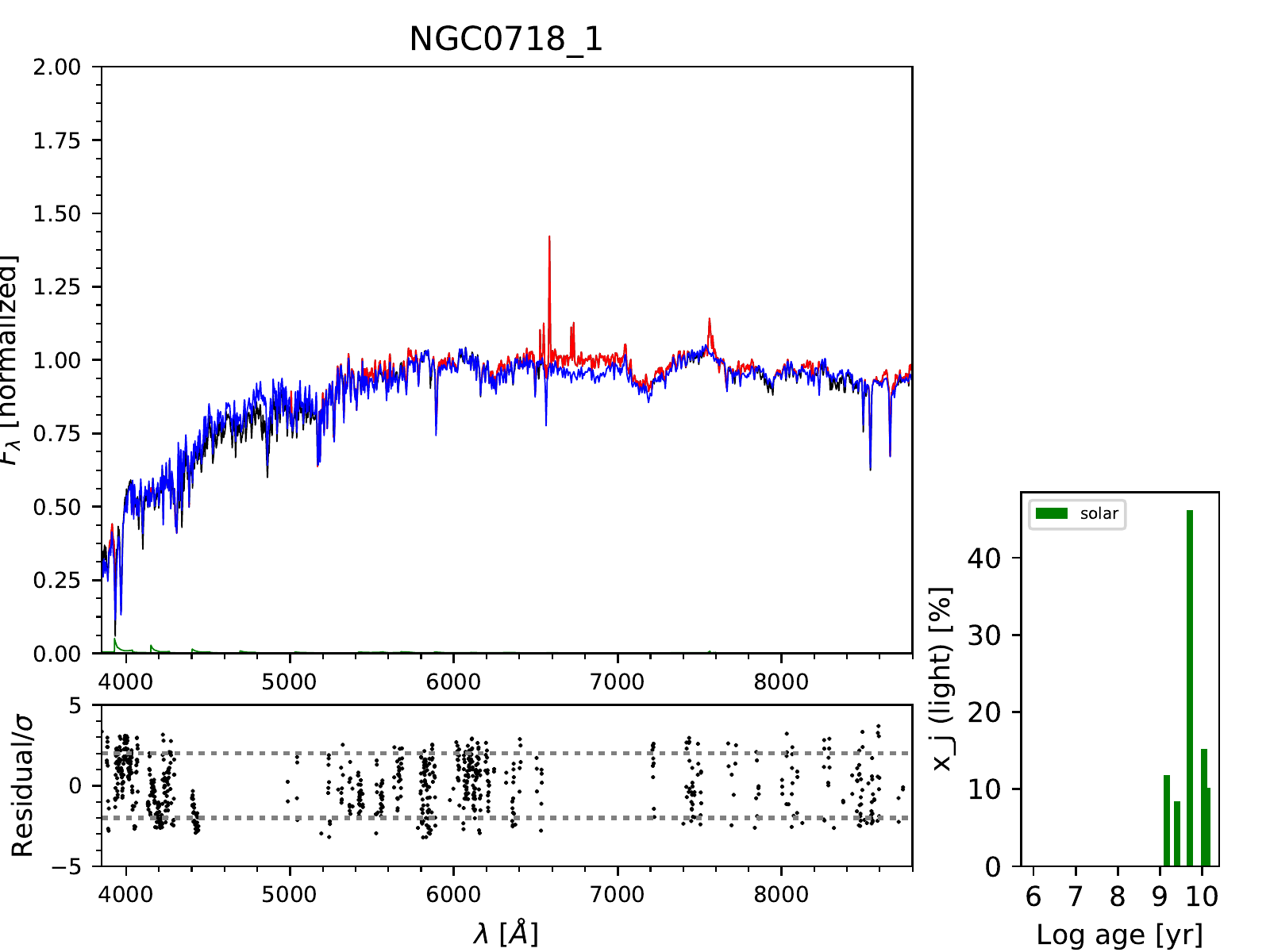}
\caption{\label{fig:SSP_NGC718_1}As Fig.~\ref{fig:SSP_MCG523_1}, but for NGC 718}
\end{figure*}
\begin{figure*}
\includegraphics[width=\textwidth]{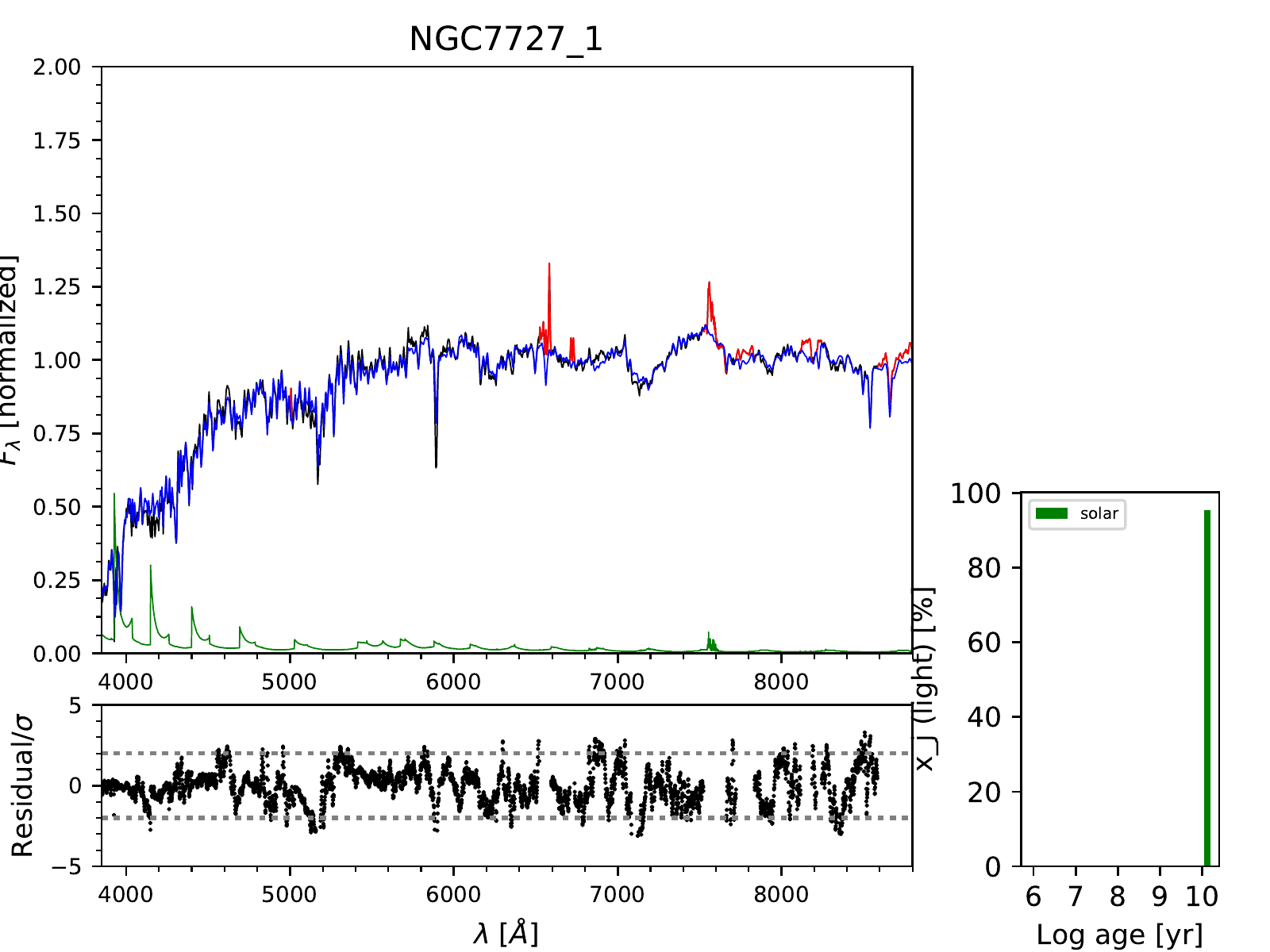}
\caption{\label{fig:SSP_NGC7727_1}As Fig.~\ref{fig:SSP_MCG523_1}, but for NGC 7727}
\end{figure*}

\clearpage
\subsection{AGNs}
\begin{figure*}
\includegraphics[width=\textwidth]{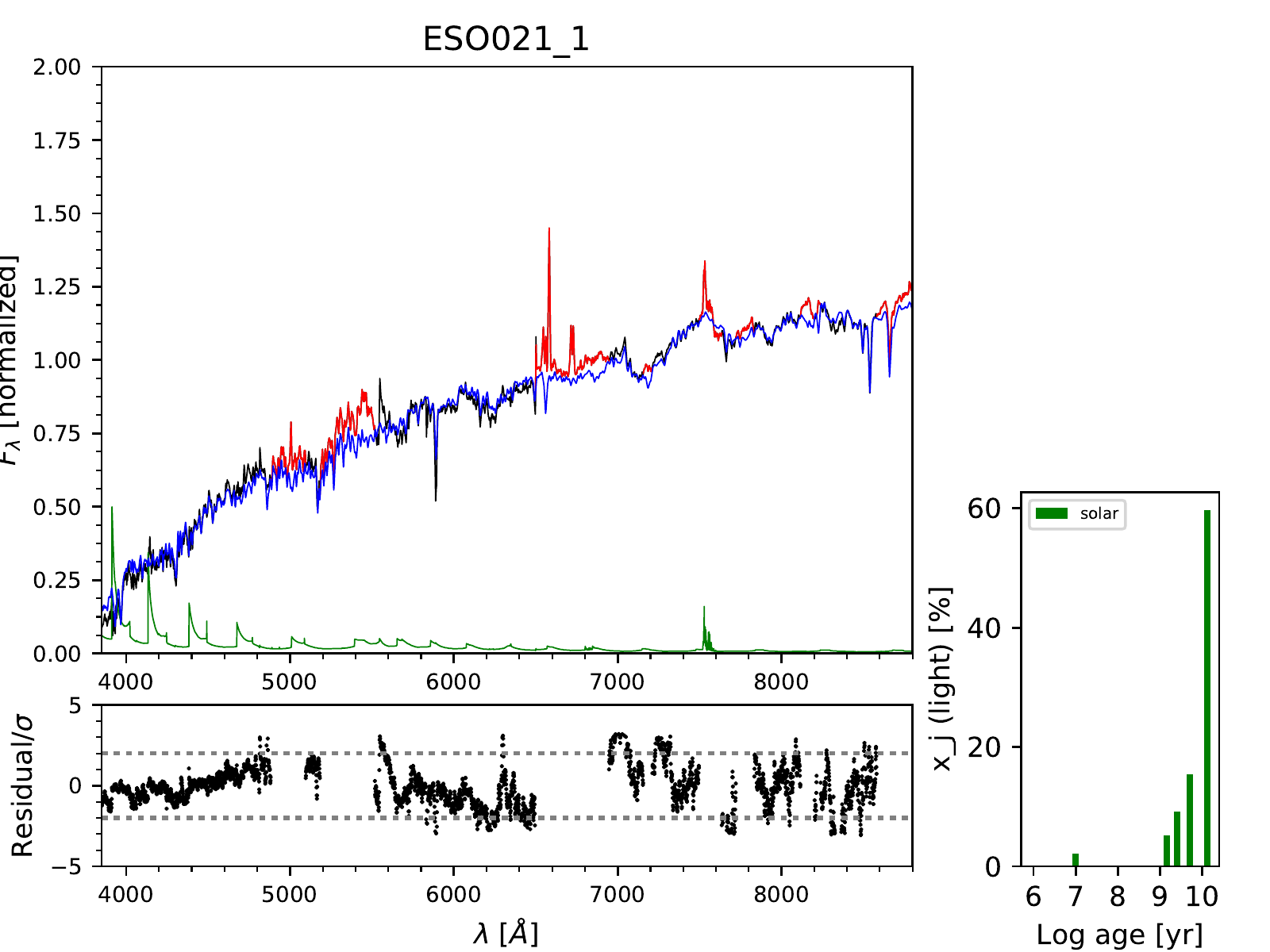}
\caption{\label{fig:SSP_ESO021_1}As Fig.~\ref{fig:SSP_MCG523_1}, but for ESO 021-G004}
\end{figure*}
\begin{figure*}
\includegraphics[width=\textwidth]{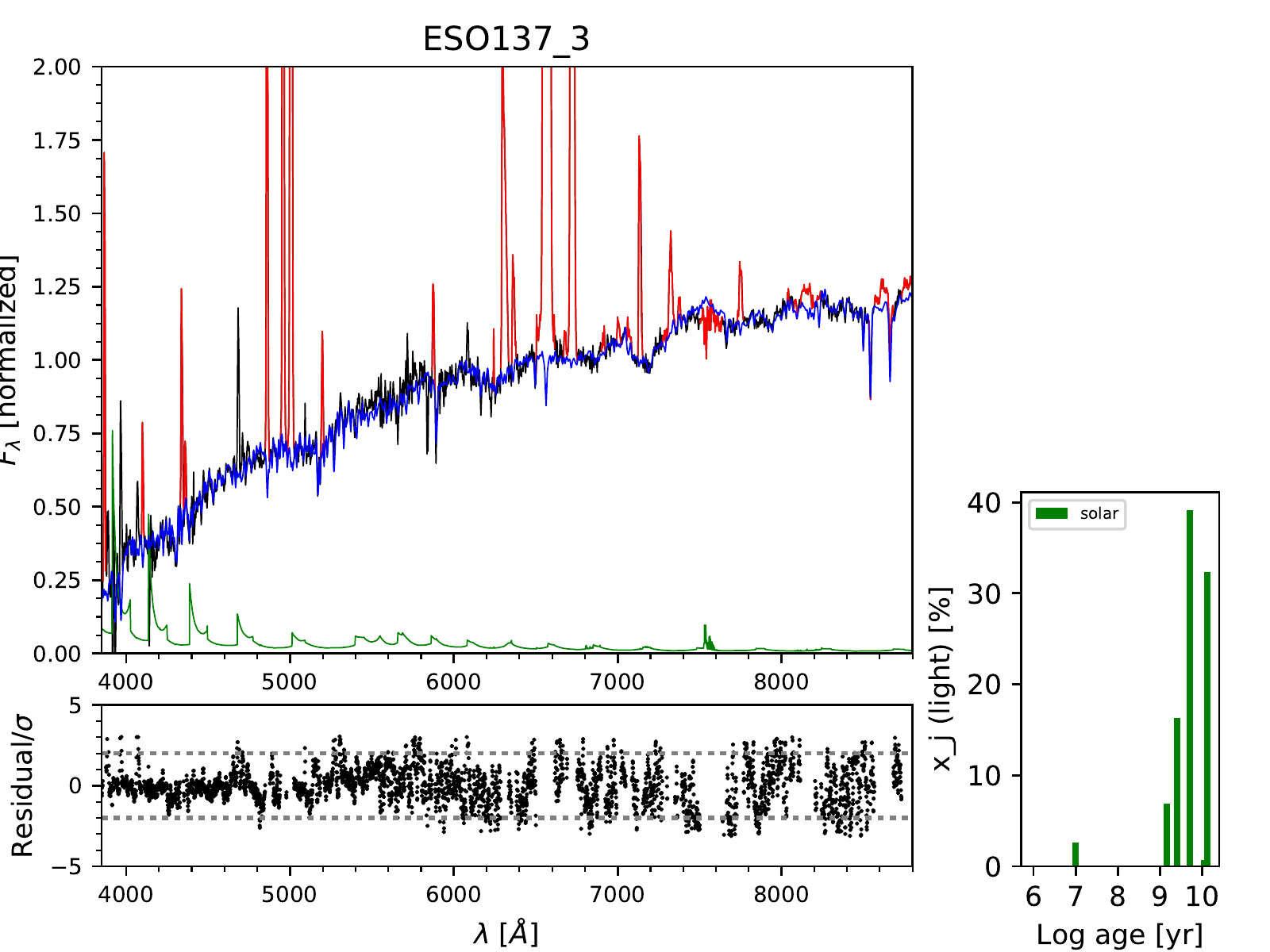}
\caption{\label{fig:SSP_ESO137_3}As Fig.~\ref{fig:SSP_MCG523_1}, but for ESO 137-G034}
\end{figure*}
\begin{figure*}
\includegraphics[width=\textwidth]{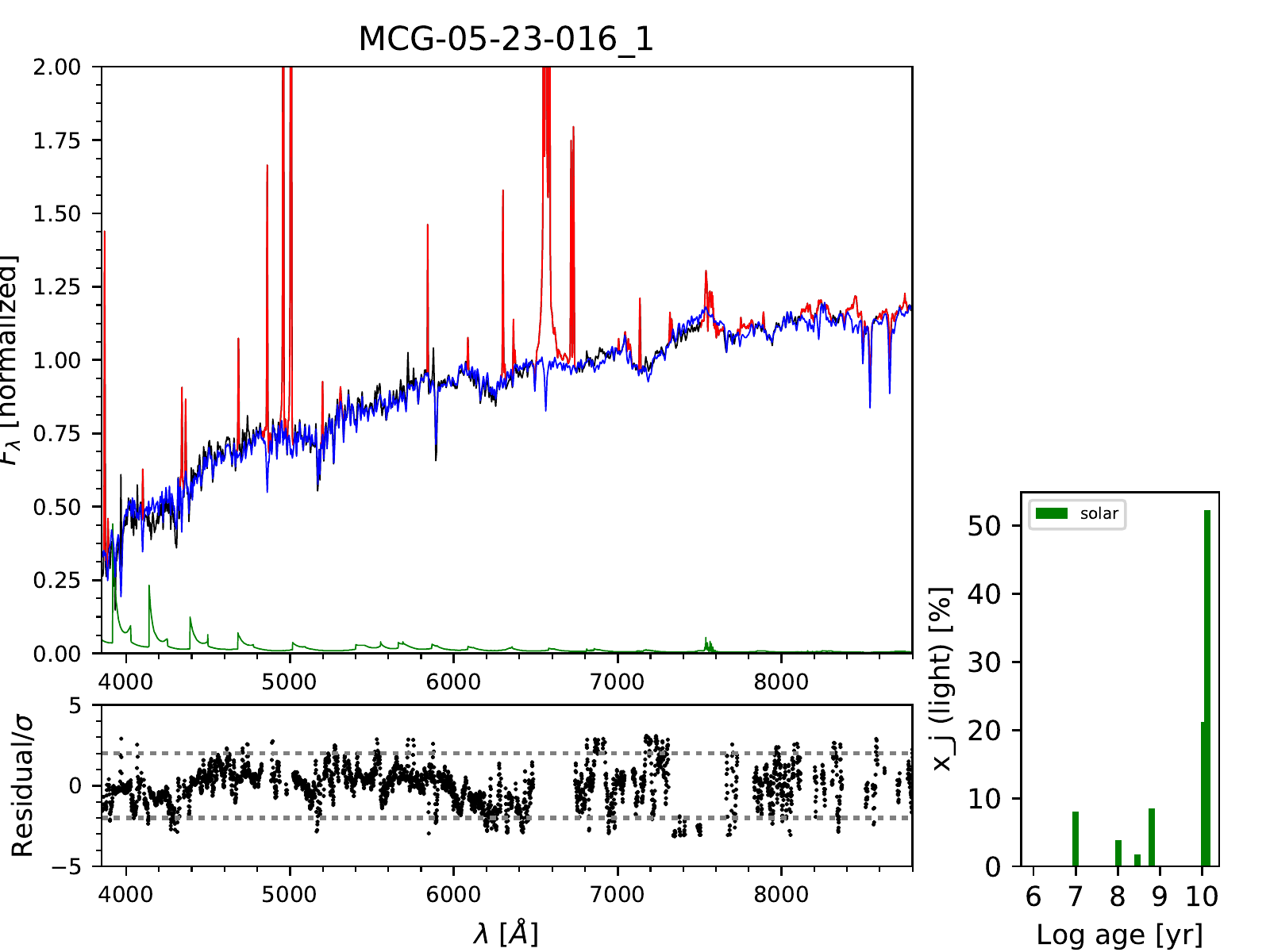}
\caption{As Fig.~\ref{fig:SSP_MCG523_1}, duplicated here for completeness.}
\end{figure*}
\begin{figure*}
\includegraphics[width=\textwidth]{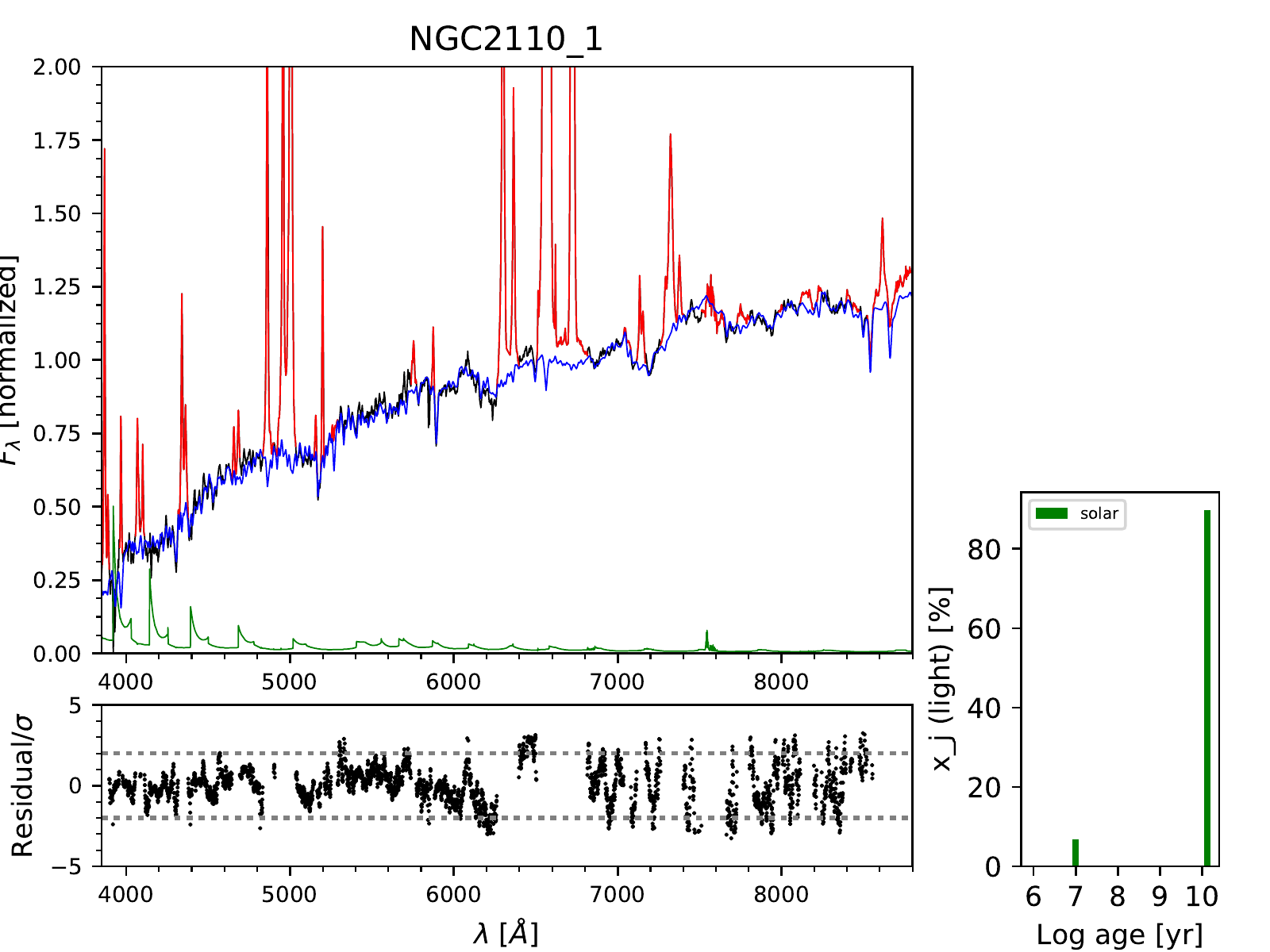}
\caption{\label{fig:SSP_NGC2110_1}As Fig.~\ref{fig:SSP_MCG523_1}, but for NGC 2110}
\end{figure*}
\begin{figure*}
\includegraphics[width=\textwidth]{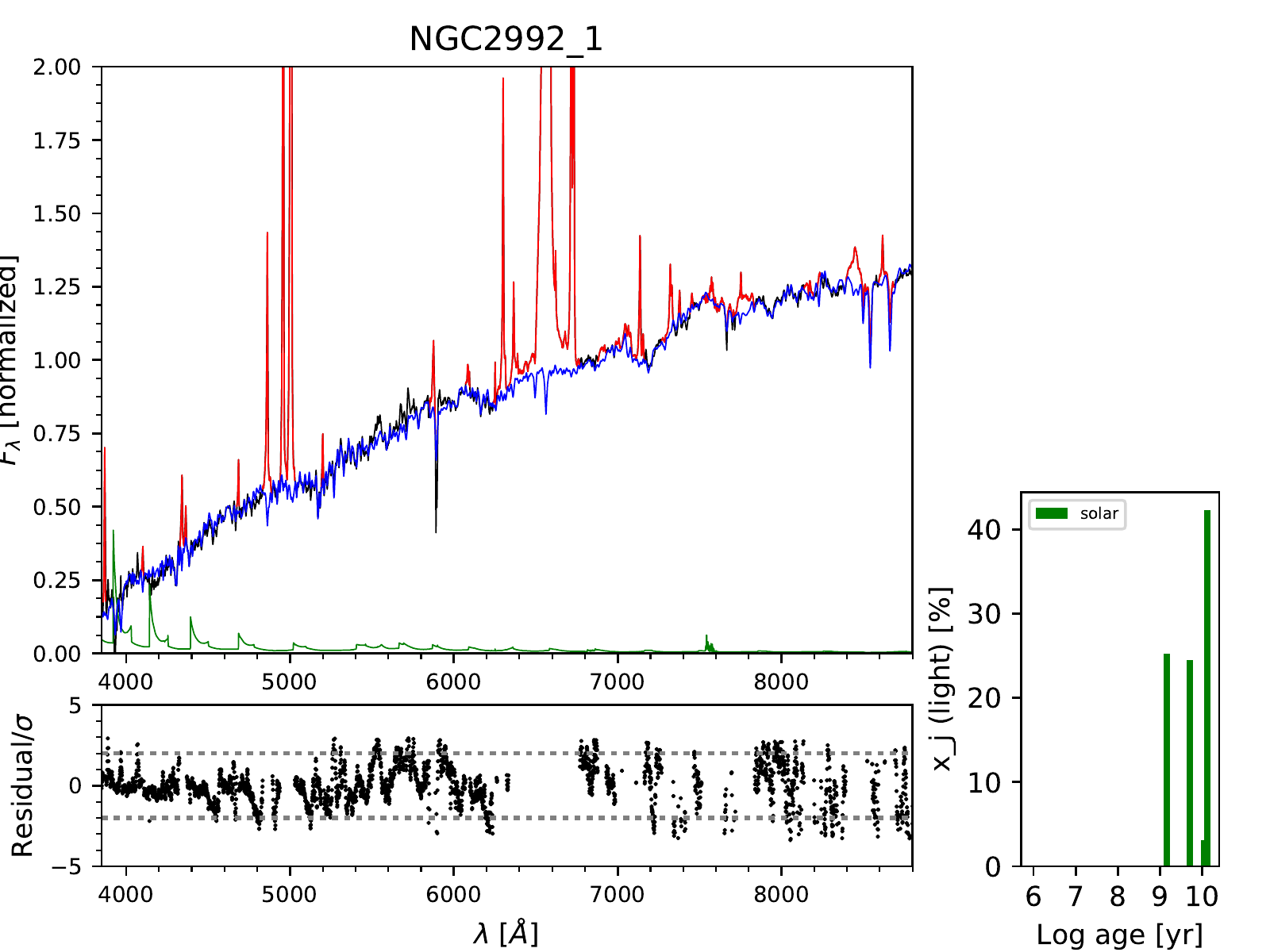}
\caption{\label{fig:SSP_NGC2992_1}As Fig.~\ref{fig:SSP_MCG523_1}, but for NGC 2992}
\end{figure*}
\begin{figure*}
\includegraphics[width=\textwidth]{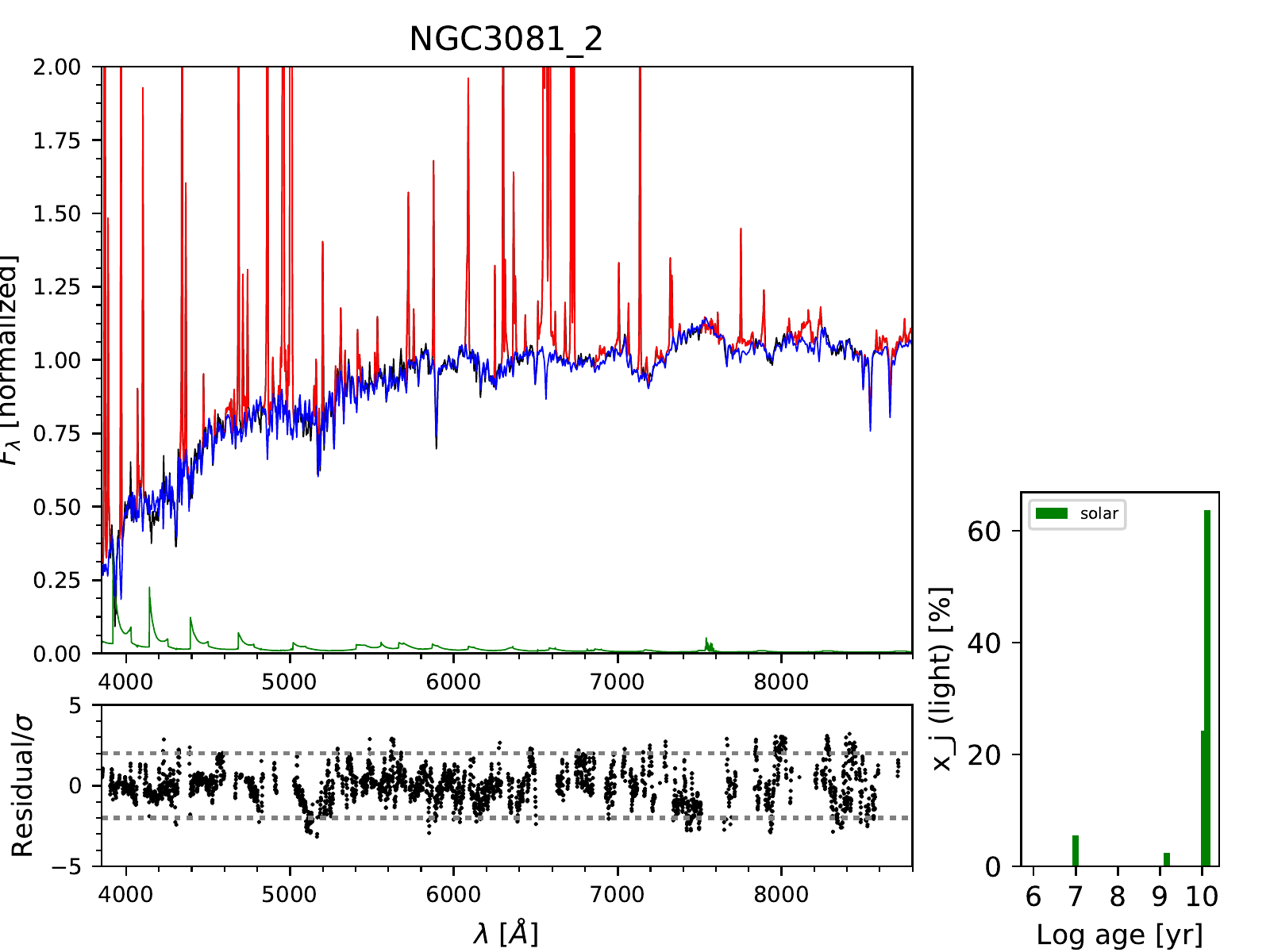}
\caption{\label{fig:SSP_NGC3081_2}As Fig.~\ref{fig:SSP_MCG523_1}, but for NGC 3081}
\end{figure*}
\begin{figure*}
\includegraphics[width=\textwidth]{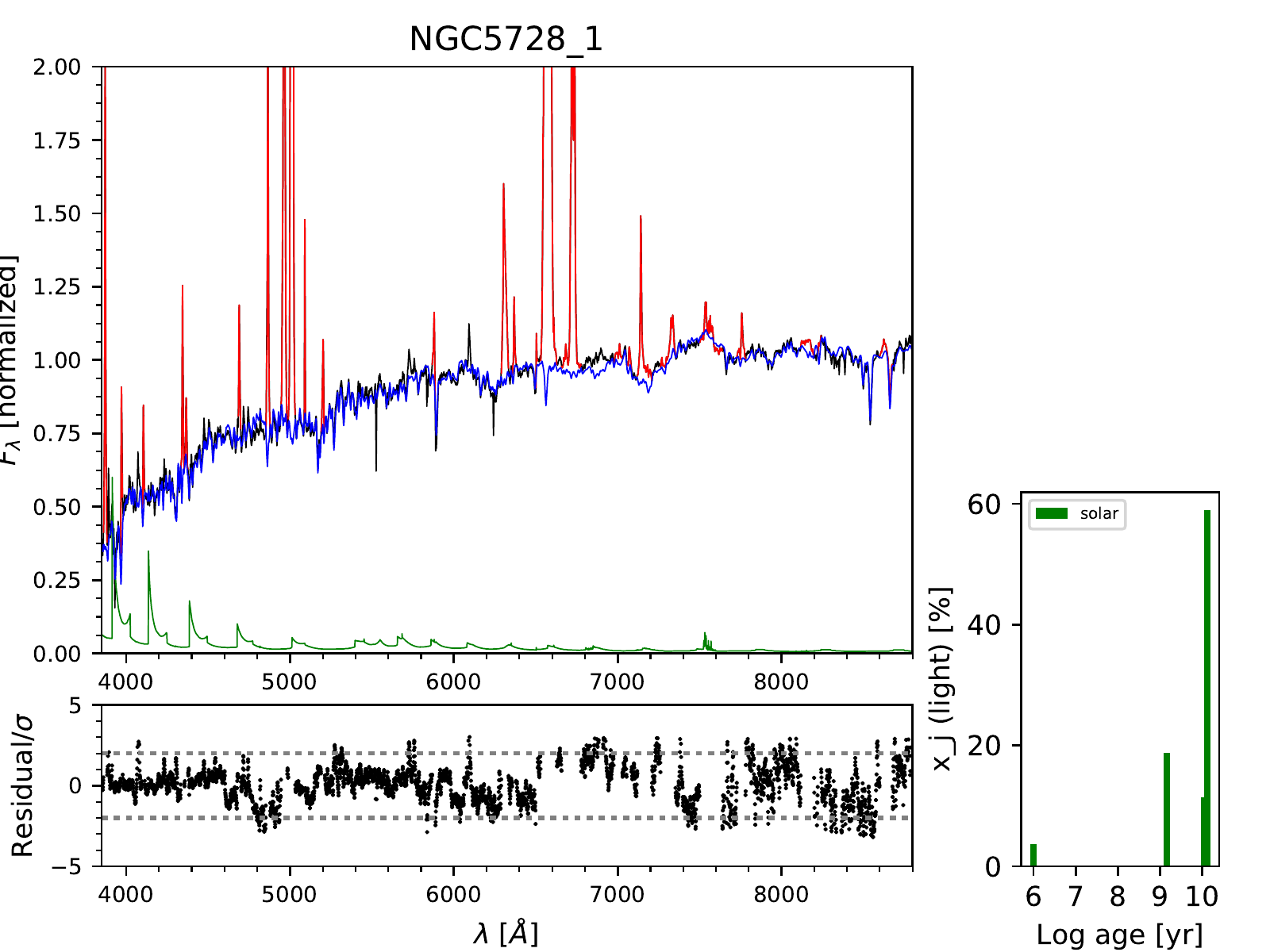}
\caption{\label{fig:SSP_NGC5728_1}As Fig.~\ref{fig:SSP_MCG523_1}, but for NGC 5728}
\end{figure*}
\begin{figure*}
\includegraphics[width=\textwidth]{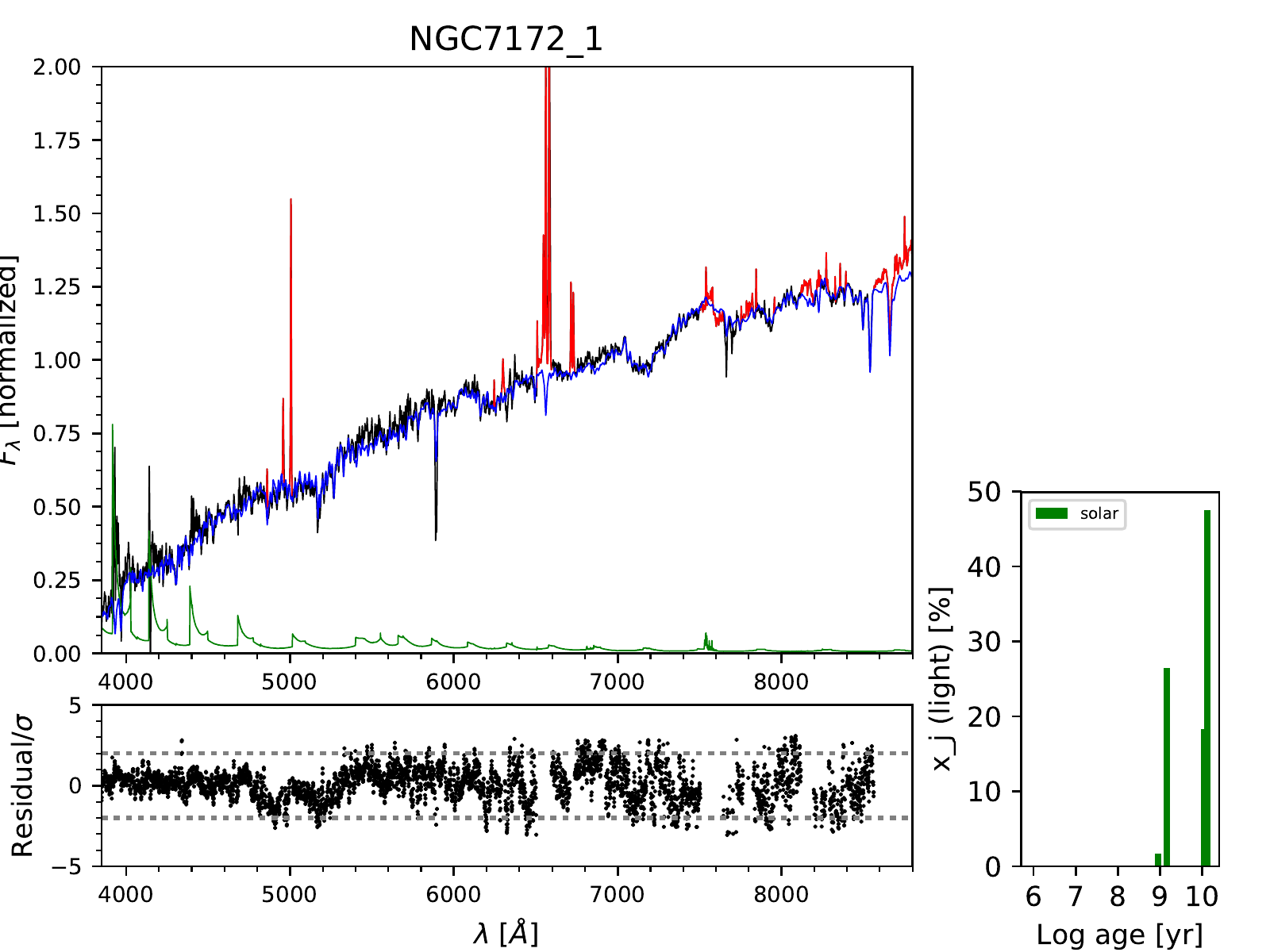}
\caption{\label{fig:SSP_NGC7172_1}As Fig.~\ref{fig:SSP_MCG523_1}, but for NGC 7172}
\end{figure*}
\begin{figure*}
\includegraphics[width=\textwidth]{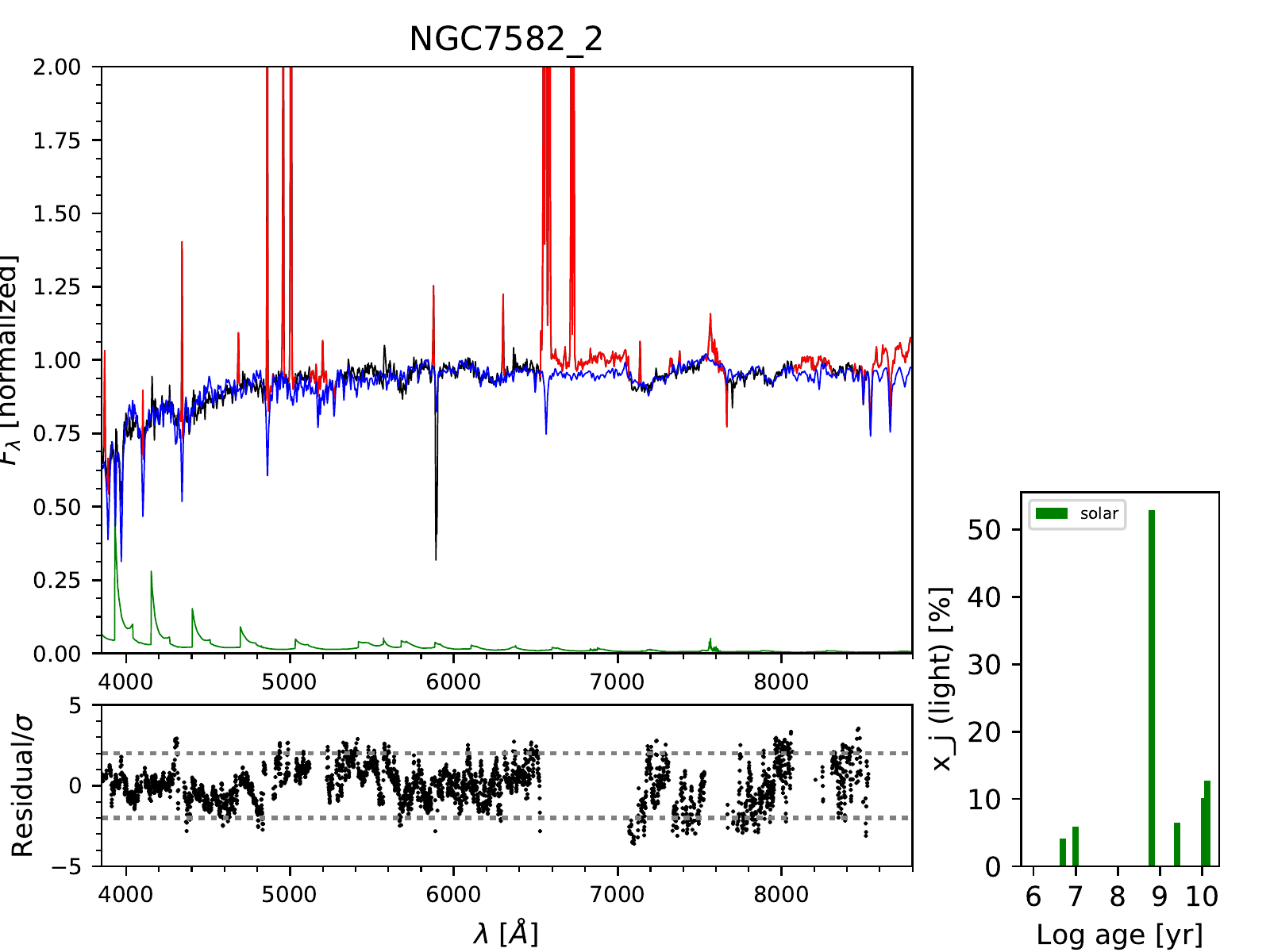}
\caption{\label{fig:SSP_NGC7582_2}As Fig.~\ref{fig:SSP_MCG523_1}, but for NGC 7582}
\end{figure*}

%% file: main.bbl
\begin{thebibliography}{188}
\expandafter\ifx\csname natexlab\endcsname\relax\def\natexlab#1{#1}\fi

\bibitem[{{Alexander} \& {Hickox}(2012)}]{alexander2012}
{Alexander}, D.~M. \& {Hickox}, R.~C. 2012, \nar, 56, 93

\bibitem[{{Allen} {et~al.}(2008){Allen}, {Groves}, {Dopita}, {Sutherland}, \&
  {Kewley}}]{allen2008}
{Allen}, M.~G., {Groves}, B.~A., {Dopita}, M.~A., {Sutherland}, R.~S., \&
  {Kewley}, L.~J. 2008, \apjs, 178, 20

\bibitem[{{Alonso-Herrero} {et~al.}(2020){Alonso-Herrero}, {Pereira-Santaella},
  {Rigopoulou}, {Garc{\'\i}a-Bernete}, {Garc{\'\i}a-Burillo},
  {Dom{\'\i}nguez-Fern{\'a}ndez}, {Combes}, {Davies}, {D{\'\i}az-Santos},
  {Esparza-Arredondo}, {Gonz{\'a}lez-Mart{\'\i}n}, {Hern{\'a}n-Caballero},
  {Hicks}, {H{\"o}nig}, {Levenson}, {Ramos Almeida}, {Roche}, \&
  {Rosario}}]{alonsoherrero2020}
{Alonso-Herrero}, A., {Pereira-Santaella}, M., {Rigopoulou}, D., {et~al.} 2020,
  \aap, 639, A43

\bibitem[{{Anders} \& {Grevesse}(1989)}]{anders1989}
{Anders}, E. \& {Grevesse}, N. 1989, \gca, 53, 197

\bibitem[{{Andrews} \& {Martini}(2013)}]{andrews2013}
{Andrews}, B.~H. \& {Martini}, P. 2013, \apj, 765, 140

\bibitem[{{Antonucci} \& {Miller}(1985)}]{antonucci1985}
{Antonucci}, R.~R.~J. \& {Miller}, J.~S. 1985, \apj, 297, 621

\bibitem[{{Asmus} {et~al.}(2011){Asmus}, {Gandhi}, {Smette}, {H{\"o}nig}, \&
  {Duschl}}]{asmus2011}
{Asmus}, D., {Gandhi}, P., {Smette}, A., {H{\"o}nig}, S.~F., \& {Duschl}, W.~J.
  2011, \aap, 536, A36

\bibitem[{{Asplund} {et~al.}(2009){Asplund}, {Grevesse}, {Sauval}, \&
  {Scott}}]{asplund2009}
{Asplund}, M., {Grevesse}, N., {Sauval}, A.~J., \& {Scott}, P. 2009, \araa, 47,
  481

\bibitem[{{Astropy Collaboration} {et~al.}(2018){Astropy Collaboration},
  {Price-Whelan}, {Sip{\H o}cz}, {G{\"u}nther}, {Lim}, {Crawford}, {Conseil},
  {Shupe}, {Craig}, {Dencheva}, {Ginsburg}, {VanderPlas}, {Bradley},
  {P{\'e}rez-Su{\'a}rez}, {de Val-Borro}, {Aldcroft}, {Cruz}, {Robitaille},
  {Tollerud}, {Ardelean}, {Babej}, {Bach}, {Bachetti}, {Bakanov}, {Bamford},
  {Barentsen}, {Barmby}, {Baumbach}, {Berry}, {Biscani}, {Boquien}, {Bostroem},
  {Bouma}, {Brammer}, {Bray}, {Breytenbach}, {Buddelmeijer}, {Burke},
  {Calderone}, {Cano Rodr{\'{\i}}guez}, {Cara}, {Cardoso}, {Cheedella},
  {Copin}, {Corrales}, {Crichton}, {D'Avella}, {Deil}, {Depagne}, {Dietrich},
  {Donath}, {Droettboom}, {Earl}, {Erben}, {Fabbro}, {Ferreira}, {Finethy},
  {Fox}, {Garrison}, {Gibbons}, {Goldstein}, {Gommers}, {Greco}, {Greenfield},
  {Groener}, {Grollier}, {Hagen}, {Hirst}, {Homeier}, {Horton}, {Hosseinzadeh},
  {Hu}, {Hunkeler}, {Ivezi{\'c}}, {Jain}, {Jenness}, {Kanarek}, {Kendrew},
  {Kern}, {Kerzendorf}, {Khvalko}, {King}, {Kirkby}, {Kulkarni}, {Kumar},
  {Lee}, {Lenz}, {Littlefair}, {Ma}, {Macleod}, {Mastropietro}, {McCully},
  {Montagnac}, {Morris}, {Mueller}, {Mumford}, {Muna}, {Murphy}, {Nelson},
  {Nguyen}, {Ninan}, {N{\"o}the}, {Ogaz}, {Oh}, {Parejko}, {Parley}, {Pascual},
  {Patil}, {Patil}, {Plunkett}, {Prochaska}, {Rastogi}, {Reddy Janga},
  {Sabater}, {Sakurikar}, {Seifert}, {Sherbert}, {Sherwood-Taylor}, {Shih},
  {Sick}, {Silbiger}, {Singanamalla}, {Singer}, {Sladen}, {Sooley},
  {Sornarajah}, {Streicher}, {Teuben}, {Thomas}, {Tremblay}, {Turner},
  {Terr{\'o}n}, {van Kerkwijk}, {de la Vega}, {Watkins}, {Weaver}, {Whitmore},
  {Woillez}, {Zabalza}, \& {Astropy Contributors}}]{astropy2018}
{Astropy Collaboration}, {Price-Whelan}, A.~M., {Sip{\H o}cz}, B.~M., {et~al.}
  2018, \aj, 156, 123

\bibitem[{{Astropy Collaboration} {et~al.}(2013){Astropy Collaboration},
  {Robitaille}, {Tollerud}, {Greenfield}, {Droettboom}, {Bray}, {Aldcroft},
  {Davis}, {Ginsburg}, {Price-Whelan}, {Kerzendorf}, {Conley}, {Crighton},
  {Barbary}, {Muna}, {Ferguson}, {Grollier}, {Parikh}, {Nair}, {Unther},
  {Deil}, {Woillez}, {Conseil}, {Kramer}, {Turner}, {Singer}, {Fox}, {Weaver},
  {Zabalza}, {Edwards}, {Azalee Bostroem}, {Burke}, {Casey}, {Crawford},
  {Dencheva}, {Ely}, {Jenness}, {Labrie}, {Lim}, {Pierfederici}, {Pontzen},
  {Ptak}, {Refsdal}, {Servillat}, \& {Streicher}}]{astropy2013}
{Astropy Collaboration}, {Robitaille}, T.~P., {Tollerud}, E.~J., {et~al.} 2013,
  \aap, 558, A33

\bibitem[{{Baldwin} {et~al.}(2018){Baldwin}, {McDermid}, {Kuntschner},
  {Maraston}, \& {Conroy}}]{baldwin2018}
{Baldwin}, C., {McDermid}, R.~M., {Kuntschner}, H., {Maraston}, C., \&
  {Conroy}, C. 2018, \mnras, 473, 4698

\bibitem[{{Baldwin} {et~al.}(1981){Baldwin}, {Phillips}, \&
  {Terlevich}}]{baldwin1981}
{Baldwin}, J.~A., {Phillips}, M.~M., \& {Terlevich}, R. 1981, \pasp, 93, 5

\bibitem[{{Baumgartner} {et~al.}(2013){Baumgartner}, {Tueller}, {Markwardt},
  {Skinner}, {Barthelmy}, {Mushotzky}, {Evans}, \& {Gehrels}}]{baumgartner2013}
{Baumgartner}, W.~H., {Tueller}, J., {Markwardt}, C.~B., {et~al.} 2013, \apjs,
  207, 19

\bibitem[{{Bing} {et~al.}(2019){Bing}, {Shi}, {Chen}, {S{\'a}nchez},
  {Maiolino}, {Riffel}, {Riffel}, {Wylezalek}, {Bizyaev}, {Pan}, \&
  {Drory}}]{bing2019}
{Bing}, L., {Shi}, Y., {Chen}, Y., {et~al.} 2019, \mnras, 482, 194

\bibitem[{B{\"o}nsch \& Potulski(1998)}]{boensch1998}
B{\"o}nsch, G. \& Potulski, E. 1998, Metrologia, 35, 133

\bibitem[{{Brinchmann} {et~al.}(2004){Brinchmann}, {Charlot}, {White},
  {Tremonti}, {Kauffmann}, {Heckman}, \& {Brinkmann}}]{brinchmann2004}
{Brinchmann}, J., {Charlot}, S., {White}, S.~D.~M., {et~al.} 2004, \mnras, 351,
  1151

\bibitem[{{Bruzual} \& {Charlot}(2003)}]{bruzual2003}
{Bruzual}, G. \& {Charlot}, S. 2003, \mnras, 344, 1000

\bibitem[{{Burtscher} {et~al.}(2016){Burtscher}, {Davies}, {Graci{\'a}-Carpio},
  {Koss}, {Lin}, {Lutz}, {Nandra}, {Netzer}, {Orban de Xivry}, {Ricci},
  {Rosario}, {Veilleux}, {Contursi}, {Genzel}, {Schnorr-M{\"u}ller},
  {Sternberg}, {Sturm}, \& {Tacconi}}]{burtscher2016}
{Burtscher}, L., {Davies}, R.~I., {Graci{\'a}-Carpio}, J., {et~al.} 2016, \aap,
  586, A28

\bibitem[{{Burtscher} {et~al.}(2013){Burtscher}, {Meisenheimer}, {Tristram},
  {Jaffe}, {H{\"o}nig}, {Davies}, {Kishimoto}, {Pott}, {R{\"o}ttgering},
  {Schartmann}, {Weigelt}, \& {Wolf}}]{burtscher2013}
{Burtscher}, L., {Meisenheimer}, K., {Tristram}, K.~R.~W., {et~al.} 2013, \aap,
  558

\bibitem[{{Burtscher} {et~al.}(2015){Burtscher}, {Orban de Xivry}, {Davies},
  {Janssen}, {Lutz}, {Rosario}, {Contursi}, {Genzel}, {Graci{\'a}-Carpio},
  {Lin}, {Schnorr-M{\"u}ller}, {Sternberg}, {Sturm}, \&
  {Tacconi}}]{burtscher2015}
{Burtscher}, L., {Orban de Xivry}, G., {Davies}, R.~I., {et~al.} 2015, \aap,
  578, A47

\bibitem[{{Caglar} {et~al.}(2020){Caglar}, {Burtscher}, {Brandl}, {Brinchmann},
  {Davies}, {Hicks}, {Koss}, {Lin}, {Maciejewski}, {M{\"u}ller-S{\'a}nchez},
  {Riffel}, {Riffel}, {Rosario}, {Schartmann}, {Schnorr-M{\"u}ller}, {Taro
  Shimizu}, {Storchi-Bergmann}, {Veilleux}, {de Xivry}, \&
  {Bennert}}]{caglar2020}
{Caglar}, T., {Burtscher}, L., {Brandl}, B., {et~al.} 2020, \aap, 634, A114

\bibitem[{{Calzetti}(1997)}]{calzetti1997}
{Calzetti}, D. 1997, \aj, 113, 162

\bibitem[{{Calzetti} {et~al.}(2000){Calzetti}, {Armus}, {Bohlin}, {Kinney},
  {Koornneef}, \& {Storchi-Bergmann}}]{calzetti2000}
{Calzetti}, D., {Armus}, L., {Bohlin}, R.~C., {et~al.} 2000, \apj, 533, 682

\bibitem[{{Calzetti} {et~al.}(1994){Calzetti}, {Kinney}, \&
  {Storchi-Bergmann}}]{calzetti1994}
{Calzetti}, D., {Kinney}, A.~L., \& {Storchi-Bergmann}, T. 1994, \apj, 429, 582

\bibitem[{{Cappellari} \& {Emsellem}(2004)}]{cappellari2004}
{Cappellari}, M. \& {Emsellem}, E. 2004, \pasp, 116, 138

\bibitem[{{Chabrier}(2003)}]{chabrier2003}
{Chabrier}, G. 2003, \pasp, 115, 763

\bibitem[{{Cid Fernandes} \& {Terlevich}(1995)}]{cidfernandes1995}
{Cid Fernandes}, Roberto, J. \& {Terlevich}, R. 1995, \mnras, 272, 423

\bibitem[{{Cid Fernandes}(2018)}]{cidfernandes2018}
{Cid Fernandes}, R. 2018, \mnras, 480, 4480

\bibitem[{{Cid Fernandes} {et~al.}(2014){Cid Fernandes}, {Gonz{\'a}lez
  Delgado}, {Garc{\'\i}a Benito}, {P{\'e}rez}, {de Amorim}, {S{\'a}nchez},
  {Husemann}, {Falc{\'o}n Barroso}, {L{\'o}pez-Fern{\'a}ndez}, \&
  {S{\'a}nchez-Bl{\'a}zquez}}]{cidfernandes2014}
{Cid Fernandes}, R., {Gonz{\'a}lez Delgado}, R.~M., {Garc{\'\i}a Benito}, R.,
  {et~al.} 2014, \aap, 561, A130

\bibitem[{{Cid Fernandes} {et~al.}(2004){Cid Fernandes}, {Gu}, {Melnick},
  {Terlevich}, {Terlevich}, {Kunth}, {Rodrigues Lacerda}, \&
  {Joguet}}]{cidfernandes2004}
{Cid Fernandes}, R., {Gu}, Q., {Melnick}, J., {et~al.} 2004, \mnras, 355, 273

\bibitem[{{Cid Fernandes} {et~al.}(2001){Cid Fernandes}, {Heckman}, {Schmitt},
  {Gonz{\'a}lez Delgado}, \& {Storchi-Bergmann}}]{cidfernandes2001b}
{Cid Fernandes}, R., {Heckman}, T., {Schmitt}, H., {Gonz{\'a}lez Delgado},
  R.~M., \& {Storchi-Bergmann}, T. 2001, \apj, 558, 81

\bibitem[{{Cid Fernandes} {et~al.}(2005){Cid Fernandes}, {Mateus}, {Sodr{\'e}},
  {Stasi{\'n}ska}, \& {Gomes}}]{cidfernandes2005}
{Cid Fernandes}, R., {Mateus}, A., {Sodr{\'e}}, L., {Stasi{\'n}ska}, G., \&
  {Gomes}, J.~M. 2005, \mnras, 358, 363

\bibitem[{{Cid Fernandes} {et~al.}(2011){Cid Fernandes}, {Stasi{\'n}ska},
  {Mateus}, \& {Vale Asari}}]{cidfernandes2011}
{Cid Fernandes}, R., {Stasi{\'n}ska}, G., {Mateus}, A., \& {Vale Asari}, N.
  2011, \mnras, 413, 1687

\bibitem[{{Cid Fernandes} {et~al.}(2010){Cid Fernandes}, {Stasi{\'n}ska},
  {Schlickmann}, {Mateus}, {Vale Asari}, {Schoenell}, \&
  {Sodr{\'e}}}]{cidfernandes2010}
{Cid Fernandes}, R., {Stasi{\'n}ska}, G., {Schlickmann}, M.~S., {et~al.} 2010,
  \mnras, 403, 1036

\bibitem[{{Cisternas} {et~al.}(2011){Cisternas}, {Jahnke}, {Inskip},
  {Kartaltepe}, {Koekemoer}, {Lisker}, {Robaina}, {Scodeggio}, {Sheth},
  {Trump}, {Andrae}, {Miyaji}, {Lusso}, {Brusa}, {Capak}, {Cappelluti},
  {Civano}, {Ilbert}, {Impey}, {Leauthaud}, {Lilly}, {Salvato}, {Scoville}, \&
  {Taniguchi}}]{cisternas2011}
{Cisternas}, M., {Jahnke}, K., {Inskip}, K.~J., {et~al.} 2011, \apj, 726, 57

\bibitem[{{Cisternas} {et~al.}(2015){Cisternas}, {Sheth}, {Salvato}, {Knapen},
  {Civano}, \& {Santini}}]{cisternas2015}
{Cisternas}, M., {Sheth}, K., {Salvato}, M., {et~al.} 2015, \apj, 802, 137

\bibitem[{{Colina} {et~al.}(1997){Colina}, {Vargas}, {Delgado}, {Mas-Hesse},
  {P{\'e}rez}, {Alberdi}, \& {Krabbe}}]{colina1997}
{Colina}, L., {Vargas}, M. L.~G., {Delgado}, R. M.~G., {et~al.} 1997, \apjl,
  488, L71

\bibitem[{{Combes} {et~al.}(2019){Combes}, {Garc{\'\i}a-Burillo}, {Audibert},
  {Hunt}, {Eckart}, {Aalto}, {Casasola}, {Boone}, {Krips}, {Viti}, {Sakamoto},
  {Muller}, {Dasyra}, {van der Werf}, \& {Martin}}]{combes2019}
{Combes}, F., {Garc{\'\i}a-Burillo}, S., {Audibert}, A., {et~al.} 2019, \aap,
  623, A79

\bibitem[{{Comer{\'o}n} {et~al.}(2010){Comer{\'o}n}, {Knapen}, {Beckman},
  {Laurikainen}, {Salo}, {Mart{\'\i}nez-Valpuesta}, \& {Buta}}]{comeron2010}
{Comer{\'o}n}, S., {Knapen}, J.~H., {Beckman}, J.~E., {et~al.} 2010, \mnras,
  402, 2462

\bibitem[{{Conroy}(2013)}]{conroy2013}
{Conroy}, C. 2013, \araa, 51, 393

\bibitem[{{Corwin} {et~al.}(1994){Corwin}, {Buta}, \& {de
  Vaucouleurs}}]{corwin1994}
{Corwin}, Harold~G., J., {Buta}, R.~J., \& {de Vaucouleurs}, G. 1994, \aj, 108,
  2128

\bibitem[{{Dahmer-Hahn} {et~al.}(2018){Dahmer-Hahn}, {Riffel},
  {Rodr{\'{\i}}guez-Ardila}, {Martins}, {Kehrig}, {Heckman}, {Pastoriza}, \&
  {Dametto}}]{dahmer_hahn2018}
{Dahmer-Hahn}, L.~G., {Riffel}, R., {Rodr{\'{\i}}guez-Ardila}, A., {et~al.}
  2018, \mnras, 476, 4459

\bibitem[{{Davies} {et~al.}(2020){Davies}, {Baron}, {Shimizu}, {Netzer},
  {Burtscher}, {de Zeeuw}, {Genzel}, {Hicks}, {Koss}, {Lin}, {Lutz},
  {Maciejewski}, {M{\"u}ller-S{\'a}nchez}, {Orban de Xivry}, {Ricci}, {Riffel},
  {Riffel}, {Rosario}, {Schartmann}, {Schnorr-M{\"u}ller}, {Shangguan},
  {Sternberg}, {Sturm}, {Storchi-Bergmann}, {Tacconi}, \&
  {Veilleux}}]{davies2020}
{Davies}, R., {Baron}, D., {Shimizu}, T., {et~al.} 2020, \mnras, 498, 4150

\bibitem[{{Davies} {et~al.}(2012){Davies}, {Burtscher}, {Dodds-Eden}, \& {Orban
  de Xivry}}]{davies2012}
{Davies}, R., {Burtscher}, L., {Dodds-Eden}, K., \& {Orban de Xivry}, G. 2012,
  Journal of Physics Conference Series, 372, 012046

\bibitem[{{Davies} {et~al.}(2015){Davies}, {Burtscher}, {Rosario},
  {Storchi-Bergmann}, {Contursi}, {Genzel}, {Graci{\'a}-Carpio}, {Hicks},
  {Janssen}, {Koss}, {Lin}, {Lutz}, {Maciejewski}, {M{\"u}ller-S{\'a}nchez},
  {Orban de Xivry}, {Ricci}, {Riffel}, {Riffel}, {Schartmann},
  {Schnorr-M{\"u}ller}, {Sternberg}, {Sturm}, {Tacconi}, \&
  {Veilleux}}]{davies2015}
{Davies}, R.~I., {Burtscher}, L., {Rosario}, D., {et~al.} 2015, \apj, 806, 127

\bibitem[{{Davies} {et~al.}(2017){Davies}, {Hicks}, {Erwin}, {Burtscher},
  {Contursi}, {Genzel}, {Janssen}, {Koss}, {Lin}, {Lutz}, {Maciejewski},
  {M{\"u}ller-S{\'a}nchez}, {Orban de Xivry}, {Ricci}, {Riffel}, {Riffel},
  {Rosario}, {Schartmann}, {Schnorr-M{\"u}ller}, {Shimizu}, {Sternberg},
  {Sturm}, {Storchi-Bergmann}, {Tacconi}, \& {Veilleux}}]{davies2017}
{Davies}, R.~I., {Hicks}, E.~K.~S., {Erwin}, P., {et~al.} 2017, \mnras, 466,
  4917

\bibitem[{{Davies} {et~al.}(2014{\natexlab{a}}){Davies}, {Maciejewski},
  {Hicks}, {Emsellem}, {Erwin}, {Burtscher}, {Dumas}, {Lin}, {Malkan},
  {M{\"u}ller-S{\'a}nchez}, {Orban de Xivry}, {Rosario}, {Schnorr-M{\"u}ller},
  \& {Tran}}]{davies2014}
{Davies}, R.~I., {Maciejewski}, W., {Hicks}, E.~K.~S., {et~al.}
  2014{\natexlab{a}}, \apj, 792, 101

\bibitem[{{Davies} {et~al.}(2007){Davies}, {Mueller S{\'a}nchez}, {Genzel},
  {Tacconi}, {Hicks}, {Friedrich}, \& {Sternberg}}]{davies2007}
{Davies}, R.~I., {Mueller S{\'a}nchez}, F., {Genzel}, R., {et~al.} 2007, \apj,
  671, 1388

\bibitem[{{Davies} {et~al.}(2014{\natexlab{b}}){Davies}, {Kewley}, {Ho}, \&
  {Dopita}}]{davies_rl2014b}
{Davies}, R.~L., {Kewley}, L.~J., {Ho}, I.~T., \& {Dopita}, M.~A.
  2014{\natexlab{b}}, \mnras, 444, 3961

\bibitem[{{Davis} {et~al.}(2019){Davis}, {Graham}, \& {Cameron}}]{davis2019}
{Davis}, B.~L., {Graham}, A.~W., \& {Cameron}, E. 2019, \apj, 873, 85

\bibitem[{{de Vaucouleurs}(1948)}]{devaucouleur1948}
{de Vaucouleurs}, G. 1948, Annales d'Astrophysique, 11, 247

\bibitem[{{Diamond-Stanic} \& {Rieke}(2010)}]{diamondstanic2010}
{Diamond-Stanic}, A.~M. \& {Rieke}, G.~H. 2010, \apj, 724, 140

\bibitem[{{Diamond-Stanic} \& {Rieke}(2012)}]{diamondstanic2012a}
{Diamond-Stanic}, A.~M. \& {Rieke}, G.~H. 2012, \apj, 746, 168

\bibitem[{{Diniz} {et~al.}(2017){Diniz}, {Riffel}, {Riffel}, {Crenshaw},
  {Storchi-Bergmann}, {Fischer}, {Schmitt}, \& {Kraemer}}]{diniz2017}
{Diniz}, M.~R., {Riffel}, R.~A., {Riffel}, R., {et~al.} 2017, \mnras, 469, 3286

\bibitem[{{Dom{\'{\i}}nguez} {et~al.}(2013){Dom{\'{\i}}nguez}, {Siana},
  {Henry}, {Scarlata}, {Bedregal}, {Malkan}, {Atek}, {Ross}, {Colbert},
  {Teplitz}, {Rafelski}, {McCarthy}, {Bunker}, {Hathi}, {Dressler}, {Martin},
  \& {Masters}}]{dominguez2013}
{Dom{\'{\i}}nguez}, A., {Siana}, B., {Henry}, A.~L., {et~al.} 2013, \apj, 763,
  145

\bibitem[{{Elvis}(2000)}]{elvis2000}
{Elvis}, M. 2000, \apj, 545, 63

\bibitem[{{Esparza-Arredondo} {et~al.}(2018){Esparza-Arredondo},
  {Gonz{\'a}lez-Mart{\'\i}n}, {Dultzin}, {Alonso-Herrero}, {Ramos Almeida},
  {D{\'\i}az-Santos}, {Garc{\'\i}a-Bernete}, {Martinez-Paredes}, \&
  {Rodr{\'\i}guez-Espinosa}}]{esparzaarredondo2018}
{Esparza-Arredondo}, D., {Gonz{\'a}lez-Mart{\'\i}n}, O., {Dultzin}, D.,
  {et~al.} 2018, \apj, 859, 124

\bibitem[{{Esquej} {et~al.}(2014){Esquej}, {Alonso-Herrero},
  {Gonz{\'a}lez-Mart{\'{\i}}n}, {H{\"o}nig}, {Hern{\'a}n-Caballero}, {Roche},
  {Ramos Almeida}, {Mason}, {D{\'{\i}}az-Santos}, {Levenson}, {Aretxaga},
  {Rodr{\'{\i}}guez Espinosa}, \& {Packham}}]{esquej2014}
{Esquej}, P., {Alonso-Herrero}, A., {Gonz{\'a}lez-Mart{\'{\i}}n}, O., {et~al.}
  2014, \apj, 780, 86

\bibitem[{{Fabian}(2012)}]{fabian2012}
{Fabian}, A.~C. 2012, ArXiv e-prints

\bibitem[{{Ferrarese} \& {Merritt}(2000)}]{ferrarese2000}
{Ferrarese}, L. \& {Merritt}, D. 2000, \apj, 539, L9

\bibitem[{{F{\"o}rster Schreiber} {et~al.}(2009){F{\"o}rster Schreiber},
  {Genzel}, {Bouch{\'e}}, {Cresci}, {Davies}, {Buschkamp}, {Shapiro},
  {Tacconi}, {Hicks}, {Genel}, {Shapley}, {Erb}, {Steidel}, {Lutz},
  {Eisenhauer}, {Gillessen}, {Sternberg}, {Renzini}, {Cimatti}, {Daddi},
  {Kurk}, {Lilly}, {Kong}, {Lehnert}, {Nesvadba}, {Verma}, {McCracken},
  {Arimoto}, {Mignoli}, \& {Onodera}}]{foersterschreiber2009}
{F{\"o}rster Schreiber}, N.~M., {Genzel}, R., {Bouch{\'e}}, N., {et~al.} 2009,
  \apj, 706, 1364

\bibitem[{{Gallagher} {et~al.}(2019){Gallagher}, {Maiolino}, {Belfiore},
  {Drory}, {Riffel}, \& {Riffel}}]{gallagher2019}
{Gallagher}, R., {Maiolino}, R., {Belfiore}, F., {et~al.} 2019, \mnras, 485,
  3409

\bibitem[{{Gao} {et~al.}(2020){Gao}, {Wang}, {Pearson}, {Gordon}, {Holwerda},
  {Hopkins}, {Brown}, {Bland-Hawthorn}, \& {Owers}}]{gao2020}
{Gao}, F., {Wang}, L., {Pearson}, W.~J., {et~al.} 2020, \aap, 637, A94

\bibitem[{{Gavazzi} {et~al.}(2018){Gavazzi}, {Consolandi}, {Belladitta},
  {Boselli}, \& {Fossati}}]{gavazzi2018}
{Gavazzi}, G., {Consolandi}, G., {Belladitta}, S., {Boselli}, A., \& {Fossati},
  M. 2018, \aap, 615, A104

\bibitem[{{Gebhardt} {et~al.}(2000){Gebhardt}, {Bender}, {Bower}, {Dressler},
  {Faber}, {Filippenko}, {Green}, {Grillmair}, {Ho}, {Kormendy}, {Lauer},
  {Magorrian}, {Pinkney}, {Richstone}, \& {Tremaine}}]{gebhardt2000}
{Gebhardt}, K., {Bender}, R., {Bower}, G., {et~al.} 2000, \apjl, 539, L13

\bibitem[{{Gonz{\'a}lez Delgado} {et~al.}(1998){Gonz{\'a}lez Delgado},
  {Heckman}, {Leitherer}, {Meurer}, {Krolik}, {Wilson}, {Kinney}, \&
  {Koratkar}}]{gonzalez_delgado1998}
{Gonz{\'a}lez Delgado}, R.~M., {Heckman}, T., {Leitherer}, C., {et~al.} 1998,
  \apj, 505, 174

\bibitem[{{Goto}(2006)}]{goto2006}
{Goto}, T. 2006, \mnras, 369, 1765

\bibitem[{{Goulding} {et~al.}(2017){Goulding}, {Matthaey}, {Greene}, {Hickox},
  {Alexander}, {Forman}, {Jones}, {Lehmer}, {Griffis}, {Kanek}, \&
  {Oulmakki}}]{goulding2017}
{Goulding}, A.~D., {Matthaey}, E., {Greene}, J.~E., {et~al.} 2017, \apj, 843,
  135

\bibitem[{{Grier} {et~al.}(2011){Grier}, {Mathur}, {Ghosh}, \&
  {Ferrarese}}]{grier2011}
{Grier}, C.~J., {Mathur}, S., {Ghosh}, H., \& {Ferrarese}, L. 2011, \apj, 731,
  60

\bibitem[{{Groves} {et~al.}(2004){Groves}, {Dopita}, \&
  {Sutherland}}]{groves2004b}
{Groves}, B.~A., {Dopita}, M.~A., \& {Sutherland}, R.~S. 2004, \apjs, 153, 75

\bibitem[{{H{\"a}ring} \& {Rix}(2004)}]{haering2004}
{H{\"a}ring}, N. \& {Rix}, H.-W. 2004, \apjl, 604, L89

\bibitem[{{Harrison}(2017)}]{harrison2017}
{Harrison}, C.~M. 2017, Nature Astronomy, 1, 0165

\bibitem[{{Heckman} \& {Best}(2014)}]{heckman2014}
{Heckman}, T.~M. \& {Best}, P.~N. 2014, \araa, 52, 589

\bibitem[{{Heller} \& {Shlosman}(1994)}]{heller1994}
{Heller}, C.~H. \& {Shlosman}, I. 1994, \apj, 424, 84

\bibitem[{{Hickox} {et~al.}(2014){Hickox}, {Mullaney}, {Alexander}, {Chen},
  {Civano}, {Goulding}, \& {Hainline}}]{hickox2014}
{Hickox}, R.~C., {Mullaney}, J.~R., {Alexander}, D.~M., {et~al.} 2014, \apj,
  782, 9

\bibitem[{{Hicks} {et~al.}(2013){Hicks}, {Davies}, {Maciejewski}, {Emsellem},
  {Malkan}, {Dumas}, {M{\"u}ller-S{\'a}nchez}, \& {Rivers}}]{hicks2013}
{Hicks}, E.~K.~S., {Davies}, R.~I., {Maciejewski}, W., {et~al.} 2013, \apj,
  768, 107

\bibitem[{{Hicks} {et~al.}(2009){Hicks}, {Davies}, {Malkan}, {Genzel},
  {Tacconi}, {M{\"u}ller S{\'a}nchez}, \& {Sternberg}}]{hicks2009}
{Hicks}, E.~K.~S., {Davies}, R.~I., {Malkan}, M.~A., {et~al.} 2009, \apj, 696,
  448

\bibitem[{{Hobbs} {et~al.}(2011){Hobbs}, {Nayakshin}, {Power}, \&
  {King}}]{hobbs2011}
{Hobbs}, A., {Nayakshin}, S., {Power}, C., \& {King}, A. 2011, \mnras, 413,
  2633

\bibitem[{{H{\"o}nig}(2019)}]{hoenig2019}
{H{\"o}nig}, S.~F. 2019, \apj, 884, 171

\bibitem[{{H{\"o}nig} {et~al.}(2013){H{\"o}nig}, {Kishimoto}, {Tristram},
  {Prieto}, {Gandhi}, {Asmus}, {Antonucci}, {Burtscher}, {Duschl}, \&
  {Weigelt}}]{hoenig2013}
{H{\"o}nig}, S.~F., {Kishimoto}, M., {Tristram}, K.~R.~W., {et~al.} 2013, \apj,
  771, 87

\bibitem[{{Hopkins} {et~al.}(2012){Hopkins}, {Hayward}, {Narayanan}, \&
  {Hernquist}}]{hopkins2012}
{Hopkins}, P.~F., {Hayward}, C.~C., {Narayanan}, D., \& {Hernquist}, L. 2012,
  \mnras, 420, 320

\bibitem[{{Hopkins} \& {Quataert}(2010)}]{hopkins2010b}
{Hopkins}, P.~F. \& {Quataert}, E. 2010, \mnras, 407, 1529

\bibitem[{{Hopkins} {et~al.}(2016){Hopkins}, {Torrey}, {Faucher-Gigu{\`e}re},
  {Quataert}, \& {Murray}}]{hopkins2016}
{Hopkins}, P.~F., {Torrey}, P., {Faucher-Gigu{\`e}re}, C.-A., {Quataert}, E.,
  \& {Murray}, N. 2016, \mnras, 458, 816

\bibitem[{{Husemann} {et~al.}(2017){Husemann}, {Davis}, {Jahnke},
  {Dannerbauer}, {Urrutia}, \& {Hodge}}]{husemann2017}
{Husemann}, B., {Davis}, T.~A., {Jahnke}, K., {et~al.} 2017, \mnras, 470, 1570

\bibitem[{{Imanishi} {et~al.}(2011){Imanishi}, {Ichikawa}, {Takeuchi},
  {Kawakatu}, {Oi}, \& {Imase}}]{imanishi2011}
{Imanishi}, M., {Ichikawa}, K., {Takeuchi}, T., {et~al.} 2011, \pasj, 63, 447

\bibitem[{{Jahnke} \& {Macci{\`o}}(2011)}]{jahnke2011}
{Jahnke}, K. \& {Macci{\`o}}, A.~V. 2011, \apj, 734, 92

\bibitem[{{Jarvis} {et~al.}(2020){Jarvis}, {Harrison}, {Mainieri}, {Calistro
  Rivera}, {Jethwa}, {Zhang}, {Alexander}, {Circosta}, {Costa}, {De Breuck},
  {Kakkad}, {Kharb}, {Lansbury}, \& {Thomson}}]{jarvis2020}
{Jarvis}, M.~E., {Harrison}, C.~M., {Mainieri}, V., {et~al.} 2020, \mnras, 498,
  1560

\bibitem[{{Jensen} {et~al.}(2017){Jensen}, {H{\"o}nig}, {Rakshit},
  {Alonso-Herrero}, {Asmus}, {Gandhi}, {Kishimoto}, {Smette}, \&
  {Tristram}}]{jensen2017}
{Jensen}, J.~J., {H{\"o}nig}, S.~F., {Rakshit}, S., {et~al.} 2017, \mnras, 470,
  3071

\bibitem[{{Kauffmann} \& {Heckman}(2009)}]{kauffmann2009}
{Kauffmann}, G. \& {Heckman}, T.~M. 2009, \mnras, 397, 135

\bibitem[{{Kauffmann} {et~al.}(2003){Kauffmann}, {Heckman}, {Tremonti},
  {Brinchmann}, {Charlot}, {White}, {Ridgway}, {Brinkmann}, {Fukugita}, {Hall},
  {Ivezi{\'c}}, {Richards}, \& {Schneider}}]{kauffmann2003c}
{Kauffmann}, G., {Heckman}, T.~M., {Tremonti}, C., {et~al.} 2003, \mnras, 346,
  1055

\bibitem[{{Kausch} {et~al.}(2015){Kausch}, {Noll}, {Smette}, {Kimeswenger},
  {Barden}, {Szyszka}, {Jones}, {Sana}, {Horst}, \& {Kerber}}]{kausch2015}
{Kausch}, W., {Noll}, S., {Smette}, A., {et~al.} 2015, \aap, 576, A78

\bibitem[{{Kawakatu} \& {Wada}(2008)}]{kawakatu2008}
{Kawakatu}, N. \& {Wada}, K. 2008, \apj, 681, 73

\bibitem[{{Kewley} \& {Dopita}(2002)}]{kewley2002}
{Kewley}, L.~J. \& {Dopita}, M.~A. 2002, \apjs, 142, 35

\bibitem[{{Kewley} {et~al.}(2001){Kewley}, {Dopita}, {Sutherland}, {Heisler},
  \& {Trevena}}]{kewley2001}
{Kewley}, L.~J., {Dopita}, M.~A., {Sutherland}, R.~S., {Heisler}, C.~A., \&
  {Trevena}, J. 2001, \apj, 556, 121

\bibitem[{{Kewley} {et~al.}(2006){Kewley}, {Groves}, {Kauffmann}, \&
  {Heckman}}]{kewley2006}
{Kewley}, L.~J., {Groves}, B., {Kauffmann}, G., \& {Heckman}, T. 2006, \mnras,
  372, 961

\bibitem[{{Knapen} {et~al.}(2000){Knapen}, {Shlosman}, \&
  {Peletier}}]{knapen2000}
{Knapen}, J.~H., {Shlosman}, I., \& {Peletier}, R.~F. 2000, \apj, 529, 93

\bibitem[{{Kocevski} {et~al.}(2012){Kocevski}, {Faber}, {Mozena}, {Koekemoer},
  {Nandra}, {Rangel}, {Laird}, {Brusa}, {Wuyts}, {Trump}, {Koo}, {Somerville},
  {Bell}, {Lotz}, {Alexander}, {Bournaud}, {Conselice}, {Dahlen}, {Dekel},
  {Donley}, {Dunlop}, {Finoguenov}, {Georgakakis}, {Giavalisco}, {Guo},
  {Grogin}, {Hathi}, {Juneau}, {Kartaltepe}, {Lucas}, {McGrath}, {McIntosh},
  {Mobasher}, {Robaina}, {Rosario}, {Straughn}, {van der Wel}, \&
  {Villforth}}]{kocevski2012}
{Kocevski}, D.~D., {Faber}, S.~M., {Mozena}, M., {et~al.} 2012, \apj, 744, 148

\bibitem[{{Kormendy} \& {Ho}(2013)}]{kormendy2013}
{Kormendy}, J. \& {Ho}, L.~C. 2013, \araa, 51, 511

\bibitem[{{Koss} {et~al.}(2011){Koss}, {Mushotzky}, {Veilleux}, {Winter},
  {Baumgartner}, {Tueller}, {Gehrels}, \& {Valencic}}]{koss2011}
{Koss}, M., {Mushotzky}, R., {Veilleux}, S., {et~al.} 2011, \apj, 739, 57

\bibitem[{{Koss} {et~al.}(2017){Koss}, {Trakhtenbrot}, {Ricci}, {Lamperti},
  {Oh}, {Berney}, {Schawinski}, {Balokovi{\'c}}, {Baronchelli}, {Crenshaw},
  {Fischer}, {Gehrels}, {Harrison}, {Hashimoto}, {Hogg}, {Ichikawa}, {Masetti},
  {Mushotzky}, {Sartori}, {Stern}, {Treister}, {Ueda}, {Veilleux}, \&
  {Winter}}]{koss2017}
{Koss}, M., {Trakhtenbrot}, B., {Ricci}, C., {et~al.} 2017, \apj, 850, 74

\bibitem[{{Koss} {et~al.}(2021){Koss}, {Strittmatter}, {Lamperti}, {Shimizu},
  {Trakhtenbrot}, {Saintonge}, {Treister}, {Cicone}, {Mushotzky}, {Oh},
  {Ricci}, {Stern}, {Ananna}, {Bauer}, {Privon}, {B{\"a}r}, {De Breuck},
  {Harrison}, {Ichikawa}, {Powell}, {Rosario}, {Sanders}, {Schawinski}, {Shao},
  {Megan Urry}, \& {Veilleux}}]{koss2021}
{Koss}, M.~J., {Strittmatter}, B., {Lamperti}, I., {et~al.} 2021, \apjs, 252,
  29

\bibitem[{{Krolik} \& {Begelman}(1988)}]{krolik1988}
{Krolik}, J.~H. \& {Begelman}, M.~C. 1988, \apj, 329, 702

\bibitem[{{Lacerda} {et~al.}(2020){Lacerda}, {S{\'a}nchez}, {Cid Fernandes},
  {L{\'o}pez-Cob{\'a}}, {Espinosa-Ponce}, \& {Galbany}}]{lacerda2020}
{Lacerda}, E. A.~D., {S{\'a}nchez}, S.~F., {Cid Fernandes}, R., {et~al.} 2020,
  \mnras, 492, 3073

\bibitem[{{Leitherer} {et~al.}(1999){Leitherer}, {Schaerer}, {Goldader},
  {Delgado}, {Robert}, {Kune}, {de Mello}, {Devost}, \&
  {Heckman}}]{leitherer1999}
{Leitherer}, C., {Schaerer}, D., {Goldader}, J.~D., {et~al.} 1999, \apjs, 123,
  3

\bibitem[{{Lin} {et~al.}(2018){Lin}, {Davies}, {Hicks}, {Burtscher},
  {Contursi}, {Genzel}, {Koss}, {Lutz}, {Maciejewski},
  {M{\"u}ller-S{\'a}nchez}, {Orban de Xivry}, {Ricci}, {Riffel}, {Riffel},
  {Rosario}, {Schartmann}, {Schnorr-M{\"u}ller}, {Shimizu}, {Sternberg},
  {Sturm}, {Storchi-Bergmann}, {Tacconi}, \& {Veilleux}}]{lin2018}
{Lin}, M.-Y., {Davies}, R.~I., {Hicks}, E.~K.~S., {et~al.} 2018, \mnras, 473,
  4582

\bibitem[{{L{\'o}pez-Gonzaga} {et~al.}(2016){L{\'o}pez-Gonzaga}, {Burtscher},
  {Tristram}, {Meisenheimer}, \& {Schartmann}}]{lopezgonzaga2016a}
{L{\'o}pez-Gonzaga}, N., {Burtscher}, L., {Tristram}, K.~R.~W., {Meisenheimer},
  K., \& {Schartmann}, M. 2016, \aap, 591, A47

\bibitem[{{Lutz} {et~al.}(2020){Lutz}, {Sturm}, {Janssen}, {Veilleux}, {Aalto},
  {Cicone}, {Contursi}, {Davies}, {Feruglio}, {Fischer}, {Fluetsch},
  {Garcia-Burillo}, {Genzel}, {Gonz{\'a}lez-Alfonso}, {Graci{\'a}-Carpio},
  {Herrera-Camus}, {Maiolino}, {Schruba}, {Shimizu}, {Sternberg}, {Tacconi}, \&
  {Wei{\ss}}}]{lutz2020}
{Lutz}, D., {Sturm}, E., {Janssen}, A., {et~al.} 2020, \aap, 633, A134

\bibitem[{{Madau} \& {Dickinson}(2014)}]{madau2014}
{Madau}, P. \& {Dickinson}, M. 2014, \araa, 52, 415

\bibitem[{{Magorrian} {et~al.}(1998){Magorrian}, {Tremaine}, {Richstone},
  {Bender}, {Bower}, {Dressler}, {Faber}, {Gebhardt}, {Green}, {Grillmair},
  {Kormendy}, \& {Lauer}}]{magorrian1998}
{Magorrian}, J., {Tremaine}, S., {Richstone}, D., {et~al.} 1998, \aj, 115, 2285

\bibitem[{{Maiolino} {et~al.}(2017){Maiolino}, {Russell}, {Fabian}, {Carniani},
  {Gallagher}, {Cazzoli}, {Arribas}, {Belfiore}, {Bellocchi}, {Colina},
  {Cresci}, {Ishibashi}, {Marconi}, {Mannucci}, {Oliva}, \&
  {Sturm}}]{maiolino2017}
{Maiolino}, R., {Russell}, H.~R., {Fabian}, A.~C., {et~al.} 2017, \nat, 544,
  202

\bibitem[{{Mallmann} {et~al.}(2018){Mallmann}, {Riffel}, {Storchi-Bergmann},
  {Rembold}, {Riffel}, {Schimoia}, {da Costa}, {{\'A}vila-Reese}, {Sanchez},
  {Machado}, {Cirolini}, {Ilha}, \& {Nascimento}}]{mallman_2018}
{Mallmann}, N.~D., {Riffel}, R., {Storchi-Bergmann}, T., {et~al.} 2018, \mnras,
  478, 5491

\bibitem[{{Marian} {et~al.}(2020){Marian}, {Jahnke}, {Andika}, {Ba{\~n}ados},
  {Bennert}, {Cohen}, {Husemann}, {Kaasinen}, {Koekemoer}, {Mechtley}, {Onoue},
  {Schindler}, {Schramm}, {Schulze}, {Silverman}, {Smirnova-Pinchukova}, {van
  der Wel}, {Villforth}, \& {Windhorst}}]{marian2020}
{Marian}, V., {Jahnke}, K., {Andika}, I., {et~al.} 2020, \apj, 904, 79

\bibitem[{{Marian} {et~al.}(2019){Marian}, {Jahnke}, {Mechtley}, {Cohen},
  {Husemann}, {Jones}, {Koekemoer}, {Schulze}, {van der Wel}, {Villforth}, \&
  {Windhorst}}]{marian2019}
{Marian}, V., {Jahnke}, K., {Mechtley}, M., {et~al.} 2019, \apj, 882, 141

\bibitem[{{Martini}(2004)}]{martini2004}
{Martini}, P. 2004, in The Interplay Among Black Holes, Stars and ISM in
  Galactic Nuclei, ed. T.~{Storchi-Bergmann}, L.~C. {Ho}, \& H.~R. {Schmitt},
  Vol. 222, 235--241

\bibitem[{{Martini} {et~al.}(2003){Martini}, {Regan}, {Mulchaey}, \&
  {Pogge}}]{martini2003b}
{Martini}, P., {Regan}, M.~W., {Mulchaey}, J.~S., \& {Pogge}, R.~W. 2003, \apj,
  589, 774

\bibitem[{{McConnell} \& {Ma}(2013)}]{mcconnell2013}
{McConnell}, N.~J. \& {Ma}, C.-P. 2013, \apj, 764, 184

\bibitem[{{Mel{\'e}ndez} {et~al.}(2014){Mel{\'e}ndez}, {Mushotzky}, {Shimizu},
  {Barger}, \& {Cowie}}]{melendez2014}
{Mel{\'e}ndez}, M., {Mushotzky}, R.~F., {Shimizu}, T.~T., {Barger}, A.~J., \&
  {Cowie}, L.~L. 2014, \apj, 794, 152

\bibitem[{{Modigliani} {et~al.}(2010){Modigliani}, {Goldoni}, {Royer},
  {Haigron}, {Guglielmi}, {Fran{\c{c}}ois}, {Horrobin}, {Bristow}, {Vernet},
  {Moehler}, {Kerber}, {Ballester}, {Mason}, \& {Christensen}}]{modigliani2010}
{Modigliani}, A., {Goldoni}, P., {Royer}, F., {et~al.} 2010, in Society of
  Photo-Optical Instrumentation Engineers (SPIE) Conference Series, Vol. 7737,
  Observatory Operations: Strategies, Processes, and Systems III, ed. D.~R.
  {Silva}, A.~B. {Peck}, \& B.~T. {Soifer}, 773728

\bibitem[{{Moehler} {et~al.}(2014){Moehler}, {Modigliani}, {Freudling},
  {Giammichele}, {Gianninas}, {Gonneau}, {Kausch}, {Lan{\c c}on}, {Noll},
  {Rauch}, \& {Vinther}}]{moehler2014}
{Moehler}, S., {Modigliani}, A., {Freudling}, W., {et~al.} 2014, \aap, 568, A9

\bibitem[{{Neumayer} {et~al.}(2020){Neumayer}, {Seth}, \&
  {B{\"o}ker}}]{neumayer2020}
{Neumayer}, N., {Seth}, A., \& {B{\"o}ker}, T. 2020, \aapr, 28, 4

\bibitem[{{Norman} \& {Scoville}(1988)}]{norman1988}
{Norman}, C. \& {Scoville}, N. 1988, \apj, 332, 124

\bibitem[{{Novak} {et~al.}(2011){Novak}, {Ostriker}, \& {Ciotti}}]{novak2011}
{Novak}, G.~S., {Ostriker}, J.~P., \& {Ciotti}, L. 2011, \apj, 737, 26

\bibitem[{{Oh} {et~al.}(2018){Oh}, {Koss}, {Markwardt}, {Schawinski},
  {Baumgartner}, {Barthelmy}, {Cenko}, {Gehrels}, {Mushotzky}, {Petulante},
  {Ricci}, {Lien}, \& {Trakhtenbrot}}]{oh2018}
{Oh}, K., {Koss}, M., {Markwardt}, C.~B., {et~al.} 2018, \apjs, 235, 4

\bibitem[{{Prieto} {et~al.}(2019){Prieto}, {Fernandez-Ontiveros}, {Bruzual},
  {Burkert}, {Schartmann}, \& {Charlot}}]{prieto2019}
{Prieto}, M.~A., {Fernandez-Ontiveros}, J.~A., {Bruzual}, G., {et~al.} 2019,
  \mnras, 485, 3264

\bibitem[{{Ramos Almeida} {et~al.}(2012){Ramos Almeida}, {Bessiere},
  {Tadhunter}, {P{\'e}rez-Gonz{\'a}lez}, {Barro}, {Inskip}, {Morganti}, {Holt},
  \& {Dicken}}]{ramosalmeida2012}
{Ramos Almeida}, C., {Bessiere}, P.~S., {Tadhunter}, C.~N., {et~al.} 2012,
  \mnras, 419, 687

\bibitem[{{Regan} \& {Teuben}(2003)}]{regan2003}
{Regan}, M.~W. \& {Teuben}, P. 2003, \apj, 582, 723

\bibitem[{{Rembold} {et~al.}(2017){Rembold}, {Shimoia}, {Storchi-Bergmann},
  {Riffel}, {Riffel}, {Mallmann}, {do Nascimento}, {Moreira}, {Ilha},
  {Machado}, {Cirolini}, {da Costa}, {Maia}, {Santiago}, {Schneider},
  {Wylezalek}, {Bizyaev}, {Pan}, \& {M{\"u}ller-S{\'a}nchez}}]{rembold2017}
{Rembold}, S.~B., {Shimoia}, J.~S., {Storchi-Bergmann}, T., {et~al.} 2017,
  \mnras, 472, 4382

\bibitem[{{Ricci} {et~al.}(2015){Ricci}, {Ueda}, {Koss}, {Trakhtenbrot},
  {Bauer}, \& {Gandhi}}]{ricci2015a}
{Ricci}, C., {Ueda}, Y., {Koss}, M.~J., {et~al.} 2015, \apjl, 815, L13

\bibitem[{{Riffel} {et~al.}(2021){Riffel}, {Mallmann}, {Ilha},
  {Storchi-Bergmann}, {Riffel}, {Rembold}, {Bizyaev}, {do Nascimento},
  {Schimoia}, {da Costa}, {Fraser Boardman}, {Boquien}, \&
  {Couto}}]{riffel_r2021}
{Riffel}, R., {Mallmann}, N.~D., {Ilha}, G.~S., {et~al.} 2021, \mnras, 501,
  4064

\bibitem[{{Riffel} {et~al.}(2015){Riffel}, {Mason}, {Martins},
  {Rodr{\'{\i}}guez-Ardila}, {Ho}, {Riffel}, {Lira}, {Gonzalez Martin},
  {Ruschel-Dutra}, {Alonso-Herrero}, {Flohic}, {McDermid}, {Ramos Almeida},
  {Thanjavur}, \& {Winge}}]{riffel_r2015}
{Riffel}, R., {Mason}, R.~E., {Martins}, L.~P., {et~al.} 2015, \mnras, 450,
  3069

\bibitem[{{Riffel} {et~al.}(2009){Riffel}, {Pastoriza},
  {Rodr{\'{\i}}guez-Ardila}, \& {Bonatto}}]{riffel_r2009}
{Riffel}, R., {Pastoriza}, M.~G., {Rodr{\'{\i}}guez-Ardila}, A., \& {Bonatto},
  C. 2009, \mnras, 400, 273

\bibitem[{{Riffel} {et~al.}(2011){Riffel}, {Riffel}, {Ferrari}, \&
  {Storchi-Bergmann}}]{riffel_r2011}
{Riffel}, R., {Riffel}, R.~A., {Ferrari}, F., \& {Storchi-Bergmann}, T. 2011,
  \mnras, 416, 493

\bibitem[{{Riffel} {et~al.}(2019){Riffel}, {Rodr{\'\i}guez-Ardila},
  {Brotherton}, {Peletier}, {Vazdekis}, {Riffel}, {Martins}, {Bonatto}, {Zanon
  Dametto}, {Dahmer-Hahn}, {Runnoe}, {Pastoriza}, {Chies-Santos}, \&
  {Trevisan}}]{riffel_r2019}
{Riffel}, R., {Rodr{\'\i}guez-Ardila}, A., {Brotherton}, M.~S., {et~al.} 2019,
  \mnras, 486, 3228

\bibitem[{{Riffel} {et~al.}(2010){Riffel}, {Storchi-Bergmann}, {Riffel}, \&
  {Pastoriza}}]{riffel_ra2010a}
{Riffel}, R.~A., {Storchi-Bergmann}, T., {Riffel}, R., \& {Pastoriza}, M.~G.
  2010, \apj, 713, 469

\bibitem[{{Rosario} {et~al.}(2018){Rosario}, {Burtscher}, {Davies}, {Koss},
  {Ricci}, {Lutz}, {Riffel}, {Alexander}, {Genzel}, {Hicks}, {Lin},
  {Maciejewski}, {M{\"u}ller-S{\'a}nchez}, {Orban de Xivry}, {Riffel},
  {Schartmann}, {Schawinski}, {Schnorr-M{\"u}ller}, {Saintonge}, {Shimizu},
  {Sternberg}, {Storchi-Bergmann}, {Sturm}, {Tacconi}, {Treister}, \&
  {Veilleux}}]{rosario2018}
{Rosario}, D.~J., {Burtscher}, L., {Davies}, R.~I., {et~al.} 2018, \mnras, 473,
  5658

\bibitem[{{Rosario} {et~al.}(2012){Rosario}, {Santini}, {Lutz}, {Shao},
  {Maiolino}, {Alexander}, {Altieri}, {Andreani}, {Aussel}, {Bauer}, {Berta},
  {Bongiovanni}, {Brandt}, {Brusa}, {Cepa}, {Cimatti}, {Cox}, {Daddi}, {Elbaz},
  {Fontana}, {F{\"o}rster Schreiber}, {Genzel}, {Grazian}, {Le Floch},
  {Magnelli}, {Mainieri}, {Netzer}, {Nordon}, {P{\'e}rez Garcia}, {Poglitsch},
  {Popesso}, {Pozzi}, {Riguccini}, {Rodighiero}, {Salvato}, {Sanchez-Portal},
  {Sturm}, {Tacconi}, {Valtchanov}, \& {Wuyts}}]{rosario2012}
{Rosario}, D.~J., {Santini}, P., {Lutz}, D., {et~al.} 2012, ArXiv e-prints

\bibitem[{{Rosario} {et~al.}(2019){Rosario}, {Togi}, {Burtscher}, {Davies},
  {Shimizu}, \& {Lutz}}]{rosario2019}
{Rosario}, D.~J., {Togi}, A., {Burtscher}, L., {et~al.} 2019, \apjl, 875, L8

\bibitem[{{Rovilos} {et~al.}(2012){Rovilos}, {Comastri}, {Gilli},
  {Georgantopoulos}, {Ranalli}, {Vignali}, {Lusso}, {Cappelluti}, {Zamorani},
  {Elbaz}, {Dickinson}, {Hwang}, {Charmandaris}, {Ivison}, {Merloni}, {Daddi},
  {Carrera}, {Brandt}, {Mullaney}, {Scott}, {Alexander}, {Del Moro},
  {Morrison}, {Murphy}, {Altieri}, {Aussel}, {Dannerbauer}, {Kartaltepe},
  {Leiton}, {Magdis}, {Magnelli}, {Popesso}, \& {Valtchanov}}]{rovilos2012}
{Rovilos}, E., {Comastri}, A., {Gilli}, R., {et~al.} 2012, \aap, 546, A58

\bibitem[{{Sabater} {et~al.}(2015){Sabater}, {Best}, \&
  {Heckman}}]{sabater2015}
{Sabater}, J., {Best}, P.~N., \& {Heckman}, T.~M. 2015, \mnras, 447, 110

\bibitem[{{Sales} {et~al.}(2010){Sales}, {Pastoriza}, \& {Riffel}}]{sales2010}
{Sales}, D.~A., {Pastoriza}, M.~G., \& {Riffel}, R. 2010, \apj, 725, 605

\bibitem[{{Santini} {et~al.}(2012){Santini}, {Rosario}, {Shao}, {Lutz},
  {Maiolino}, {Alexander}, {Altieri}, {Andreani}, {Aussel}, {Bauer}, {Berta},
  {Bongiovanni}, {Brandt}, {Brusa}, {Cepa}, {Cimatti}, {Daddi}, {Elbaz},
  {Fontana}, {F{\"o}rster Schreiber}, {Genzel}, {Grazian}, {Le Floc'h},
  {Magnelli}, {Mainieri}, {Nordon}, {P{\'e}rez Garcia}, {Poglitsch}, {Popesso},
  {Pozzi}, {Riguccini}, {Rodighiero}, {Salvato}, {Sanchez-Portal}, {Sturm},
  {Tacconi}, {Valtchanov}, \& {Wuyts}}]{santini2012}
{Santini}, P., {Rosario}, D.~J., {Shao}, L., {et~al.} 2012, \aap, 540, A109

\bibitem[{{Sarzi} {et~al.}(2007{\natexlab{a}}){Sarzi}, {Allard}, {Knapen}, \&
  {Mazzuca}}]{sarzi2007}
{Sarzi}, M., {Allard}, E.~L., {Knapen}, J.~H., \& {Mazzuca}, L.~M.
  2007{\natexlab{a}}, \mnras, 380, 949

\bibitem[{{Sarzi} {et~al.}(2007{\natexlab{b}}){Sarzi}, {Shields}, {Pogge}, \&
  {Martini}}]{sarzi2007b}
{Sarzi}, M., {Shields}, J.~C., {Pogge}, R.~W., \& {Martini}, P.
  2007{\natexlab{b}}, in Astronomical Society of the Pacific Conference Series,
  Vol. 373, The Central Engine of Active Galactic Nuclei, ed. L.~C. {Ho} \&
  J.~W. {Wang}, 643

\bibitem[{{Schartmann} {et~al.}(2010){Schartmann}, {Burkert}, {Krause},
  {Camenzind}, {Meisenheimer}, \& {Davies}}]{schartmann2010}
{Schartmann}, M., {Burkert}, A., {Krause}, M., {et~al.} 2010, \mnras, 403, 1801

\bibitem[{{Schartmann} {et~al.}(2009){Schartmann}, {Meisenheimer}, {Klahr},
  {Camenzind}, {Wolf}, \& {Henning}}]{schartmann2009}
{Schartmann}, M., {Meisenheimer}, K., {Klahr}, H., {et~al.} 2009, \mnras, 393,
  759

\bibitem[{{Schawinski} {et~al.}(2015){Schawinski}, {Koss}, {Berney}, \&
  {Sartori}}]{schawinski2015}
{Schawinski}, K., {Koss}, M., {Berney}, S., \& {Sartori}, L.~F. 2015, \mnras,
  451, 2517

\bibitem[{{Schipani} {et~al.}(2020){Schipani}, {Campana}, {Claudi}, {Aliverti},
  {Baruffolo}, {Ben-Ami}, {Biondi}, {Capasso}, {Cosentino}, {D'Alessio},
  {D'Avanzo}, {Hershko}, {Kuncarayakti}, {Landoni}, {Munari}, {Pignata},
  {Rubin}, {Scuderi}, {Vitali}, {Young}, {Achr{\'e}n}, {Araiza-Duran},
  {Arcavi}, {Brucalassi}, {Bruch}, {Cappellaro}, {Colapietro}, {Della Valle},
  {De Pascale}, {Di Benedetto}, {D'Orsi}, {Gal-Yam}, {Genoni}, {Hernandez},
  {Kotilainen}, {Li Causi}, {Mattila}, {Radhakrishnan}, {Rappaport}, {Ricci},
  {Riva}, {Salasnich}, {Savarese}, {Smartt}, {Sanchez}, {Stritzinger},
  {Ventura}, {Pasquini}, {Sch{\"o}ller}, {Ka{\"u}fl}, {Accardo}, {Mehrgan}, \&
  {Pompei}}]{schipani2020}
{Schipani}, P., {Campana}, S., {Claudi}, R., {et~al.} 2020, in Society of
  Photo-Optical Instrumentation Engineers (SPIE) Conference Series, Vol. 11447,
  Society of Photo-Optical Instrumentation Engineers (SPIE) Conference Series,
  1144709

\bibitem[{{Schnorr-M{\"u}ller} {et~al.}(2016){Schnorr-M{\"u}ller}, {Davies},
  {Korista}, {Burtscher}, {Rosario}, {Storchi-Bergmann}, {Contursi}, {Genzel},
  {Graci{\'a}-Carpio}, {Hicks}, {Janssen}, {Koss}, {Lin}, {Lutz},
  {Maciejewski}, {M{\"u}ller-S{\'a}nchez}, {Orban de Xivry}, {Riffel},
  {Riffel}, {Schartmann}, {Sternberg}, {Sturm}, {Tacconi}, {Veilleux}, \&
  {Ulrich}}]{schnorrmueller2016}
{Schnorr-M{\"u}ller}, A., {Davies}, R.~I., {Korista}, K.~T., {et~al.} 2016,
  \mnras, 462, 3570

\bibitem[{{Schulze} {et~al.}(2019){Schulze}, {Silverman}, {Daddi},
  {Rujopakarn}, {Liu}, {Schramm}, {Mainieri}, {Imanishi}, {Hirschmann}, \&
  {Jahnke}}]{schulze2019}
{Schulze}, A., {Silverman}, J.~D., {Daddi}, E., {et~al.} 2019, \mnras, 488,
  1180

\bibitem[{{Scoville} {et~al.}(1983){Scoville}, {Becklin}, {Young}, \&
  {Capps}}]{scoville1983}
{Scoville}, N.~Z., {Becklin}, E.~E., {Young}, J.~S., \& {Capps}, R.~W. 1983,
  \apj, 271, 512

\bibitem[{{Shangguan} {et~al.}(2020){Shangguan}, {Ho}, {Bauer}, {Wang}, \&
  {Treister}}]{shangguan2020}
{Shangguan}, J., {Ho}, L.~C., {Bauer}, F.~E., {Wang}, R., \& {Treister}, E.
  2020, \apj, 899, 112

\bibitem[{{Shimizu} {et~al.}(2019){Shimizu}, {Davies}, {Lutz}, {Burtscher},
  {Lin}, {Baron}, {Davies}, {Genzel}, {Hicks}, {Koss}, {Maciejewski},
  {M{\"u}ller-S{\'a}nchez}, {de Xivry}, {Price}, {Ricci}, {Riffel}, {Riffel},
  {Rosario}, {Schartmann}, {Schnorr-M{\"u}ller}, {Sternberg}, {Sturm},
  {Storchi-Bergmann}, {Tacconi}, \& {Veilleux}}]{shimizu2019}
{Shimizu}, T.~T., {Davies}, R.~I., {Lutz}, D., {et~al.} 2019, \mnras, 2449

\bibitem[{{Siebenmorgen} {et~al.}(2004){Siebenmorgen}, {Kr{\"u}gel}, \&
  {Spoon}}]{siebenmorgen2004}
{Siebenmorgen}, R., {Kr{\"u}gel}, E., \& {Spoon}, H.~W.~W. 2004, \aap, 414, 123

\bibitem[{{Silverman} {et~al.}(2009){Silverman}, {Lamareille}, {Maier},
  {Lilly}, {Mainieri}, {Brusa}, {Cappelluti}, {Hasinger}, {Zamorani},
  {Scodeggio}, {Bolzonella}, {Contini}, {Carollo}, {Jahnke}, {Kneib}, {Le
  F{\`e}vre}, {Merloni}, {Bardelli}, {Bongiorno}, {Brunner}, {Caputi},
  {Civano}, {Comastri}, {Coppa}, {Cucciati}, {de la Torre}, {de Ravel},
  {Elvis}, {Finoguenov}, {Fiore}, {Franzetti}, {Garilli}, {Gilli}, {Iovino},
  {Kampczyk}, {Knobel}, {Kova{\v{c}}}, {Le Borgne}, {Le Brun}, {Mignoli},
  {Pello}, {Peng}, {Perez Montero}, {Ricciardelli}, {Tanaka}, {Tasca},
  {Tresse}, {Vergani}, {Vignali}, {Zucca}, {Bottini}, {Cappi}, {Cassata},
  {Fumana}, {Griffiths}, {Kartaltepe}, {Koekemoer}, {Marinoni}, {McCracken},
  {Memeo}, {Meneux}, {Oesch}, {Porciani}, \& {Salvato}}]{silverman2009}
{Silverman}, J.~D., {Lamareille}, F., {Maier}, C., {et~al.} 2009, \apj, 696,
  396

\bibitem[{{Sim{\~o}es Lopes} {et~al.}(2007){Sim{\~o}es Lopes},
  {Storchi-Bergmann}, {de F{\'a}tima Saraiva}, \& {Martini}}]{simoeslopes2007}
{Sim{\~o}es Lopes}, R.~D., {Storchi-Bergmann}, T., {de F{\'a}tima Saraiva}, M.,
  \& {Martini}, P. 2007, \apj, 655, 718

\bibitem[{{Smette} {et~al.}(2015){Smette}, {Sana}, {Noll}, {Horst}, {Kausch},
  {Kimeswenger}, {Barden}, {Szyszka}, {Jones}, {Gallenne}, {Vinther},
  {Ballester}, \& {Taylor}}]{smette2015}
{Smette}, A., {Sana}, H., {Noll}, S., {et~al.} 2015, \aap, 576, A77

\bibitem[{{Smith} {et~al.}(2007){Smith}, {Draine}, {Dale}, {Moustakas},
  {Kennicutt}, {Helou}, {Armus}, {Roussel}, {Sheth}, {Bendo}, {Buckalew},
  {Calzetti}, {Engelbracht}, {Gordon}, {Hollenbach}, {Li}, {Malhotra},
  {Murphy}, \& {Walter}}]{smith2007}
{Smith}, J.~D.~T., {Draine}, B.~T., {Dale}, D.~A., {et~al.} 2007, \apj, 656,
  770

\bibitem[{{Somerville} {et~al.}(2008){Somerville}, {Hopkins}, {Cox},
  {Robertson}, \& {Hernquist}}]{somerville2008}
{Somerville}, R.~S., {Hopkins}, P.~F., {Cox}, T.~J., {Robertson}, B.~E., \&
  {Hernquist}, L. 2008, \mnras, 391, 481

\bibitem[{{Speagle} {et~al.}(2014){Speagle}, {Steinhardt}, {Capak}, \&
  {Silverman}}]{speagle2014}
{Speagle}, J.~S., {Steinhardt}, C.~L., {Capak}, P.~L., \& {Silverman}, J.~D.
  2014, \apjs, 214, 15

\bibitem[{{Stasi{\'n}ska} {et~al.}(2008){Stasi{\'n}ska}, {Vale Asari}, {Cid
  Fernandes}, {Gomes}, {Schlickmann}, {Mateus}, {Schoenell}, {Sodr{\'e}}, \&
  {Seagal Collaboration}}]{stasinska2008}
{Stasi{\'n}ska}, G., {Vale Asari}, N., {Cid Fernandes}, R., {et~al.} 2008,
  \mnras, 391, L29

\bibitem[{{Storchi-Bergmann} {et~al.}(2000){Storchi-Bergmann}, {Raimann},
  {Bica}, \& {Fraquelli}}]{storchibergmann2000}
{Storchi-Bergmann}, T., {Raimann}, D., {Bica}, E.~L.~D., \& {Fraquelli}, H.~A.
  2000, \apj, 544, 747

\bibitem[{{Storchi-Bergmann} {et~al.}(2012){Storchi-Bergmann}, {Riffel},
  {Riffel}, {Diniz}, {Borges Vale}, \& {McGregor}}]{storchibergmann2012}
{Storchi-Bergmann}, T., {Riffel}, R.~A., {Riffel}, R., {et~al.} 2012, \apj,
  755, 87

\bibitem[{{Su} {et~al.}(2019){Su}, {Hopkins}, {Hayward}, {Ma},
  {Faucher-Gigu{\`e}re}, {Kere{\v{s}}}, {Orr}, {Chan}, \& {Robles}}]{su2019}
{Su}, K.-Y., {Hopkins}, P.~F., {Hayward}, C.~C., {et~al.} 2019, \mnras, 487,
  4393

\bibitem[{{Terlevich} {et~al.}(1990){Terlevich}, {Diaz}, \&
  {Terlevich}}]{terlevich1990}
{Terlevich}, E., {Diaz}, A.~I., \& {Terlevich}, R. 1990, \mnras, 242, 271

\bibitem[{{Thompson} {et~al.}(2005){Thompson}, {Quataert}, \&
  {Murray}}]{thompson2005}
{Thompson}, T.~A., {Quataert}, E., \& {Murray}, N. 2005, \apj, 630, 167

\bibitem[{{Tran}(1995)}]{tran1995}
{Tran}, H.~D. 1995, \apj, 440, 597

\bibitem[{{Treister} {et~al.}(2012){Treister}, {Schawinski}, {Urry}, \&
  {Simmons}}]{treister2012}
{Treister}, E., {Schawinski}, K., {Urry}, C.~M., \& {Simmons}, B.~D. 2012,
  \apjl, 758, L39

\bibitem[{{Vale Asari} {et~al.}(2009){Vale Asari}, {Stasi{\'n}ska}, {Cid
  Fernandes}, {Gomes}, {Schlickmann}, {Mateus}, \& {Schoenell}}]{asari2009}
{Vale Asari}, N., {Stasi{\'n}ska}, G., {Cid Fernandes}, R., {et~al.} 2009,
  \mnras, 396, L71

\bibitem[{{Vazdekis} {et~al.}(2016){Vazdekis}, {Koleva}, {Ricciardelli},
  {R{\"o}ck}, \& {Falc{\'o}n-Barroso}}]{vazdekis2016}
{Vazdekis}, A., {Koleva}, M., {Ricciardelli}, E., {R{\"o}ck}, B., \&
  {Falc{\'o}n-Barroso}, J. 2016, \mnras, 463, 3409

\bibitem[{{Veilleux} \& {Osterbrock}(1987)}]{veilleux1987}
{Veilleux}, S. \& {Osterbrock}, D.~E. 1987, \apjs, 63, 295

\bibitem[{{Vernet} {et~al.}(2011){Vernet}, {Dekker}, {D'Odorico}, {Kaper},
  {Kjaergaard}, {Hammer}, {Randich}, {Zerbi}, {Groot}, {Hjorth}, {Guinouard},
  {Navarro}, {Adolfse}, {Albers}, {Amans}, {Andersen}, {Andersen}, {Binetruy},
  {Bristow}, {Castillo}, {Chemla}, {Christensen}, {Conconi}, {Conzelmann},
  {Dam}, {de Caprio}, {de Ugarte Postigo}, {Delabre}, {di Marcantonio},
  {Downing}, {Elswijk}, {Finger}, {Fischer}, {Flores}, {Fran{\c c}ois},
  {Goldoni}, {Guglielmi}, {Haigron}, {Hanenburg}, {Hendriks}, {Horrobin},
  {Horville}, {Jessen}, {Kerber}, {Kern}, {Kiekebusch}, {Kleszcz}, {Klougart},
  {Kragt}, {Larsen}, {Lizon}, {Lucuix}, {Mainieri}, {Manuputy}, {Martayan},
  {Mason}, {Mazzoleni}, {Michaelsen}, {Modigliani}, {Moehler}, {M{\o}ller},
  {Norup S{\o}rensen}, {N{\o}rregaard}, {P{\'e}roux}, {Patat}, {Pena}, {Pragt},
  {Reinero}, {Rigal}, {Riva}, {Roelfsema}, {Royer}, {Sacco}, {Santin},
  {Schoenmaker}, {Spano}, {Sweers}, {Ter Horst}, {Tintori}, {Tromp}, {van
  Dael}, {van der Vliet}, {Venema}, {Vidali}, {Vinther}, {Vola}, {Winters},
  {Wistisen}, {Wulterkens}, \& {Zacchei}}]{vernet2011}
{Vernet}, J., {Dekker}, H., {D'Odorico}, S., {et~al.} 2011, \aap, 536, A105

\bibitem[{{Vernet} \& {Mason}(2009)}]{vernet2009}
{Vernet}, J. \& {Mason}, E. 2009, X-shooter User Manual, 2nd edn., ESO,
  Karl-Schwarzschild-Str. 2, 85748 Garching bei M{\"u}nchen, Germany

\bibitem[{{V{\'e}ron-Cetty} \& {V{\'e}ron}(2010)}]{veroncetty2010}
{V{\'e}ron-Cetty}, M. \& {V{\'e}ron}, P. 2010, \aap, 518, A10+

\bibitem[{{Vogelsberger} {et~al.}(2014){Vogelsberger}, {Genel}, {Springel},
  {Torrey}, {Sijacki}, {Xu}, {Snyder}, {Nelson}, \&
  {Hernquist}}]{vogelsberger2014}
{Vogelsberger}, M., {Genel}, S., {Springel}, V., {et~al.} 2014, \mnras, 444,
  1518

\bibitem[{{Vollmer} {et~al.}(2008){Vollmer}, {Beckert}, \&
  {Davies}}]{vollmer2008}
{Vollmer}, B., {Beckert}, T., \& {Davies}, R.~I. 2008, \aap, 491, 441

\bibitem[{{Vollmer} \& {Davies}(2013)}]{vollmer2013}
{Vollmer}, B. \& {Davies}, R.~I. 2013, \aap, 556, A31

\bibitem[{{Volonteri} {et~al.}(2015){Volonteri}, {Capelo}, {Netzer},
  {Bellovary}, {Dotti}, \& {Governato}}]{volonteri2015}
{Volonteri}, M., {Capelo}, P.~R., {Netzer}, H., {et~al.} 2015, \mnras, 449,
  1470

\bibitem[{{Wada} \& {Norman}(2002)}]{wada2002}
{Wada}, K. \& {Norman}, C.~A. 2002, \apjl, 566, L21

\bibitem[{{Wada} {et~al.}(2009){Wada}, {Papadopoulos}, \& {Spaans}}]{wada2009}
{Wada}, K., {Papadopoulos}, P.~P., \& {Spaans}, M. 2009, \apj, 702, 63

\bibitem[{{Wegner} {et~al.}(2003){Wegner}, {Bernardi}, {Willmer}, {da Costa},
  {Alonso}, {Pellegrini}, {Maia}, {Chaves}, \& {Rit{\'e}}}]{wegner2003}
{Wegner}, G., {Bernardi}, M., {Willmer}, C.~N.~A., {et~al.} 2003, \aj, 126,
  2268

\bibitem[{{Wild} {et~al.}(2011){Wild}, {Groves}, {Heckman}, {Sonnentrucker},
  {Armus}, {Schiminovich}, {Johnson}, {Martins}, \& {Lamassa}}]{wild2011}
{Wild}, V., {Groves}, B., {Heckman}, T., {et~al.} 2011, \mnras, 410, 1593

\bibitem[{{Xue} {et~al.}(2010){Xue}, {Brandt}, {Luo}, {Rafferty}, {Alexander},
  {Bauer}, {Lehmer}, {Schneider}, \& {Silverman}}]{xue2010}
{Xue}, Y.~Q., {Brandt}, W.~N., {Luo}, B., {et~al.} 2010, \apj, 720, 368

\bibitem[{{Yesuf} \& {Ho}(2020)}]{yesuf2020}
{Yesuf}, H.~M. \& {Ho}, L.~C. 2020, \apj, 901, 42

\bibitem[{{Yuan} {et~al.}(2013){Yuan}, {Kewley}, \& {Richard}}]{yuan2013}
{Yuan}, T.~T., {Kewley}, L.~J., \& {Richard}, J. 2013, \apj, 763, 9

\bibitem[{{Zahid} {et~al.}(2011){Zahid}, {Kewley}, \& {Bresolin}}]{zahid2011}
{Zahid}, H.~J., {Kewley}, L.~J., \& {Bresolin}, F. 2011, \apj, 730, 137

\bibitem[{{Zhuang} {et~al.}(2021){Zhuang}, {Ho}, \& {Shangguan}}]{zhuang2020}
{Zhuang}, M.-Y., {Ho}, L.~C., \& {Shangguan}, J. 2021, \apj, 906, 38

\bibitem[{{Zinn} {et~al.}(2013){Zinn}, {Middelberg}, {Norris}, \&
  {Dettmar}}]{zinn2013}
{Zinn}, P.~C., {Middelberg}, E., {Norris}, R.~P., \& {Dettmar}, R.~J. 2013,
  \apj, 774, 66

\bibitem[{{Zubovas} {et~al.}(2013){Zubovas}, {Nayakshin}, {King}, \&
  {Wilkinson}}]{zubovas2013}
{Zubovas}, K., {Nayakshin}, S., {King}, A., \& {Wilkinson}, M. 2013, \mnras,
  433, 3079

\end{thebibliography}
